\pdfoutput=1
\documentclass[a4paper,11pt]{article}
\usepackage{jheppub}
\usepackage{amsmath,amssymb,booktabs,array,longtable,slashed,microtype,xurl,placeins,needspace}
\usepackage{tikz,pgfplots}
\usepackage{upgreek}
\pgfplotsset{compat=1.17}
\usepgfplotslibrary{groupplots}
\usetikzlibrary{arrows.meta,decorations.markings,decorations.pathmorphing}
\allowdisplaybreaks[2]

\newcommand{\as}{a_s}
\newcommand{\MSb}{\overline{\mathrm{MS}}}
\newcommand{\Tr}{\operatorname{tr}}
\newcommand{\eps}{\epsilon}
\let\CharmTableOfContents\tableofcontents
\renewcommand{\tableofcontents}{%
  \begingroup\renewcommand{\baselinestretch}{.95}\small
  \CharmTableOfContents\endgroup}
\title{Dimension-seven spectator effects in charm-hadron lifetimes at NLO}
\author{Chia-Wei Liu}
\affiliation{Hangzhou Institute for Advanced Study, University of Chinese Academy of Sciences, Hangzhou 310024, China}
\emailAdd{chiaweiliu@ucas.ac.cn}
\abstract{%
We compute the next-to-leading-order (NLO) QCD corrections to the
connected current--current part of the dimension-seven spectator coefficients in the heavy-quark expansion of
charm-hadron widths. The finite terms are fixed in NDR--$\MSb$, with the evanescent operators defined by the standard Greek projection and averaged over Hermitian-conjugate pairs, and with a finite subtraction that sets the finite mixing of the dimension-seven operators into the dimension-six ones to zero. In several channels the
one-loop term is comparable to  the LO coefficient, and two operator families absent at LO are generated.
Since the NLO matrix elements of the dimension-seven operators are not known, we estimate the effect of the coefficients on the widths with tree-level matrix elements from vacuum insertion for $D$ mesons and from the bag model for charmed baryons. The $1/m_c$ corrections to the matrix elements of the dimension-six QCD operators are treated by a naive expansion. Without their $O(\alpha_s)$ part the semileptonic width of the $D_s^+$ would be negative. The NLO dimension-seven correction raises the $D^+$ width by $0.78~\mathrm{ps}^{-1}$, so that its central value, slightly negative as published, becomes positive. At central inputs, it is about $1\%$ of the $D^0$ width, $10\%$ of the $D_s^+$ width and at most $2\%$ of the widths of the singly charmed baryons. Added to the published pole-scheme width, the $D^+$ correction gives $\tau(D^+)=1.41^{+\infty}_{-0.76}$~ps. The NLO term reduces the scale dependence of the nonleptonic dimension-seven contribution severalfold for the singly charmed baryons, whose matrix elements are evolved, but not uniformly for the $D$ mesons.}
\begin{document}
\maketitle
\raggedbottom
\clubpenalty=10000
\widowpenalty=10000

\section{Introduction}
\begin{figure}[htbp]
\centering
\includegraphics[width=\textwidth,trim=0 5 0 2,clip]{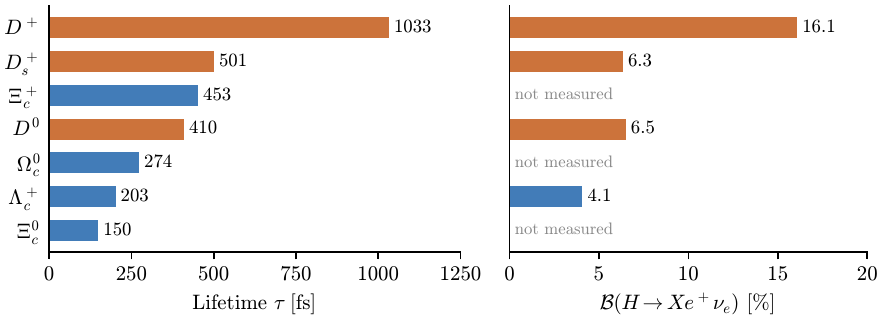}
\setlength{\abovecaptionskip}{3pt}
\caption{Measured lifetimes of the $D$ mesons (orange) and of the singly charmed baryons (blue),  and their inclusive electron branching fractions~\cite{ParticleDataGroup:2026,BESIII:2023inclusiveLc,Belle-II:2021cxx,Belle-II:2023eii,LHCb:2017knt,LHCb:2019ldj,Belle-II:2022ggx,LHCb:2021vll,LHCb:2018nfa,Belle-II:2022plj}. Uncertainties are not shown.}
\label{fig:charm-lifetime-ratios}
\end{figure}

Inclusive widths of heavy hadrons are described by the heavy-quark expansion
(HQE)~\cite{Shifman:1984wx,Shifman:1986mx,Bigi:1992su,Blok:1992hw,Blok:1992he},
an expansion in inverse powers of the heavy-quark mass, reviewed in
refs.~\cite{Lenz:2014jha,Albrecht:2024oyn}. At leading order the width of a
heavy hadron is that of the free heavy quark, independent of the light
degrees of freedom, so that all hadrons with the same single heavy quark would
have nearly the same lifetime, as is the case for the bottom
hadrons~\cite{ParticleDataGroup:2026}. For the charmed hadrons the
measured lifetimes differ by up to a factor of seven (figure~\ref{fig:charm-lifetime-ratios}), from about $150$~fs for
the $\Xi_c^0$ to about $1$~ps for the $D^+$~\cite{ParticleDataGroup:2026}.
In the HQE these differences are attributed to the spectator contributions,
$\mathbf C_6\cdot\langle\boldsymbol Q\rangle$, with $\boldsymbol Q$ the
four-quark operators of dimension six  and $\mathbf C_6$
their short-distance coefficients. They start at $1/m_c^3$, but they
are enhanced by the two-body phase space, a factor of order $16\pi^2$
relative to the three-body free-quark decay~\cite{Neubert:1996we}, and
destructive Pauli interference makes the $D^+$
long-lived~\cite{Guberina:1979xw,Bilic:1984nq,Shifman:1984wx}. The
interference is in fact so large that the
calculated
 $D^+$ width comes out negative~\cite{Gratrex:2022xpm}.

For the $\Omega_c^0$, by contrast, the HQE has proved predictive. At
dimension six the strange quark produced in $c\to s$ transitions interferes
constructively with the valence strange quarks of the $\Omega_c^0$, which was
therefore expected to be the shortest-lived charmed baryon,
$\tau(\Xi_c^+)>\tau(\Lambda_c^+)>\tau(\Xi_c^0)>\tau(\Omega_c^0)$~\cite{Guberina:1986gd,Cheng:1997xba},
and the fixed-target experiments WA89, E687 and FOCUS found central values
between $55$ and $86$~fs~\cite{WA89:1995lbz,E687:1995cvt,FOCUS:2003ylx}. In
March 2018 Cheng showed that the $1/m_c^4$ corrections to the spectator terms
reduce this interference and could make the $\Omega_c^0$ live longer than
the $\Lambda_c^+$~\cite{Cheng:2018talk,Cheng:2018rkz}. A few months later LHCb
measured $\tau(\Omega_c^0)=268\pm24\pm10\pm2$~fs in semileptonic $\Omega_b^-$
decays~\cite{LHCb:2018nfa}, nearly four times the previous world average, a
result confirmed with promptly produced $\Omega_c^0$~\cite{LHCb:2021vll}{,}
by Belle~II~\cite{Belle-II:2022plj}{ and in hadronic $\Omega_b^-$ decays~\cite{LHCb:2025oww}}. With the $\Xi_c^0$
lifetime also measured to be longer~\cite{LHCb:2019ldj,LHCb:2021vll,LHCb:2025oww}, the hierarchy is now
$\tau(\Xi_c^+)>\tau(\Omega_c^0)>\tau(\Lambda_c^+)>\tau(\Xi_c^0)$. Later
HQE analyses also    reproduce it~\cite{Gratrex:2022xpm, Cheng:2023jpz}, and the framework
has been extended to doubly charmed baryons~\cite{Cheng:2018mwu,Dulibic:2023jeu,Cheng:2026doubly}.

The two cases seem to tell very different stories. The $D^+$ suggests that
the HQE fails for charm, the $\Omega_c^0$ that it has real predictive power.
The subtlety lies at dimension seven. The dimension-seven four-quark
operators $\boldsymbol P$~\cite{Gabbiani:2003pq,Gabbiani:2004tp} contribute
$\mathbf C_7\cdot\langle\boldsymbol P\rangle$, which in some channels is as
large as $\mathbf C_6\cdot\langle\boldsymbol Q\rangle$ and, for both the $D^+$
and the $\Omega_c^0$, of opposite sign. For the $\Omega_c^0$ this produces the
observed hierarchy. For the $D^+$ it compensates more than a third of the
Pauli interference but leaves a negative central value,
$-0.07~\mathrm{ps}^{-1}$~\cite{Gratrex:2022xpm}. The two terms are not known to
the same accuracy. $\mathbf C_6$ is known at
NLO~\cite{Ciuchini:2001vx,Beneke:2002rj,Franco:2002fc,Lenz:2013aua,King:2021xqp},
and at NNLO for the meson lifetime ratios~\cite{Moretti:2026lifetimeRatios},
whereas $\mathbf C_7$ has been taken at LO in all analyses that include it. This mismatch is a
plausible origin of the negative $D^+$ width, and leaves the LO explanation of the $\Omega_c^0$ success untested at NLO. It is therefore important to take
$\mathbf C_7$ to NLO.

In this paper we compute the connected current--current part of the
dimension-seven spectator coefficients at NLO in QCD. Section~\ref{sec:method} summarizes the calculation and maps the coefficients onto bottom decays, and
section~\ref{sec:scope} compares the coefficients with LO and combines them with hadronic
inputs for the meson, baryon and semileptonic results.   The dimension-seven Wilson coefficients are given in
appendix~\ref{app:charm-analytic-coefficients} and in the ancillary
Mathematica file \texttt{charm\_wilson\_coefficients.m}. The ancillary files also contain a notebook with examples of its use and the page \texttt{charm\_nlo\_diagrams.html}, which calculates the two-loop forward amplitudes of the full-QCD side and the one-loop operator insertions diagram by diagram, for massless quarks in strictly forward kinematics.\footnote{The ancillary codes were organized with ChatGPT Astra, Claude Opus~5.5 and Claude Fable~5.2.}

\section{Outline of the NLO calculation}
\label{sec:method}
For a current--current channel with CKM factor $\lambda$, use
${\rm L}=1-\gamma_5$ and ${\rm R}=1+\gamma_5$ to write
\begin{equation}
\begin{aligned}
 \mathcal H_{\rm eff}&=\frac{G_F\lambda}{\sqrt2}(c_1W_1+c_2W_2),\\
 W_1&=(\bar q_{1a}\gamma_\mu{\rm L} c_a)(\bar q_{2b}\gamma^\mu{\rm L} q_{3b}),\qquad
 W_2=(\bar q_{1a}\gamma_\mu{\rm L} c_b)(\bar q_{2b}\gamma^\mu{\rm L} q_{3a}).
\end{aligned}
\label{eq:weak}
\end{equation}
The width is the normalized forward matrix element of
$2\,\mathrm{Im}\,i\int d^4z\,T\{ \mathcal H_{\rm eff} ^\dagger(z) \mathcal H_{\rm eff}  (0)\}$.
With $M=m_c^{\rm pole}$,  $\langle O\rangle_H=\langle H|O|H\rangle/(2M_H)$ and $H$ the charmed hadron, the spectator contribution to the width is
\begin{eqnarray}
 \Gamma_{\rm spec}(H)&=&\sum_{\lambda,q}\frac{G_F^2|\lambda|^2M^2}{6\pi}
 \sum_{w=22,12,11}\omega_w
 \left[\mathbf C_{6,w}^q\langle\boldsymbol Q^q\rangle_H+
 \mathbf C_{7,w}^q\langle\boldsymbol P^q\rangle_H\right],\label{eq:width}
 \\
 (\omega_{22},\omega_{12},\omega_{11})&=& (c_1^2,c_1c_2,c_2^2),\qquad
 \mathbf C_n=\mathbf C_n^{(0)}+\as\mathbf C_n^{(1)},\qquad
 \as=\frac{\alpha_s(\mu)}{4\pi},\qquad l=\log\frac{\mu^2}{M^2}. \nonumber 
\end{eqnarray}
Here $\lambda=V_{cq_1}^*V_{q_2q_3}$ runs over the nonleptonic channels
$c\to q_1q_2\bar q_3$, namely $c\to su\bar d$, $su\bar s$, $du\bar d$ and
$du\bar s$, whose CKM factors are all different, and $q$ is the light quark
retained as an operator field. Retaining $q_1$,
$q_2$ or $q_3$ leaves the cut-mass pairs $(m_2,m_3)$, $(m_1,m_3)$ or
$(m_1,m_2)$, respectively. The mixed row contains both weak insertion orders. The retained-$q_2$
coefficients follow from those of $q_1$ by exchanging $c_1\leftrightarrow c_2$ and
$m_{q_1}\leftrightarrow m_{q_2}$. The same coefficients
apply to mesons and baryons. The coefficients ${\bf C}_n$ depend on the channel, the weak row $w$ and the retained flavour $q$, and the operators on $q$. We suppress these labels where they are not needed. The semileptonic channels $c\to q\ell^+\nu_\ell$ enter in the same way, with $\lambda=V_{cq}^*$, $\omega_w\to1$. 
The cut-mass ratio $z=(m_s/M)^2$ is kept exact, while $m_u=m_d=0$. The terms quadratic in
the retained light-quark mass are dropped, however, so for $q=s$ the $z$ dependence of the
coefficients is consistent at dimension seven but not complete.

\begin{figure}[!b]
\centering
\scalebox{0.918}{%
\begin{tikzpicture}[x=1bp,y=1bp,line cap=round,line join=round,
  fermion/.style={postaction={decorate},decoration={markings,
    mark=at position #1 with {\arrow{Stealth[length=2.15pt,width=1.8pt,inset=0.55pt]}}}},
  fermion/.default=0.54]
\definecolor{faheavy}{RGB}{36,78,110}
\definecolor{fagluon}{RGB}{172,48,38}
\definecolor{favertex}{RGB}{160,102,25}
\node[inner sep=0pt,anchor=south west] at (4.06,93.9)
  {\includegraphics[viewport=0 93.9 128.37 160.9,clip]{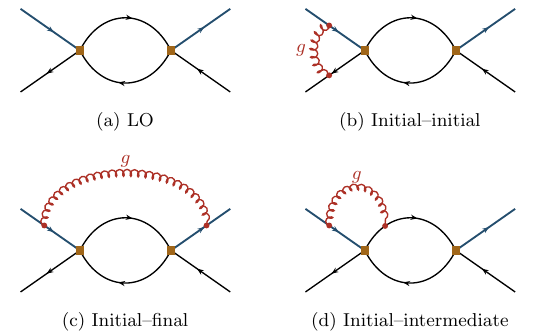}};
\node[inner sep=0pt,anchor=south west] at (124.23,93.9)
  {\includegraphics[viewport=128.37 93.9 256.74 160.9,clip]{figures/forward_amplitudes.pdf}};
\node[inner sep=0pt,anchor=south west] at (-60.14,0)
  {\includegraphics[viewport=0 0 128.37 93.9,clip]{figures/forward_amplitudes.pdf}};
\node[inner sep=0pt,anchor=south west] at (60.03,0)
  {\includegraphics[viewport=128.37 0 256.74 93.9,clip]{figures/forward_amplitudes.pdf}};
\begin{scope}[shift={(256.8,40.79)}]
 \draw[faheavy,line width=0.9bp,fermion] (-50.33,19.94) -- (-21.84,0);
 \draw[faheavy,line width=0.9bp,fermion] (21.84,0) -- (50.33,19.94);
 \draw[line width=0.65bp,fermion] (-21.84,0) -- (-50.33,-19.94);
 \draw[line width=0.65bp,fermion] (50.33,-19.94) -- (21.84,0);
 \draw[line width=0.65bp,fermion=0.3] (-21.84,0) .. controls (-11.40,20.89) and (11.40,20.89) .. (21.84,0);
 \draw[line width=0.65bp,fermion=0.3] (21.84,0) .. controls (11.40,-20.89) and (-11.40,-20.89) .. (-21.84,0);
 \draw[fagluon,line width=0.65bp,decorate,
   decoration={coil,amplitude=2.6bp,segment length=3.9bp,aspect=0.5}] (0,15.67) -- (0,-15.67);
 \fill[fagluon] (0,15.67) circle (1.41bp) (0,-15.67) circle (1.41bp);
 \node[fagluon,anchor=west,font=\fontsize{9}{11}\selectfont] at (4,0) {$g$};
 \fill[favertex] (-23.83,-1.99) rectangle (-19.85,1.99) (19.85,-1.99) rectangle (23.83,1.99);
 \node[anchor=base,font=\fontsize{9}{11}\selectfont] at (0,-36.6) {(e) Intermediate--intermediate};
\end{scope}
\end{tikzpicture}}
\caption{Representative uncut forward amplitudes: (a) the LO amplitude
  and (b,c,d,e) two-loop amplitudes. Fermion flavours are
left unspecified. Squares denote effective weak vertices and red dots
denote gluon vertices. Other attachments and self-energies are not shown.}
\label{fig:sl-amplitude-topologies}
\end{figure}

The operator vectors in eq.~\eqref{eq:width} form the full-QCD basis, renormalized at scale $\mu$ in the NDR--$\MSb$ scheme with the finite subtraction described below and with $m_q=\overline m_q(\mu)$:
\begin{equation}
 \boldsymbol Q=(Q,Q_S,T,T_S)^T,\qquad
 \boldsymbol P=(P_1^1,\ldots,P_5^1,P_1^8,\ldots,P_5^8)^T.
 \label{eq:uv-basis}
\end{equation}
\enlargethispage{2\baselineskip}
With $\sigma^{\mu\nu}=i[\gamma^\mu,\gamma^\nu]/2$, the singlet operators are
\begin{equation}
\begin{aligned}
 Q^q&=(\bar c\gamma_\mu{\rm L} q)(\bar q\gamma^\mu{\rm L} c),&
 Q_S^q&=(\bar c{\rm L} q)(\bar q{\rm R}c),\\
 P_1^{q,1}&=\frac{1}{2M^2}\bigl[(\bar c\overleftarrow D_\rho\gamma_\mu{\rm L} D^\rho q)
                    (\bar q\gamma^\mu{\rm L} c)+\mathrm{h.c.}\bigr],\\
 P_2^{q,1}&=-\frac{1}{M^2}\bigl[(\bar c\overleftarrow D_\rho{\rm L} D^\rho q)
                    (\bar q{\rm R}c)+\mathrm{h.c.}\bigr],\\
 P_3^{q,1}&=\frac{m_q}{M}\bigl[(\bar c{\rm L} q)(\bar q{\rm L} c)+(\bar c{\rm R}q)(\bar q{\rm R}c)\bigr],\\
 P_4^{q,1}&=\frac{m_q}{M}\bigl[(\bar c_i{\rm L} c_j)(\bar q_j{\rm L} q_i)
                         +(\bar c_i{\rm R}c_j)(\bar q_j{\rm R}q_i)\bigr],\\
 P_5^{q,1}&=\frac1M\bigl[(\bar c\overleftarrow D_\mu)\sigma^{\mu\nu}c+
              \bar c\sigma^{\mu\nu}D_\mu c\bigr](\bar q\gamma_\nu{\rm L} q).
\end{aligned}
\label{eq:operators}
\end{equation}
The octet operators, with $\Tr(t^At^B)=\delta^{AB}/2$, $N_c=3$ and $C_F=4/3$, are
\begin{equation}
\begin{aligned}
 T^q&=(\bar c\gamma_\mu{\rm L}\,t^Aq)(\bar q\gamma^\mu{\rm L}\,t^Ac),\qquad
 T_S^q=(\bar c{\rm L}\,t^Aq)(\bar q{\rm R}\,t^Ac),\\
 P_1^{q,8}&=\frac{1}{2M^2}\bigl[(\bar c\overleftarrow D_\rho\gamma_\mu{\rm L}\,t^AD^\rho q)
                    (\bar q\gamma^\mu{\rm L}\,t^Ac)+\mathrm{h.c.}\bigr],\\
 P_2^{q,8}&=-\frac{1}{M^2}\bigl[(\bar c\overleftarrow D_\rho{\rm L}\,t^AD^\rho q)
                    (\bar q{\rm R}\,t^Ac)+\mathrm{h.c.}\bigr],\\
 P_3^{q,8}&=\frac{m_q}{M}\bigl[(\bar c{\rm L}\,t^Aq)(\bar q{\rm L}\,t^Ac)+(\bar c{\rm R}\,t^Aq)(\bar q{\rm R}\,t^Ac)\bigr],\\
 P_4^{q,8}&=\frac{m_q}{M}\bigl[(\bar c_i{\rm L} c_l)(\bar q_k{\rm L} q_j)
                         +(\bar c_i{\rm R}c_l)(\bar q_k{\rm R}q_j)\bigr]t^A_{ij}t^A_{kl},\\
 P_5^{q,8}&=\frac1M\bigl[(\bar c\overleftarrow D_\mu)\sigma^{\mu\nu}t^Ac+
              \bar c\sigma^{\mu\nu}t^AD_\mu c\bigr](\bar q\gamma_\nu{\rm L}\,t^Aq).
\end{aligned}
\label{eq:octet-operators}
\end{equation}%
Only $P_{1,2,3}$ have nonzero LO coefficients. 
At LO the loop of the two cut quarks (figure~\ref{fig:sl-amplitude-topologies}a) gives only $g^{\alpha\beta}K^2$ and $K^\alpha K^\beta$, with $K$ the sum of the heavy- and light-quark momenta. With the equations of motion (EOM), $\bar c\slashed K{\rm L} q\to M\bar c{\rm L} q-m_q\bar c{\rm R}q$, these give
 only the structures of $Q$, $Q_S$ and
$P_{1,2,3}$.   Hence $P_4$ and $P_5$ are
absent at LO.
With the factors of $1/M$, all $P_i$ have mass
dimension six, like $Q$.\footnote{In the original definition~\cite{Gabbiani:2003pq}, the mass term is written in the $(\bar bb)(\bar qq)$ order as ${\rm L}\otimes{\rm L}$, together with its colour-rearranged partner. We recombine it into $P_3$, in the heavy--light order of $Q$, and $P_4$, in the other order, and add the Hermitian conjugate ${\rm R}\otimes {\rm R}$, which has the same forward matrix element. We keep the $P_i$ at mass dimension six because they mix with the dimension-six operators, as shown in section~\ref{sec:evolution}, and include $1/M$ in them to save symbols.}  

\begin{figure}[!htb]
\centering
\includegraphics[width=.62\textwidth]{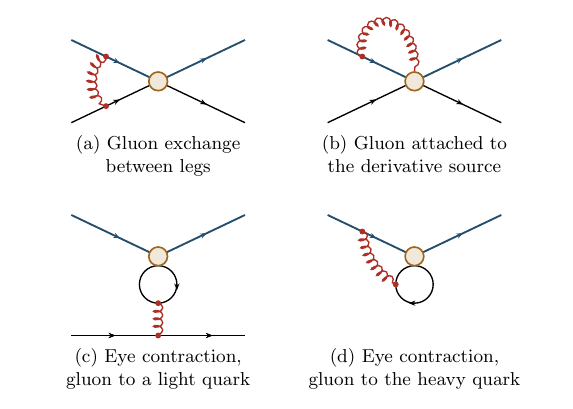}
\caption{(a, b) One-loop insertions of the Born sources (circles):
(a) gluon exchange between quark legs, (b) gluon attached to the
covariant-derivative source, new at dimension seven. (c, d) Eye contractions
of $P_i^q$, not calculated here. They belong to the dimension-seven two-quark
operators of Darwin type. Thick lines denote the heavy quark.}
\label{fig:born-source-contact}
\end{figure}

The NLO matching compares two-loop forward amplitudes with one-loop
insertions of the Born sources, the operator vertices of the $d$-dimensional LO
amplitude. In $d$ dimensions the full-theory amplitude is matched onto
\begin{equation}
{\cal A}_d={\bf C}\langle\boldsymbol O\rangle+{\bf C}_E\langle{\bf E}\rangle,
\label{eq:matching-full}
\end{equation}
where ${\cal A}_d$ is the forward amplitude of
$2\,\mathrm{Im}\,i\int d^4z\,T\{\mathcal H_{\rm eff}^\dagger(z)\mathcal H_{\rm eff}(0)\}$
between on-shell heavy- and light-quark states in $d=4-2\eps$ dimensions.
On the right-hand side,
$\boldsymbol O=(\boldsymbol Q,\boldsymbol P)$ are the physical operators of
eq.~\eqref{eq:uv-basis} with coefficients ${\bf C}=({\bf C}_6,{\bf C}_7)$, and
$\langle\,\rangle$ are their matrix elements between the same quark states for the matching. The evanescent operators ${\bf E}$, with
coefficients ${\bf C}_E$, collect the Dirac structures of ${\cal A}_d$ that do
not reduce to $\boldsymbol O$ in $d$ dimensions but vanish at $d=4$ at LO. We expand
each term in $\as$, ${\cal A}_d={\cal A}_d^{(0)}+\as{\cal A}_d^{(1)}$,
${\bf C}={\bf C}^{(0)}+\as{\bf C}^{(1)}$ and
$\langle\boldsymbol O\rangle=\langle\boldsymbol O\rangle^{(0)}+\as\langle\boldsymbol O\rangle^{(1)}$,
  and likewise ${\bf C}_E$ and
$\langle{\bf E}\rangle$. At $O(\as^0)$ eq.~\eqref{eq:matching-full} gives the Born
amplitude ${\cal A}_d^{(0)}={\bf C}^{(0)}\langle\boldsymbol O\rangle^{(0)}
+{\bf C}_E^{(0)}\langle{\bf E}\rangle^{(0)}$, and at $O(\as)$
\begin{equation}
{\cal A}_d^{(1)}={\bf C}^{(1)}\langle\boldsymbol O\rangle^{(0)}
+{\bf C}^{(0)}\langle\boldsymbol O\rangle^{(1)}
+{\bf C}_E^{(0)}\langle{\bf E}\rangle^{(1)}+{\bf C}_E^{(1)}\langle{\bf E}\rangle^{(0)}.
\label{eq:matching-as}
\end{equation} 
Figure~\ref{fig:sl-amplitude-topologies}
shows representative full-theory graphs, with gluons connecting
initial, final and intermediate fermion lines, and
figure~\ref{fig:born-source-contact} the local source insertions.
Covariant derivatives at dimension seven generate an additional
source-to-leg gluon contact (figure~\ref{fig:born-source-contact}b), absent for the nondifferentiated
dimension-six sources.

The full-theory side is the discontinuity of the forward amplitude
between on-shell heavy- and light-quark states,
figure~\ref{fig:sl-amplitude-topologies}. At LO, panel~(a), it is the cut through
the two intermediate quarks. At NLO the gluon is virtual or real depending on
where it is attached. In panel~(b) it connects the two initial lines at one weak
vertex and is a virtual correction to that vertex. In panel~(c) it connects an initial and a final line and spans
both weak vertices, so a cut through the intermediate quarks also crosses the
gluon. This three-particle cut is real-gluon emission. When the gluon is
attached to the light quark, a cut can instead cross a light-quark propagator
with the gluon uncut. Appendix~\ref{app:real-example} works out such a graph. In
panel~(d) it connects an initial line with an intermediate quark and has both
cuts. The two-particle cut is a virtual vertex correction, and the three-particle
cut is the interference of gluon emission from the two lines. In panel~(e) it
connects the two intermediate quarks inside the quark loop. Its
two-particle cuts are virtual exchanges between the two quarks, and its
three-particle cut is real emission from them. Mirror graphs and self-energies
complete the set. Each graph
is Taylor-expanded about vanishing retained light-quark momentum and mass $m_q$.
The constant term gives dimension six and the terms linear in either give
dimension seven. Since $P_5$ vanishes in strictly forward kinematics, a small
momentum transfer between the external legs is kept to first order and set to
zero only after projection. Real and virtual full-theory cuts and local effective-theory insertions
are expanded to the same soft order. We use  an
anticommuting $\gamma_5$ on open chains and $\MSb$ pole subtraction. The
weak and dimension-six prescriptions are those of
refs.~\cite{Buras:1989xd,Beneke:2002rj}. 

The one-loop matrix elements $\langle\boldsymbol O\rangle^{(1)}$ include the $\MSb$
field factors of the external quarks and the counterterm of the retained mass. With both UV and
IR divergences regulated dimensionally, they are 
\begin{equation}
\begin{aligned}
\langle\boldsymbol O\rangle^{(1)}={}&\bigl({\bf J}+{\bf F}\bigr)\langle\boldsymbol O\rangle^{(0)}
-\frac{1}{2\eps_{\rm UV}}\Bigl[\boldsymbol\gamma^{(0)}\langle\boldsymbol O\rangle^{(0)}
+\boldsymbol\gamma^{(0)}_E\langle{\bf E}\rangle^{(0)}\Bigr]\\
&+\frac{1}{\eps_{\rm IR}}\Bigl[\bigl({\bf J}_{\rm IR}+\eps\,{\bf J}_{\rm IR}'\bigr)
\langle\boldsymbol O\rangle^{(0)}+{\bf J}_{\rm IR}^E\langle{\bf E}\rangle^{(0)}\Bigr]+O(\eps).
\end{aligned}
\label{eq:insert-Laurent}
\end{equation}
Here $\boldsymbol\gamma^{(0)}$ is the one-loop anomalous-dimension matrix,
$d\langle\boldsymbol O\rangle/d\log\mu=-\as\boldsymbol\gamma^{(0)}\langle\boldsymbol O\rangle+O(\as^2)$~\cite{Buras:1989xd,Buras:2000if}.
The matrix $\boldsymbol\gamma^{(0)}_E$ is its mixing into ${\bf E}$, and ${\bf F}$ is a finite
subtraction.  The trees in the pole terms are
four-dimensional, so the $O(\eps)$ parts that multiply the UV poles are finite and belong
to ${\bf J}$. The
$\MSb$ counterterms, including the evanescent ones, remove the $1/\eps_{\rm UV}$ terms.
The IR term is the IR pole of the soft region of the graphs, in which the gluon momentum
is small compared with $M$, with the $O(\eps)$ part of its residue kept. Its terms with
evanescent trees appear identically in the full amplitude below and do not enter the
matching.
In blocks, the finite part is
\begin{equation}
\langle\boldsymbol O\rangle_{\rm fin}^{(1)}
 = 
\left[ 
\begin{array}{c}
\langle \boldsymbol Q\rangle_{\rm fin}^{(1)}\\
\langle \boldsymbol P\rangle_{\rm fin}^{(1)}
\end{array}
\right] 
= 
\left[
\begin{array}{cc}
{\bf J}_{66} & {\bf J}_{67} \\ 
{\bf J}_{76} & {\bf J}_{77}  
\end{array}
\right] 
\left[ 
\begin{array}{c}
	\langle \boldsymbol Q  \rangle  ^{(0)}\\
	\langle \boldsymbol P    \rangle ^{(0 )} 
\end{array}
\right]
=
{\bf J}
\langle
\boldsymbol O \rangle ^{(0 )}  
\,,
 \label{eq:J-definition}
\end{equation}
where 
$ 	\langle \boldsymbol Q  \rangle  ^{(0)} 
= [ 
	\langle Q \rangle  ^{(0)} , 
 \langle Q_S  \rangle  ^{(0)}  , 
 	\langle T  \rangle  ^{(0)} , 
 \langle T_S   \rangle  ^{(0)}   
] ^T 
$
and 
$\langle\boldsymbol P\rangle^{(0)}$ lists $\langle P_i^1\rangle^{(0)}$ and then
$\langle P_i^8\rangle^{(0)}$, $i=1,\ldots,5$, so ${\bf J}$ is a $14\times14$ matrix, with rows for
the inserted operators and columns for the trees.
The subtraction ${\bf F}$ has a single nonzero block, ${\bf F}_{76}$, fixed by
${\bf J}_{76}=0$, which sets to zero the finite mixing of the dimension-seven operators into the
dimension-six ones. This mixing would violate the power counting~\cite{Hovhannisyan:2025zev}. The $76$ block has no IR term, so its finite part
depends on $\mu$ only through the UV pole, and we set
$d{\bf F}_{76}/d\log\mu=-\boldsymbol\gamma^{(0)}_{76}$, so that ${\bf J}_{76}$ remains
zero at every $\mu$.

On the full-theory side, ${\cal A}_n^{(1)}$ is the renormalized two-loop forward
amplitude in Feynman gauge, including the weak counterterm, the field and mass counterterms of the
cut quarks and the $\MSb$ field factors $Z_cZ_q$ of the external heavy and retained
quarks. Here ${\cal A}_6^{(1)}$ and ${\cal A}_7^{(1)}$ are the zeroth-order and
first-order terms of the soft expansion. For the outgoing light quark $q$ the mass
is written in terms of the same retained mass $m_q=\overline m_q(\mu)$ as in $P_3$
and $P_4$. After the real--virtual cancellation it has no UV pole and takes the form
\begin{equation}
{\cal A}_n^{(1)}=\frac{1}{\eps_{\rm IR}}
\left({\cal A}_{n,{\rm IR}}^{(1)}+\eps\,{\cal A}_{n,{\rm IR}}^{(1)\prime}\right)
+{\cal A}_{n,{\rm fin}}^{(1)}+O(\eps).
\label{eq:amp-Laurent}
\end{equation}
The retained light-quark mass is kept to first order, so its one-loop self-energy
is scaleless and, in this expansion, $m_q^{\rm pole}=\overline m_q(\mu)\,(1-3C_F\as/\eps_{\rm IR})$. Writing
$m_q$ in ${\cal A}_7^{(0)}$ in this way adds to ${\cal A}_7^{(1)}$ the purely IR term
$-(3C_F/\eps_{\rm IR})\,C_{P_3}^{(0)}\langle P_3\rangle^{(0)}$. We have checked that, with this
term included, ${\cal A}_{n,{\rm IR}}^{(1)}$ and ${\cal A}_{n,{\rm IR}}^{(1)\prime}$ are cancelled
exactly by the corresponding IR terms of
${\bf C}^{(0)}\langle\boldsymbol O\rangle^{(1)}+{\bf C}_E^{(0)}\langle{\bf E}\rangle^{(1)}$.

The evanescent operators enter the EFT side in two roles. As inputs they are
inserted with the Born coefficients ${\bf C}_E^{(0)}$, and the UV poles of their one-loop
insertions, ${\bf C}_E^{(0)}\langle{\bf E}\rangle^{(1)}$, leave finite terms.
On the on-shell states used here these terms are local and
reduce to physical operators~\cite{Herrlich:1994kh}. We therefore insert
$\widehat{\bf E}={\bf E}+\as\,{\bf e}\,\boldsymbol O$ in place of ${\bf E}$, with the finite
matrix ${\bf e}$ chosen such that the non-IR finite part of the renormalized
${\bf C}_E^{(0)}\langle\widehat{\bf E}\rangle^{(1)}$ vanishes for $d\to4$.  With dimensional IR regularization,
${\bf C}_E^{(0)}\langle\widehat{\bf E}\rangle^{(1)}$ then contains only
$\eps/\eps_{\rm IR}$ terms, each an $O(\eps)$ part of the Born amplitude times an IR pole,
so it enters only the IR terms above and ${\bf C}_E$ does not enter the matching.
In practice this amounts to dropping the non-IR finite part of ${\bf C}_E^{(0)}\langle{\bf E}\rangle^{(1)}$ in the matching. 

As outputs, the evanescent operators appear as the trees
$\langle{\bf E}\rangle^{(0)}$ of eq.~\eqref{eq:insert-Laurent}, produced by the insertions of
the physical operators. These trees are $O(\eps)$, but they multiply the UV pole, so their $O(\eps)$ definition is part of the scheme, and NDR--$\MSb$ alone does not fix ${\bf J}$
uniquely~\cite{Herrlich:1994kh}. The replacement
\begin{equation} 
 {\bf E}\longrightarrow{\bf E}+\eps\,{\bf d}\,\boldsymbol O,
 \label{eq:evanescent-freedom}
\end{equation}
with a constant matrix ${\bf d}$, still gives $\langle{\bf E}\rangle^{(0)}=0$ at $d=4$, but the UV
pole of eq.~\eqref{eq:insert-Laurent} multiplies the $O(\eps)$ term and shifts ${\bf J}$ by
$\tfrac12\boldsymbol\gamma^{(0)}_E{\bf d}$. 
In $d$ dimensions an operator and its Fierz rearrangement are different operators, and
their difference is evanescent. For each of the ten inputs $P_i$, $i=1,\ldots,5$, in
both colours, we therefore use two representatives, the literal operator $P_i$ and its
literal Fierz form $\mathcal F[P_i]$. The two have the same physical UV pole, since
$\mathcal F[P_i]=P_i$ in four dimensions. We write $E(Z)$ for the evanescent part of the UV
pole of the one-loop insertion of $Z=P_i$ or $\mathcal F[P_i]$, and $d_Z$ for its freedom
under eq.~\eqref{eq:evanescent-freedom}, $E(Z)\to E(Z)+\eps\,d_ZO_Z$.
We fix the difference $d_{P_i}-d_{\mathcal F[P_i]}$ by Fierz symmetry and the common part by a rule for the
$d$-dimensional Dirac algebra, an ordered map $\Pi$ that writes a $d$-dimensional Dirac structure as
four-dimensional ones with coefficients through $O(\eps)$, the remainder being evanescent. We use it for every
$d$-dimensional Dirac structure that multiplies a pole, UV or IR, on both sides of the matching. All open-index gamma matrices are first
moved to the right of a contracted core using exact Clifford algebra. The momentum-free core
is then reduced by the Greek rules, with the right tails treated as part of the
spinors and restored afterwards. Only then are momenta, spinors and the external EOM
evaluated. Since the Greek projection does not commute with the EOM reduction, this order is
part of the definition. Explicitly, we demand
\begin{align}
 \langle P_i-\mathcal F[P_i]\rangle_{\rm fin}^{(1)}&=0,
 \label{eq:scheme-plus-minus}\\
 \frac12\left[\Pi({\cal E}_i)+
 \bigl(\Pi({\cal E}_i^{\dagger})\bigr)^{\dagger}\right]
 &=O(\eps^2),\qquad
 {\cal E}_i=\langle E(P_i)\rangle^{(0)}
 +\langle E(\mathcal F[P_i])\rangle^{(0)}. 
 \label{eq:scheme-residue}
\end{align}  
A common shift, $d_{P_i}=d_{\mathcal F[P_i]}$, cancels
in eq.~\eqref{eq:scheme-plus-minus} but moves the finite insertions of
both representatives equally. The second condition fixes it by setting to zero the
$O(\eps)$ physical image of ${\cal E}_i$. The average is needed because $\Pi$
alone is not Hermitian, eq.~\eqref{eq:pi-not-hermitian}.   The Greek rules and the
explicit insertions are given in appendices~\ref{sec:evanescent-conventions} and~\ref{app:uv-poles}. The dimension-six
inputs follow the same convention,  reproducing the standard NDR finite terms of ${\bf J}_{66}$.

With ${\bf J}$ fixed in this way and the IR terms removed from both sides, the
remaining finite parts, denoted by the subscript fin, give the matching equations
\begin{equation}
\begin{aligned}
 \mathbf C_6^{(1)}  \cdot  
 \langle \boldsymbol Q \rangle  ^{(0)} 
 &={\cal A}_{6,{\rm fin}}^{(1)}
 -
 \mathbf C_6^{(0)}
 \cdot 
 {\bf J}_{66} \cdot  \langle \boldsymbol Q \rangle  ^{(0)}   
 ,\\
 \mathbf C_7^{(1)}\cdot 	\langle \boldsymbol P    \rangle ^{(0 )}  
 &={\cal A}_{7,{\rm fin}}^{(1)}
 -
 \left( 
 \mathbf C_6^{(0)}
 \cdot 
 {\bf J}_{67} 	\cdot \langle \boldsymbol P   
 \rangle ^{(0 )}   
 +\mathbf C_7^{(0)}
 \cdot 
 {\bf J}_{77} \cdot 	\langle \boldsymbol P   
 \rangle ^{(0 )}  
 \right) .
\end{aligned}
\label{eq:matching}
\end{equation}
Eq.~\eqref{eq:matching} fixes the NLO coefficients $\mathbf C_6^{(1)}$ and $\mathbf C_7^{(1)}$. Its first line is the
usual dimension-six matching, with ${\bf J}_{66}$ in the standard NDR scheme. Since
${\bf J}_{76}=0$, no dimension-seven input enters it. In the second line the
dimension-six Born coefficients enter through ${\bf J}_{67}$, which is finite and has no
UV pole. The connected dimension-six coefficients from the same calculation agree with
refs.~\cite{Beneke:2002rj,Franco:2002fc}. We have checked that regulating the IR divergences with a gluon mass instead gives the same $\mathbf C^{(1)}$. 

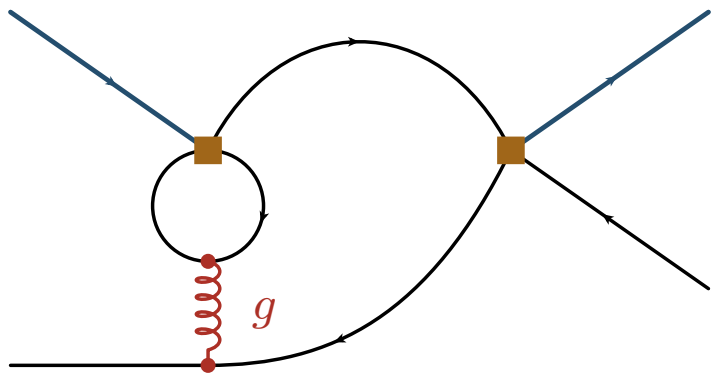
\begin{figure}[!b]
\centering
\begin{tikzpicture}[x=1.3bp,y=1.3bp,line cap=round,line join=round,
  fermion/.style={postaction={decorate},decoration={markings,
    mark=at position #1 with {\arrow{Stealth[length=2.15pt,width=1.8pt,inset=0.55pt]}}}},
  fermion/.default=0.54]
\definecolor{faheavy}{RGB}{36,78,110}
\definecolor{fagluon}{RGB}{172,48,38}
\definecolor{favertex}{RGB}{160,102,25}
 \draw[faheavy,line width=0.9bp,fermion] (-50.33,19.94) -- (-21.84,0);
 \draw[faheavy,line width=0.9bp,fermion] (21.84,0) -- (50.33,19.94);
 \draw[line width=0.65bp,fermion=0.5] (-21.84,0) .. controls (-11.40,20.89) and (11.40,20.89) .. (21.84,0);
 \draw[line width=0.65bp,fermion] (50.33,-19.94) -- (21.84,0);
 \draw[line width=0.65bp,fermion=0.45] (21.84,0) .. controls (10,-24) and (-4,-31) .. (-21.84,-31)
   -- (-50.33,-31);
 \draw[line width=0.65bp,fermion=0.3] (-21.84,0) arc[start angle=90,end angle=-270,radius=8];
 \draw[fagluon,line width=0.65bp,decorate,
   decoration={coil,amplitude=2.2bp,segment length=3.2bp,aspect=0.5}] (-21.84,-16) -- (-21.84,-31);
 \fill[fagluon] (-21.84,-16) circle (1.41bp) (-21.84,-31) circle (1.41bp);
 \node[fagluon,anchor=west,font=\fontsize{9}{11}\selectfont] at (-18.5,-23.5) {$g$};
 \fill[favertex] (-23.83,-1.99) rectangle (-19.85,1.99) (19.85,-1.99) rectangle (23.83,1.99);
\end{tikzpicture}
\caption{Eye contraction at a weak vertex, not computed here. A quark--antiquark pair from
one weak vertex closes into a loop, and the gluon from the loop attaches to the light-quark
line. Notation as in figure~\ref{fig:sl-amplitude-topologies}.}
\label{fig:weak-eye}
\end{figure}

We compute the connected current--current contractions between on-shell quark
states at first soft order, figures~\ref{fig:sl-amplitude-topologies}
and~\ref{fig:born-source-contact}a,~b. The penguin-like contributions, in which a quark pair
closes into a loop that emits a gluon, are omitted in charm. First, the eye contractions of the
operators $P_i^q$, figure~\ref{fig:born-source-contact}c,~d, are matrix elements of the
dimension-seven Darwin-type two-quark operators. Second, in the eye contractions at the weak vertex,
figure~\ref{fig:weak-eye}, the $s\bar s$ and $d\bar d$ loops cancel by the GIM mechanism
up to terms proportional to $V_{cb}^*V_{ub}$ or suppressed by powers of $z$. Third, the
penguin operators of ${\cal H}_{\rm eff}$ come with $V_{cb}^*V_{ub}$. All three have a different flavour structure, with a closed quark loop or a sum over light flavours, and are matched by operators of the same structure. The connected contractions thus form a closed set, whose IR poles cancel on their own.

The same coefficients describe bottom decays. With $c\to b$ in the heavy fields and
$q_1=c$, the channels $b\to c\bar ud$, $c\bar us$, $c\bar cs$ and $c\bar cd$ have
$(q_2,q_3)=(d,u),(s,u),(s,c),(d,c)$ and $\lambda=V_{cb}V_{q_3q_2}^*$. The cut pairs,
$(m_1,m_3)$ and $(m_1,m_2)$, are $(m_c,0)$ or $(m_c,m_c)$, that is, the charm
configurations $(z,0)$ and $(z,z)$. In the notation of appendix~\ref{sec:conventions}, we have 
\begin{equation}
 \begin{aligned}
  \bigl[C^{d\mid c\bar u}\bigr]_{b\text{-decays}}=\bigl[C^{s\mid c\bar u}\bigr]_{b\text{-decays}}
  &=\bigl[C^{u\mid s\bar d}\bigr]_{c\text{-decays}},\\
  \bigl[C^{u\mid cd}\bigr]_{b\text{-decays}}=\bigl[C^{u\mid cs}\bigr]_{b\text{-decays}}
  &=\bigl[C^{d\mid su}\bigr]_{c\text{-decays}},\\
  \bigl[C^{d\mid c\bar c}\bigr]_{b\text{-decays}}=\bigl[C^{s\mid c\bar c}\bigr]_{b\text{-decays}}
  &=\bigl[C^{u\mid s\bar s}\bigr]_{c\text{-decays}},
 \end{aligned}
\label{eq:bottom-map}
\end{equation}
for $P_{1,2,5}$, every weak row and colour, now with $M=m_b^{\rm pole}$, also
in the scale logarithms, and $z=(m_c^{\rm pole}/M)^2$. The coefficients are exact in $z$.
The equal-mass configuration requires $z<1/4$. The light quarks are massless, so
$P_{3,4}$ drop out. The coefficients $c_{1,2}$
and $\alpha_s$ are taken with five flavours, and $\mathbf C_7^{(1)}$ itself contains no
closed quark loop. The map is complete for $b\to c\bar ud$ and $c\bar us$, and hence for $B^-$ and, up to
$|V_{cd}|^2$ terms, for $\bar B^0$. For $b\to c\bar cs$ and $c\bar cd$ the $c\bar c$ eye
contraction and the penguin operators are also needed. The complete penguin-like contributions will be presented in a
forthcoming work.

\section{From coefficients to widths}
\label{sec:scope}
Section~\ref{sec:coefficients} compares the NLO coefficients with LO and evolves the hadronic matrix elements to $\mu=M$. Sections~\ref{sec:charm} and~\ref{sec:baryons} contract the coefficients with tree-level matrix elements, in the vacuum insertion approximation (VIA) for $D$ mesons and in the bag model for charmed baryons. For $D$ mesons we add the NLO correction to the published widths of ref.~\cite{Gratrex:2022xpm}, and for baryons we recalculate all terms from the inputs of ref.~\cite{Cheng:2023jpz}. Without the $O(\alpha_s)$ matrix elements of $P_{1,2,3}$ these are coefficient-level estimates. The hadronic inputs are given in appendix~\ref{app:hadronic}.

\subsection[NLO coefficients and evolution of the matrix elements]{\texorpdfstring{NLO coefficients and evolution of the matrix elements}{NLO coefficients and evolution of the matrix elements}}
\label{sec:coefficients}\label{sec:evolution}
\begin{figure}[!htb]
\centering
\includegraphics[width=\textwidth,height=0.83\textheight,keepaspectratio]{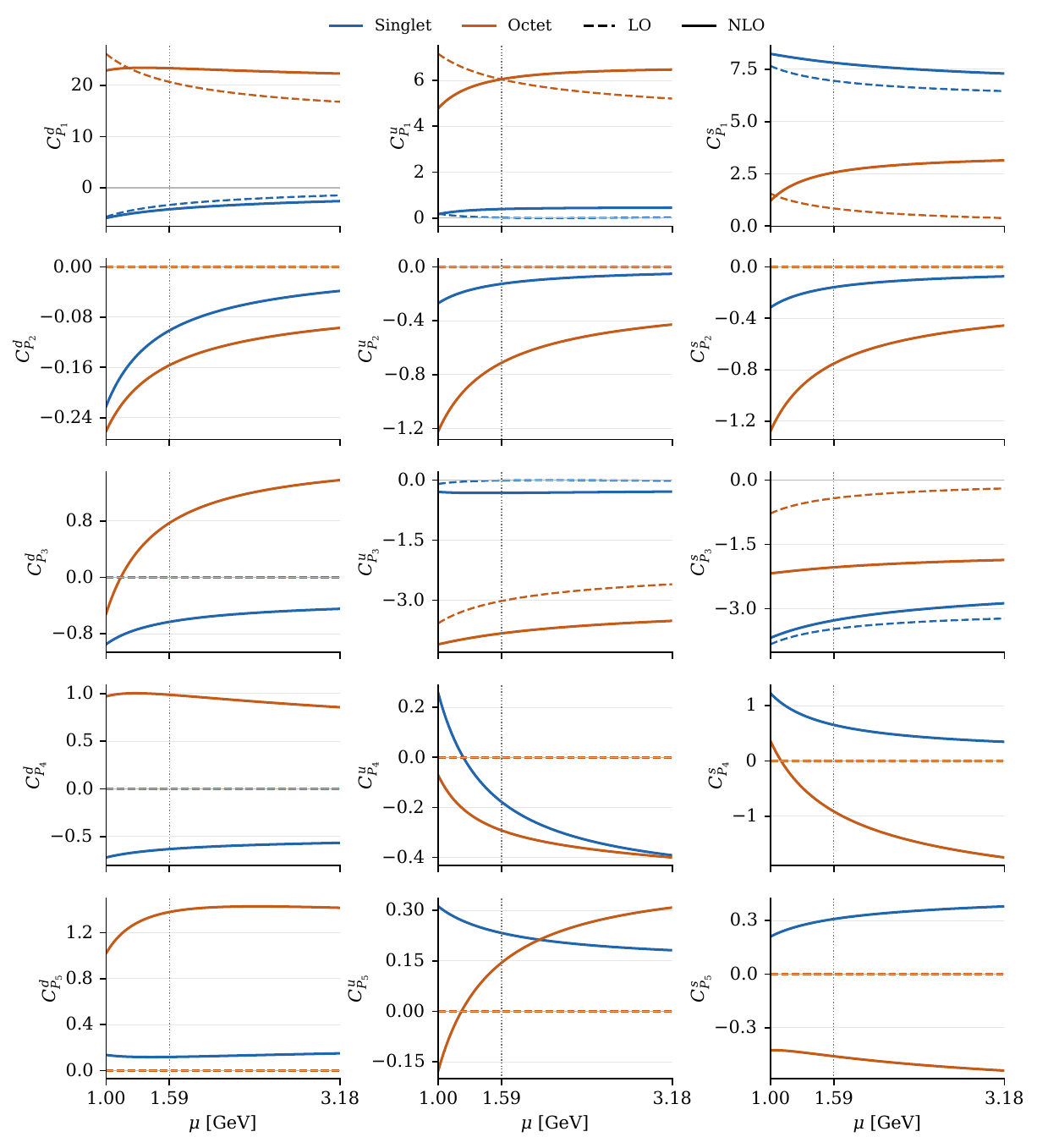}
\caption{Dimension-seven LO coefficients (dashed) and NLO coefficients
(solid) for $c\to su\bar d$, for both colours and all five families.}
\label{fig:new-coefficient-scales-dim7}
\end{figure}

We compare the NLO coefficients with LO for the Cabibbo-favoured channel $c\to su\bar d$.
With the weights also expanded, $\omega_w=\omega_w^{(0)}+\as\omega_w^{(1)}$, the summed coefficient reads 
\begin{equation}
 C_{P_i}^{(0)}=\sum_w\omega_w^{(0)}C_{P_i,w}^{(0)},\qquad
 C_{P_i}^{(1)}=\sum_w\bigl[\omega_w^{(1)}C_{P_i,w}^{(0)}+\omega_w^{(0)}C_{P_i,w}^{(1)}\bigr].
 \label{eq:weak-summed}
\end{equation}
For example, $\as\omega_{12}^{(1)}=c_1^{\rm NLO}c_2^{\rm NLO}-c_1^{\rm LO}c_2^{\rm LO}$. The product $\as^2\omega_w^{(1)}C_{P_i,w}^{(1)}$ is beyond NLO. The Born term keeps the products of the NLO coefficients, including the terms quadratic in $c_i^{\rm NLO}-c_i^{\rm LO}$. With the inputs of section~\ref{sec:baryons}, $M=1.59$~GeV, $\overline m_s(2~\mathrm{GeV})=0.09$~GeV, $\alpha_s(M)=0.34017$, $(c_1,c_2)_{\rm LO}=(1.22932,-0.45991)$ and
$(c_1,c_2)_{\rm NLO}=(1.17025,-0.35078)$,\footnote{Five-loop $\alpha_s$ from $\alpha_s(M_Z)=0.1181$ with RunDec~\cite{Herren:2017osy}, with $n_f=4$ below $\overline m_b(\overline m_b)=4.18$~GeV. The coefficients $c_{1,2}$ are evaluated in NDR~\cite{Buras:1989xd,Buchalla:1995vs} from $M_W=80.379$~GeV with this $\alpha_s$ at LO and NLO.} figure~\ref{fig:new-coefficient-scales-dim7} shows the scale
dependence of the weak-summed coefficients at LO and NLO, and table~\ref{tab:d-meson-coefficient-comparison} gives the values at $\mu=M$.

The NLO correction is largest where LO is colour suppressed. For $P_1^{u,1}$, $C_{P_1}^{u,(0)}$,
proportional to $(c_1+3c_2)^2$, is $0.015$ and $\as C_{P_1}^{u,(1)}=0.377$. For $P_1^{d,1}$
and $P_1^{s,1}$ the ratio $\as C_{P_1}^{(1)}/C_{P_1}^{(0)}$ is $0.26$ and $0.12$, respectively. $P_4$ and $P_5$ first
appear at NLO, and $C_{P_2}^{(0)}$ starts at $O(z^2)$, in agreement with
refs.~\cite{Gratrex:2022xpm,Cheng:2023jpz}.
Individual coefficient variation does not measure the scale uncertainty
of a physical width, which also depends on the hadronic matrix elements.

\begin{table}[tbp]
\centering\small
\setlength{\tabcolsep}{5pt}
\begin{tabular}{c r r r r r r}
\toprule
 & \multicolumn{3}{c}{Singlet, $P_i^{q,1}$} & \multicolumn{3}{c}{Octet, $P_i^{q,8}$} \\
\cmidrule(lr){2-4}\cmidrule(lr){5-7}
Operator & $C^{(0)}$ & $\as C^{(1)}$ & $C^{(0)}+\as C^{(1)}$ & $C^{(0)}$ & $\as C^{(1)}$ & $C^{(0)}+\as C^{(1)}$ \\
\midrule
$P_{1}^{d}$ & $-3.34$ & $-0.854$ & $-4.19$ & $20.7$ & $2.67$ & $23.3$ \\
$P_{2}^{d}$ & $0$ & $-0.101$ & $-0.101$ & $0$ & $-0.157$ & $-0.157$ \\
$P_{3}^{d}$ & $0$ & $-0.631$ & $-0.631$ & $0$ & $0.771$ & $0.771$ \\
$P_{4}^{d}$ & $0$ & $-0.632$ & $-0.632$ & $0$ & $0.988$ & $0.988$ \\
$P_{5}^{d}$ & $0$ & $0.119$ & $0.119$ & $0$ & $1.38$ & $1.38$ \\
\midrule
$P_{1}^{u}$ & $0.0151$ & $0.377$ & $0.392$ & $6.04$ & $0.00778$ & $6.05$ \\
$P_{2}^{u}$ & $O(z^2)\simeq0$ & $-0.125$ & $-0.125$ & $O(z^2)\simeq0$ & $-0.712$ & $-0.712$ \\
$P_{3}^{u}$ & $-0.00754$ & $-0.313$ & $-0.320$ & $-3.02$ & $-0.810$ & $-3.83$ \\
$P_{4}^{u}$ & $0$ & $-0.178$ & $-0.178$ & $0$ & $-0.292$ & $-0.292$ \\
$P_{5}^{u}$ & $0$ & $0.233$ & $0.233$ & $0$ & $0.144$ & $0.144$ \\
\midrule
$P_{1}^{s}$ & $6.95$ & $0.866$ & $7.81$ & $0.846$ & $1.72$ & $2.56$ \\
$P_{2}^{s}$ & $0$ & $-0.158$ & $-0.158$ & $0$ & $-0.753$ & $-0.753$ \\
$P_{3}^{s}$ & $-3.47$ & $0.201$ & $-3.27$ & $-0.423$ & $-1.61$ & $-2.03$ \\
$P_{4}^{s}$ & $0$ & $0.651$ & $0.651$ & $0$ & $-0.913$ & $-0.913$ \\
$P_{5}^{s}$ & $0$ & $0.309$ & $0.309$ & $0$ & $-0.461$ & $-0.461$ \\
\bottomrule
\end{tabular}
\caption{Dimension-seven coefficients at $\mu=M=1.59\,\mathrm{GeV}$, to three significant figures. Entries $O(z^2)\simeq0$ vanish for massless cut quarks. The LO coefficient of $P_2$ starts at second order in $z$, the squared cut-quark mass over $M^2$, and is below $2\times10^{-4}$ here. $C^{(0)}$ and $C^{(1)}$ as in eq.~\eqref{eq:weak-summed}. All rows are for the Cabibbo-favoured channel $c\to su\bar d$.}
\label{tab:d-meson-coefficient-comparison}
\end{table}

Matrix elements given at a hadronic scale $\mu_H$ have to be evolved to the scale of the coefficients, $\mu=M$. We evolve them in full QCD, at NLO at dimension six and at one loop at dimension seven,
neglecting mixing with penguin operators, as ref.~\cite{Cheng:2023jpz} does for $\langle Q\rangle$, and write $\eta=\alpha_s(\mu_H)/\alpha_s(M)$, $a_H=a_s(\mu_H)$, $a_M=a_s(M)$ and $\delta a=a_H-a_M$. In this paper only the baryon matrix elements are evolved, section~\ref{sec:baryons}. The meson matrix elements of section~\ref{sec:charm} are used at the scale of the coefficients.

At dimension six the two-loop anomalous dimensions are known~\cite{Buras:1989xd,Buras:2000if} in the scheme of $\mathbf C_6^{(1)}$, and they compensate its scheme
dependence at $\mu=M$. The pairs $(Q,T)$ and $(Q_S,T_S)$ do not mix. The valence input $\langle T\rangle=-\frac23\langle Q\rangle$ is an eigenvector of the $(Q,T)$ evolution at both orders, so $Q$ and $T$ carry the same factor
\begin{eqnarray}
\lefteqn{\langle\boldsymbol Q(M)\rangle_H={\bf U}\,\langle\boldsymbol Q(\mu_H)\rangle_H,}
 \label{eq:baryon-QS-evolution}\\
{\bf U}&=&\begin{pmatrix}
\eta^{4/9}\bigl[1-\frac{181}{81}\delta a\bigr] & 0 & 0 & 0\\[3pt]
 0 & \eta^{8/9}\bigl[1+\frac{544}{81}\delta a\bigr] & 0 & \frac{67}{12}\bigl(a_H\eta^{8/9}-a_M\eta^{-1/9}\bigr)\\[3pt]
 0 & 0 & \eta^{4/9}\bigl[1-\frac{181}{81}\delta a\bigr] & 0\\[3pt]
 0 & -\frac{5}{27}a_H\eta^{-1/9}\log\eta & 0 & \eta^{-1/9}\bigl[1-\frac{641}{324}\delta a\bigr]
 \end{pmatrix}.\nonumber
\end{eqnarray}%
At dimension seven the running follows from the anomalous dimensions $\boldsymbol\gamma^{(0)}_{76}$ and $\boldsymbol\gamma^{(0)}_{77}$, eqs.~\eqref{eq:uv-J76-div} and~\eqref{eq:uv-J77-div}, which contain no penguin contractions. $\boldsymbol Q$ does not run into
$\boldsymbol P$. Although ${\bf J}_{67}$ is nonzero, it has no UV pole, $\boldsymbol\gamma^{(0)}_{67}=0$, since UV poles are polynomial in the
masses and every $P_i$ carries an inverse power of $M$. The terms of ${\bf J}_{67}$ in $l$ accompany its IR pole
and define no anomalous dimension. $\boldsymbol P$ does have a UV pole in $\boldsymbol Q$,
$\boldsymbol\gamma^{(0)}_{76}\neq0$. However,  the subtraction
${\bf F}_{76}$ removes the running by keeping ${\bf J}_{76}=0$ at every $\mu$. With $n_f=3$ this gives
\begin{equation}
 \langle\boldsymbol P(M)\rangle_H={\eta}^{-\boldsymbol\gamma^{(0)}_{77}/18}\,\langle\boldsymbol P(\mu_H)\rangle_H.
 \label{eq:baryon-P-evolution}
\end{equation}
This includes the running of $\overline m_s(\mu)$ in $P_{3,4}$.

\subsection{\texorpdfstring{$D$}{D} mesons}
\label{sec:charm}
Let $\Gamma_{\rm ref}$ denote the pole-scheme widths in
ref.~\cite{Gratrex:2022xpm}, table~10, whose dimension-six
matrix elements are the HQET sum-rule bag and eye parameters of
refs.~\cite{King:2021xqp,King:2021jsq}. A lattice determination of the dimension-six bag parameters of the $D_s^+$ is now available~\cite{Black:2026dzp,Black:2026prl} but we do not use it here. We keep these widths as published and add the NLO update of their dimension-seven term
$\Gamma_7$, which includes the $1/m_c$ corrections to the matrix elements of ${\bf Q}$:
\begin{equation}
 \delta\Gamma_7=\Gamma_{7,\rm NLO}-\Gamma_{7,\rm LO},
 \qquad \Gamma_{\rm updated}=\Gamma_{\rm ref}+\delta\Gamma_7.
 \label{eq:meson-literature-replacement}
\end{equation}
For each channel, through first soft order, the correction reads
\begin{eqnarray}
&&  \delta\Gamma_7=\frac{G_F^2|\lambda|^2M^2}{6\pi}
 \sum_w\Big\{(\omega_w^{\rm NLO}-\omega_w^{\rm LO})
 \bigl(\mathbf C_{6,w}^{(0)}\langle\boldsymbol Q\rangle^{(0)}_H|_{M^{-1}}+\mathbf C_{7,w}^{(0)}\langle\boldsymbol P\rangle^{(0)}_H\bigr)\label{eq:meson-replacement-contraction}\\
&&\hspace{13mm}+\as\omega_w^{\rm LO}
 \bigl[\mathbf C_{6,w}^{(1)}\langle\boldsymbol Q\rangle^{(0)}_H|_{M^{-1}}
  +\mathbf C_{7,w}^{(1)}\langle\boldsymbol P\rangle^{(0)}_H+\kappa\,C_{Q_S,w}^{(0)}\langle Q_S\rangle^{(0)}_H|_{M^{-1}}\bigr]\Big\}
 +\delta\Gamma_{\rm mass}, \nonumber 
\end{eqnarray}
where all matrix elements are the tree-level VIA ones and $\langle\boldsymbol Q\rangle^{(0)}_H|_{M^{-1}}$ is the first-power part of $\langle\boldsymbol Q\rangle^{(0)}_H$ in a naive expansion in $1/m_c$. The first term
changes the weak products of the Born part from LO to NLO. The second contains the NLO coefficients with LO weak products
and the $\alpha_s/m_c$ term of $\langle Q_S\rangle$, given by
\begin{equation}
 \langle Q_S\rangle=\langle Q_S\rangle^{(0)}+\kappa\as\langle Q_S\rangle^{(0)}+O(\as^2),\qquad
 \kappa=2C_F(4+3l).
 \label{eq:meson-kappa}
\end{equation}
The factor $\kappa$ follows from VIA. For a massless light quark the axial Ward identity gives
$\langle Q_S\rangle=B_HM_H^2/\overline m_c^2(\mu)$, with $B_H=f_H^2M_H/2$, and $\langle Q_S\rangle^{(0)}=B_HM_H^2/M^2$,
so that eq.~\eqref{eq:meson-kappa} follows from $\overline m_c(\mu)=M[1-\as C_F(4+3l)]$. Its first-power part is
the $\kappa$ term of eq.~\eqref{eq:meson-replacement-contraction}. The term $\delta\Gamma_{\rm mass}$ converts the strange cut mass of the coefficients from the
on-shell to the $\MSb$ mass of the reference, with the matrix elements held fixed. It changes the widths by less than $10^{-4}~\mathrm{ps}^{-1}$.

As in ref.~\cite{Gratrex:2022xpm}, we use $M=m_c^{\rm pole}=1.49$~GeV, $\mu=1.5$~GeV,
$m_u=m_d=0$ and $\overline m_s(\mu)=0.1$~GeV, the decay constants and meson masses of its table~16,
$\alpha_s(\mu)$ and the LO and NLO weak coefficients of its tables~20 and 21, with $c_1=C_{2,\rm lit}$ and
$c_2=C_{1,\rm lit}$, and the CKM entries of its eq.~(A.1).
The dimension-seven matrix elements are the VIA ones of ref.~\cite{Gratrex:2022xpm} with its inputs,
including $\bar\Lambda_{u,d}=0.5$~GeV and $\bar\Lambda_s=0.6$~GeV of its eq.~(4.6), to which we assign $\pm0.1$~GeV, except that the
light-quark mass in them is set to zero, which leaves the LO term unchanged. They are given in
appendix~\ref{app:meson-literature-bridge}, together with those of $P_4,P_5$.
As in the reference, the doubly Cabibbo-suppressed channel $c\to du\bar s$ is not included.
Semileptonic annihilation includes the electron and the massive muon,
$m_\mu=0.105658$~GeV. For $D_s^+$ we keep the published reduced widths $\widetilde\Gamma$,
which exclude $D_s^+\to\tau^+\nu_\tau$, and obtain the lifetime from
ref.~\cite{Gratrex:2022xpm}, eq.~(2.2), with $\mathcal B_{\tau\nu}=0.0539\pm0.0009$ from ref.~\cite{ParticleDataGroup:2026}.

\begin{figure}[!t]
 \centering
 \includegraphics[width=\textwidth]{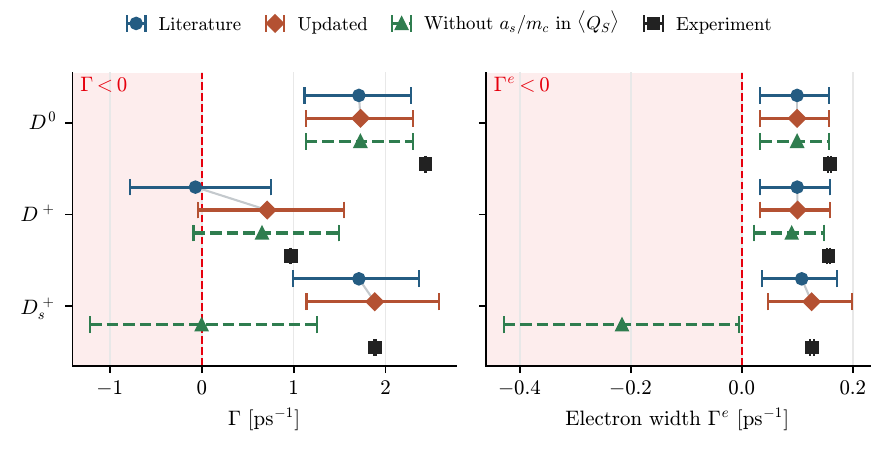}
 \caption{$D$-meson widths (left) and electron-channel semileptonic widths, with VIA semileptonic matrix elements (right). Blue circles: published widths
 $\Gamma_{\rm ref}$. Orange diamonds: $\Gamma_{\rm ref}+\delta\Gamma_7$. Green triangles with dashed bars: without the $\alpha_s/m_c$ term of $\langle Q_S\rangle$. Black squares: PDG 2026
 widths~\cite{ParticleDataGroup:2026} (for $D_s^+$: $(1-\mathcal B_{\tau\nu})/\tau$). Error bars combine the input
 and scale components in quadrature. Red bands: negative width.}
 \label{fig:literature-meson-widths}
\end{figure}

Table~\ref{tab:literature-meson-widths} and figure~\ref{fig:literature-meson-widths} give the results. The reference electron widths are the branching fractions of
ref.~\cite{Gratrex:2022xpm}, table~11 and eq.~(4.8), divided by the lifetimes used there. The scale is
varied over $1\le\mu\le3$~GeV, with $\alpha_s$ and the LO and NLO $c_{1,2}$ evolved from their values in ref.~\cite{Gratrex:2022xpm} at $1.5$~GeV
at five loops, LO and NLO, respectively, with $n_f=3$ below and $n_f=4$ above $1.5$~GeV. The errors of $\delta\Gamma_7$ combine this variation and
the input errors of $\bar\Lambda_q$, $f_H$, $M_H$ and $G_F$ in quadrature. Those of $\tau$ and $\Gamma^e_{\rm updated}$ add them in quadrature
to the errors of $\Gamma_{\rm ref}$, whose correlations are not available. The NLO term does not uniformly reduce this scale dependence (table~\ref{tab:scale-dependence}), because the VIA matrix elements
are not evolved and the logarithms of $\mathbf C_7^{(1)}$ that compensate the running of $\langle\boldsymbol P\rangle$ remain.

\begin{table}[!t]
\centering\small
\setlength{\tabcolsep}{4.5pt}
\renewcommand{\arraystretch}{1.35}
\begin{tabular}{lcccccc}
\toprule
Meson & $\Gamma_{\rm ref}$ & $\delta\Gamma_7$ & $\tau$ [ps] & $\Gamma^e_{\rm ref}$ & $\delta\Gamma^e$ & $\Gamma^e_{\rm updated}$ \\
\midrule
$D^0$ & $1.71^{+0.41}_{-0.47}\,{}^{+0.39}_{-0.36}$ & $+0.02^{+0.06}_{-0.04}$ & $0.58^{+0.30}_{-0.14}$ & $0.099^{+0.054}_{-0.062}\,{}^{+0.020}_{-0.024}$ & $0$ & $0.099^{+0.058}_{-0.066}$ \\
$D^+$ & $-0.07^{+0.76}_{-0.68}\,{}^{+0.31}_{-0.20}$ & $+0.78^{+0.18}_{-0.26}$ & $1.41^{+\infty}_{-0.76}$ & $0.099^{+0.055}_{-0.063}\,{}^{+0.020}_{-0.023}$ & $0.001^{+0.001}_{-0.001}$ & $0.100^{+0.058}_{-0.067}$ \\
$D_s^+$ & $1.71^{+0.49}_{-0.60}\,{}^{+0.44}_{-0.40}$ & $+0.17^{+0.23}_{-0.18}$ & $0.50^{+0.33}_{-0.14}$ & $0.108^{+0.060}_{-0.068}\,{}^{+0.019}_{-0.022}$ & $0.018^{+0.038}_{-0.033}$ & $0.125^{+0.073}_{-0.079}$ \\
\bottomrule
\end{tabular}
\caption{Reference widths,
NLO dimension-seven corrections, lifetimes and inclusive widths for one electron species, with widths in ps$^{-1}$. $\Gamma_{\rm ref}$ is the pole-scheme width of
ref.~\cite{Gratrex:2022xpm}, table~10. Its errors are input and
scale variations, as for $\Gamma^e_{\rm ref}$. The $D_s^+$ widths exclude $\tau\nu$, and the $D_s^+$ lifetime
is $(1-\mathcal B_{\tau\nu})/\widetilde\Gamma$. An upper lifetime limit
$\infty$ means that the width interval reaches zero.  }
\label{tab:literature-meson-widths}
\end{table}

\begin{table}[!htb]
\centering\small
\begin{tabular*}{0.8\textwidth}{@{\extracolsep{\fill}}l r r r r r r r}\toprule
 & \multicolumn{7}{c}{\makebox[0pt]{Scale variation of the nonleptonic dimension-seven term [\% of $1/\tau$]}}\\\cmidrule(lr){2-8}
 & $D^0$ & $D^+$ & $D_s^+$ & $\Lambda_c^+$ & $\Xi_c^+$ & $\Xi_c^0$ & $\Omega_c^0$\\\midrule
LO & $0.0$ & $140.7$ & $6.5$ & $7.0$ & $6.5$ & $5.9$ & $12.0$\\
NLO & $3.9$ & $111.1$ & $20.7$ & $2.3$ & $0.8$ & $2.0$ & $3.0$\\
\bottomrule\end{tabular*}
\caption{Maximum minus minimum over $1\le\mu\le3$~GeV, with $M$ fixed. $\tau$ is the measured lifetime~\cite{ParticleDataGroup:2026}; for $D_s^+$, $1/\tau$ is replaced by $(1-\mathcal B_{\tau\nu})/\tau$.}
\label{tab:scale-dependence}
\end{table}

At $O(\alpha_s)$ the width contains
$\mathbf C_w^{(1)}\langle\boldsymbol P\rangle^{(0)}+\mathbf C_w^{(0)}\langle\boldsymbol P\rangle^{(1)}$,
eq.~\eqref{eq:matching-as}. We include only the first term, and $\mathbf C_w^{(0)}\langle P_i\rangle^{(1)}$ is missing.
VIA is inconsistent with the real-gluon class of figure~\ref{fig:sl-amplitude-topologies}c. Since it keeps only the vacuum
between the two currents, this class vanishes by colour, while the subtraction of its infrared region from
$\mathbf C^{(1)}$ stays. The estimate is therefore a partial NLO result, without $\langle\boldsymbol Q\rangle^{(1)}_H|_{M^{-1}}$
and $\langle\boldsymbol P\rangle^{(1)}_H$, except for the $\alpha_s/m_c$ term of $\langle Q_S\rangle$ in
eq.~\eqref{eq:meson-kappa}. 
Without this term (green in figure~\ref{fig:literature-meson-widths}), $\Gamma^e(D_s^+)$ is
negative and the total $D_s^+$ width vanishes, which shows a substantial model dependence. For $D^+$, $\delta\Gamma_7=0.78~\mathrm{ps}^{-1}$ consists of $0.44$ from $P_5$, $0.74$ from the NLO coefficient of $P_1^{d,1}$, $-0.46$ from the weak products and $0.06$ from the other terms, in ps$^{-1}$.

\subsection{Charmed baryons}
\label{sec:baryons}
We recalculate the baryon widths $\Gamma_{\rm ref}$ from the bag-model inputs of ref.~\cite{Cheng:2023jpz} at the
common scale $\mu=M=1.59$~GeV, with $\delta\Gamma_7$ and $\Gamma_{\rm updated}$ as in
eq.~\eqref{eq:meson-literature-replacement}. The strange cut mass is $\overline m_s(2~\mathrm{GeV})=0.09$~GeV, converted with the one-loop on-shell--$\MSb$ relation at $\mu=2$~GeV. Our widths are
roughly
 $20 \%$   below the published ones, mostly through $\alpha_s$ and the evolution.
Ref.~\cite{Cheng:2023jpz} uses the LO weak coefficients $(c_1,c_2)=(1.298,-0.565)$ at $1.5$~GeV. With the updated
$\alpha_s$ we use those of section~\ref{sec:coefficients}, $(1.2293,-0.4599)$ at LO and $(1.1703,-0.3508)$ at NLO. We also evolve the matrix elements with
eqs.~\eqref{eq:baryon-QS-evolution} and~\eqref{eq:baryon-P-evolution} in place of the source's common evolution factor. The
leading and kinetic terms use their common coefficient $c_3$ through NLO, with the kernels of refs.~\cite{Egner:2024partonic} (nonleptonic)
and~\cite{Moreno:2022semitauonic} (semileptonic). Its NNLO correction~\cite{Egner:2024partonic,vanRitbergen:1999gs},
about $-0.21~\mathrm{ps}^{-1}$ in the massless-light limit and mostly semileptonic, is not included.  

The four-quark matrix elements of the published bag model are recomputed, and those of $P_4$ and $P_5$ calculated
from the same wavefunctions (appendix~\ref{app:baryon-inputs}). The published kinetic and magnetic matrix elements,
bag parameters and CKM inputs are retained. We take the bag-model matrix elements as the complete ones at
$\mu_H=1.0\pm0.2$~GeV, in the scheme of this paper, and evolve them to $\mu=M$ with five-loop running for
$n_f=3$~\cite{Herren:2017osy} from $\alpha_s(1~\mathrm{GeV})=0.495$~\cite{Cheng:2023jpz}, used only in
the evolution. $\delta\Gamma_7$ contracts the NLO coefficients, the semileptonic ones of
appendix~\ref{app:sl-massive-coefficients} with $\rho_e=0$ and $\rho_\mu=m_\mu^2/M^2$, with the evolved tree-level
matrix elements.

\begingroup\setlength{\extrarowheight}{3pt}
\begin{table}[!b]
\centering\small
\begin{tabular}{l r r r r r r}\toprule
Baryon & $\Gamma_{\rm ref}$ & $\delta\Gamma_7$ & $\tau$ [fs] & $\Gamma^e_{\rm ref}$ & $\delta\Gamma^e$ & $\mathcal B^e$ [\%]\\\midrule
$\Lambda_c^+$ & $3.66^{+0.96}_{-0.79}$ & $0.050^{+0.070}_{-0.161}$ & $269^{+68}_{-51}$ & $0.242^{+0.086}_{-0.073}$ & $0$ & $6.5^{+1.7}_{-1.7}$\\
$\Xi_c^+$ & $2.39^{+0.80}_{-0.62}$ & $-0.001^{+0.049}_{-0.124}$ & $418^{+141}_{-100}$ & $0.293^{+0.111}_{-0.094}$ & $0.010^{+0.005}_{-0.010}$ & $12.7^{+2.7}_{-3.1}$\\
$\Xi_c^0$ & $4.38^{+1.10}_{-0.89}$ & $0.081^{+0.085}_{-0.207}$ & $224^{+53}_{-42}$ & $0.297^{+0.113}_{-0.095}$ & $0.010^{+0.005}_{-0.010}$ & $6.9^{+1.7}_{-1.8}$\\
$\Omega_c^0$ & $2.87^{+1.09}_{-0.82}$ & $0.029^{+0.110}_{-0.313}$ & $344^{+127}_{-85}$ & $0.164^{+0.125}_{-0.100}$ & $0.048^{+0.014}_{-0.030}$ & $7.3^{+2.9}_{-3.2}$\\
\bottomrule\end{tabular}
\caption{Recalculated baseline
widths $\Gamma_{\rm ref}$, the NLO dimension-seven
correction (nonleptonic, electron and muon), both in ps$^{-1}$, updated
lifetimes, inclusive widths for one electron species in ps$^{-1}$, and the electron branching fraction
$\mathcal B^e=\Gamma^e_{\rm updated}/\Gamma_{\rm updated}$. Central values use the common scale $\mu=M=1.59$~GeV.
The bands propagate the same input variations and the scale variation $1\le\mu\le3$~GeV, with $M$ fixed, through the baseline,
correction, full width and branching fraction.}
\label{tab:charm-baryon-widths}
\end{table}
\endgroup

\begin{figure}[!htb]
\centering
\includegraphics[width=.91\textwidth]{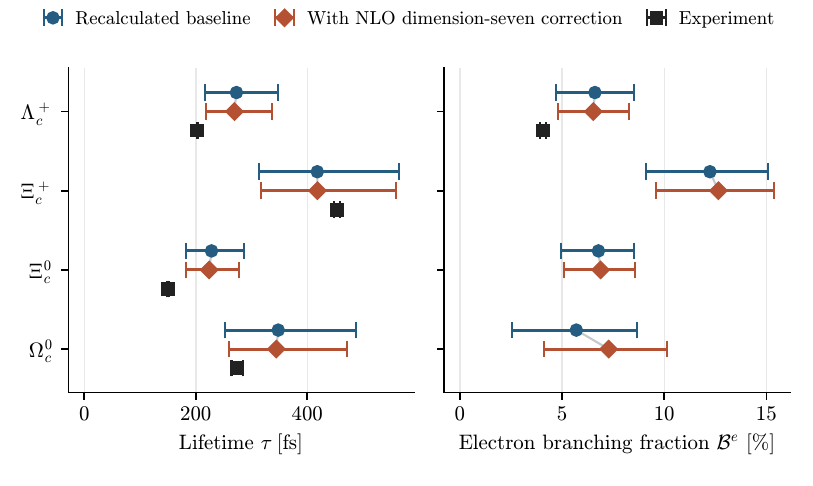}
\caption{Charmed-baryon lifetimes from the recalculated baseline and
the full width $\Gamma_{\rm ref}+\delta\Gamma_7$ (left), and the electron branching fractions $\mathcal B^e$ (right). Each band propagates
common input and scale variations through its complete width. Black squares show the PDG 2026
lifetimes~\cite{ParticleDataGroup:2026}{, and on the right the
inclusive $\Lambda_c^+$ measurement of
ref.~\cite{BESIII:2023inclusiveLc}}.}
\label{fig:baryons}
\end{figure}

At central inputs, the correction changes the singly charmed baryon widths by at most $2\%$ (table~\ref{tab:charm-baryon-widths} and figure~\ref{fig:baryons}), and the central lifetimes keep the measured hierarchy
$\tau(\Xi_c^+)>\tau(\Omega_c^0)>\tau(\Lambda_c^+)>\tau(\Xi_c^0)$. For the electron width, table~\ref{tab:baryon-electron-ratios}, the correction is $-8\%$ to $-11\%$ of the LO
dimension-seven term $\Gamma^e_{7,\rm LO}$ for every baryon. For $\Omega_c^0$, $\Gamma^e_{7,\rm LO}$ cancels $78\%$ of the lower-dimensional terms, which makes $\delta\Gamma^e$ $29\%$ of $\Gamma^e_{\rm ref}$, against $3\%$ for $\Xi_c^{+,0}$ and
$0.1\%$ for $\Lambda_c^+$. For $\Lambda_c^+$ the NNLO
term would lower $\mathcal B^e$ to $4.3^{+2.0}_{-2.4}\%$, close to the measured $(4.06\pm0.13)\%$~\cite{BESIII:2023inclusiveLc}. The values of
$\mathcal B^e$ exceed those of ref.~\cite{Cheng:2023jpz}, $4.57$, $8.57$, $4.40$ and $1.88\%$, mainly through the smaller
total widths. For $\Omega_c^0$, $\Gamma^e_{\rm ref}$ is also twice as large, and $\mathcal B^e$ comes close to the
quark-model value $7.59\%$ quoted there.

For the singly charmed baryons, whose matrix elements are evolved with the operators, the NLO term reduces the scale
dependence of the nonleptonic dimension-seven term (table~\ref{tab:scale-dependence}). The bands of the
updated widths in table~\ref{tab:charm-baryon-widths} are dominated by $M=1.59\pm0.09$~GeV. The scale, the four-quark
matrix elements and $\mu_H$ also contribute. That of $\delta\Gamma_7$,
the NLO minus the LO term at the same $\mu$, is dominated by the scale.

\begin{table}[!htb]
\centering\small
\begin{tabular}{l *{4}{>{\raggedleft\arraybackslash}p{2.5cm}}}\toprule
 & \multicolumn{2}{c}{Relative to $\Gamma_{\rm ref}-\Gamma_{7,\rm LO}$} & \multicolumn{2}{c}{Relative to $\Gamma^e_{\rm ref}-\Gamma^e_{7,\rm LO}$}\\
\cmidrule(lr){2-3}\cmidrule(lr){4-5}
Baryon & $\Gamma_{7,\rm LO}$ & $\delta\Gamma_7$ & $\Gamma^e_{7,\rm LO}$ & $\delta\Gamma^e$\\\midrule
$\Lambda_c^+$ & $16.6\%$ & $1.6\%$ & $-1.32\%$ & $0.14\%$\\
$\Xi_c^+$ & $-16.6\%$ & $-0.04\%$ & $-28.7\%$ & $2.4\%$\\
$\Xi_c^0$ & $-3.6\%$ & $1.8\%$ & $-28.9\%$ & $2.4\%$\\
$\Omega_c^0$ & $-49.8\%$ & $0.5\%$ & $-78.2\%$ & $6.3\%$\\
\bottomrule\end{tabular}
\caption{Central values of the LO dimension-seven term and its NLO correction, relative to the width up to
dimension six, $\Gamma_{\rm ref}-\Gamma_{7,\rm LO}$ for the total width and $\Gamma^e_{\rm ref}-\Gamma^e_{7,\rm LO}$ for the electron width.}
\label{tab:baryon-electron-ratios}
\end{table}

\section{Conclusions}
We have computed the NLO QCD corrections to the connected current--current part of the dimension-seven spectator
coefficients of charm-hadron widths, for all five operator families in both colour structures, in NDR--$\MSb$ with the
Hermitian-averaged Greek projection for the evanescent operators.
The finite subtraction  
removes the   mixing of ${\bf P}$ into ${\bf Q}$. In several channels the one-loop term is as large as the LO one. It
dominates for $P_1^{u,1}$, whose LO coefficient is proportional to $(c_1+3c_2)^2$, and is twice the LO value for
$P_1^{s,8}$ (table~\ref{tab:d-meson-coefficient-comparison}). The operators $P_4$ and $P_5$ first appear at NLO.
The coefficients of $P_{1,2,5}$ also apply to the $b\to c\bar ud$ and $c\bar us$ channels of bottom decays.

\enlargethispage{-2\baselineskip}The main result is for the $D^+$. At central inputs, the NLO dimension-seven correction raises the published
$D^+$ width from $-0.07$ to $0.71~\mathrm{ps}^{-1}$. The correction, $0.78~\mathrm{ps}^{-1}$, is about two thirds of
the LO dimension-seven term. With the
published pole-scheme width it gives $\tau(D^+)=1.41^{+\infty}_{-0.76}$~ps, compatible with the measured
$1.033$~ps~\cite{ParticleDataGroup:2026}. The correction is small for $D^0$ and about $10\%$ for $D_s^+$. At central inputs, it changes the
singly charmed baryon widths by less than $2\%$ and preserves their lifetime hierarchy.

Without the NLO matrix elements the result is not complete. At NLO the scale dependence of the nonleptonic dimension-seven term
drops severalfold for the singly charmed baryons, but not uniformly for the $D$ mesons, whose
matrix elements are not evolved.
Reliable dimension-seven contributions
require the $O(\alpha_s)$ hadronic matrix elements of $P_{1,2,3}$ in
the same scheme, the first-power part of $\langle\boldsymbol Q\rangle^{(1)}$ beyond VIA and the two-loop evolution of all dimension-seven matrix elements.

\acknowledgments

{We thank  Hai-Yang Cheng, Maria Laura Piscopo, and Peng-Yuan Zheng for valuable discussions.
This work was supported in part by the National Natural Science Foundation
of China (NSFC) under Grant No.~12575096.
We thank the Mainz Institute for Theoretical Physics (MITP) of the Cluster
of Excellence PRISMA+ (Project ID~390831469) for its hospitality and support.}

\appendix
\section{Details of the calculation}
\label{app:calculation}
\begingroup
\let\section\subsection
\let\subsection\subsubsection
{
\section{Spinor conventions and Fierz representatives}
\label{sec:evanescent-conventions}
For ordered fields $(\bar c_{i\alpha},q_{j\beta},\bar q_{k\gamma},c_{l\delta})$ the spinor
contractions are
\begin{equation}
\begin{aligned}
 D(A,B)_{\alpha\beta;\gamma\delta}&=A_{\alpha\delta}B_{\gamma\beta},&
 X(A,B)_{\alpha\beta;\gamma\delta}&=-A_{\alpha\beta}B_{\gamma\delta}.
\end{aligned}
\label{app:new-pairings}
\end{equation}
The minus sign in $X$ includes the Grassmann interchange. For $P_1,\ldots,P_4$,
the singlet and octet colour tensors are $\mathcal C_1=\delta_{ij}\delta_{kl}$ and
$\mathcal C_8=t^A_{ij}t^A_{kl}=\tfrac12(\delta_{il}\delta_{kj}-\delta_{ij}\delta_{kl}/N_c)$,
respectively. Colour is held fixed during the spinor rearrangement.
The endpoint momenta $(p_f,r_i,r_f,p_i)$ of the heavy and light lines, with $i$ and $f$ for initial and final, follow this field order. The momentum transfer is $\Delta=p_f-p_i=r_i-r_f$.
  At tree level the derivatives act on the external lines, $\bar c\overleftarrow D_\rho\to ip_{f\rho}$ and $D^\rho q\to-ir_i^\rho$, and on $p_i$ and $r_f$ in the Hermitian conjugates, so that, with the colour suppressed, the operators of eq.~\eqref{eq:operators} give
\begin{equation}
\begin{aligned}
 \langle P_1\rangle^{(0)}&=\tfrac12(p_f\cdot r_i+p_i\cdot r_f)\,X(\gamma_\mu{\rm L},\gamma^\mu{\rm L}),&
 \langle P_2\rangle^{(0)}&=-(p_f\cdot r_i+p_i\cdot r_f)\,X({\rm L},{\rm R}),\\
 \langle P_3\rangle^{(0)}&=\tfrac{m_q}{M}\bigl[X({\rm L},{\rm L})+X({\rm R},{\rm R})\bigr],&
 \langle P_4\rangle^{(0)}&=\tfrac{m_q}{M}\bigl[D({\rm L},{\rm L})+D({\rm R},{\rm R})\bigr],\\
 \langle P_5\rangle^{(0)}&=i\Delta_\mu D(\sigma^{\mu\nu},\gamma_\nu{\rm L}).
\end{aligned}
\label{app:new-tree-P}
\end{equation}
The map $\Pi$ of section~\ref{sec:method} writes a $d$-dimensional Dirac structure as four-dimensional ones with
coefficients through $O(\epsilon)$. It acts on the contracted Dirac tensors
\begin{equation}
 C_k^{\chi\chi'}
 =\gamma_{\mu_1}\cdots\gamma_{\mu_k}\chi
  \otimes\gamma^{\mu_1}\cdots\gamma^{\mu_k}\chi',
 \qquad \chi,\chi'\in\{{\rm L},{\rm R}\}.
 \label{app:new-core-basis}
\end{equation}
The reduction uses the rules of table~\ref{tab:dirac-first-core-rules}.

\begin{table}[!htbp]
\centering\small
\begin{tabular}{@{}cll@{}}
\toprule
$k$ & $\chi'=\chi$ & $\chi'\neq\chi$\\
\midrule
0 & $C_{0}$ & $C_{0}$\\
1 & $C_{1}$ & $C_{1}$\\
2 & $C_2-2\epsilon C_0$ & $4(1+\epsilon)C_0$\\
3 & $(16-4\epsilon)C_1$ & $4(1+\epsilon)C_1$\\
4 & $(16-8\epsilon)C_2-64\epsilon C_0$ & $(16+128\epsilon)C_0$\\
5 & $(256-224\epsilon)C_1$ & $(16+128\epsilon)C_1$\\
6 & $(256-352\epsilon)C_2-1536\epsilon C_0$ & $(64+2880\epsilon)C_0$\\
\bottomrule
\end{tabular}
\caption{Standard NDR Greek prescription~\cite{Tracas:1982gp} for the contracted tensors $C_k^{\chi\chi'}$ through $O(\epsilon)$, following appendix~A of Aebischer et~al.~\cite{Aebischer:2025LEFT} and expressed in the $C_k$ basis. In both columns, $C_0$, $C_1$ and $C_2$ are the four-dimensional tensors.}
\label{tab:dirac-first-core-rules}
\end{table}

With the colour label suppressed, the tree-level amputated matrix elements of the Fierz forms
$\mathcal F[P_i]$ are
\begin{equation}
\begin{aligned}
 \langle\mathcal F[P_1]\rangle^{(0)}
 &=\frac{p_f\cdot r_i+p_i\cdot r_f}{2}\,D(\gamma_\mu{\rm L},\gamma^\mu{\rm L}),\\
 \langle\mathcal F[P_2]\rangle^{(0)}
 &=\frac{p_f\cdot r_i+p_i\cdot r_f}{2}\,D(\gamma_\mu {\rm R},\gamma^\mu{\rm L}),\\
 \langle\mathcal F[P_3]\rangle^{(0)}&=\frac{m_q}{8M}\sum_{\chi={\rm L},{\rm R}}
       D(\gamma_\mu\gamma_\nu\chi,\gamma^\mu\gamma^\nu\chi)
       -(1-\epsilon/4)\langle P_4\rangle^{(0)},\\
 \langle\mathcal F[P_4]\rangle^{(0)}&=\frac{m_q}{8M}\sum_{\chi={\rm L},{\rm R}}
       X(\gamma_\mu\gamma_\nu\chi,\gamma^\mu\gamma^\nu\chi)
       -(1-\epsilon/4)\langle P_3\rangle^{(0)}.
\end{aligned}
\label{app:new-fierz14}
\end{equation}
For $P_5$, with the density-colour tensor of
eqs.~\eqref{eq:operators} and~\eqref{eq:octet-operators}, the Fierz form reads
\begin{equation}
\begin{aligned}
 \langle\mathcal F[P_5]\rangle^{(0)}
 &=\tfrac12X({\slashed\Delta}{\rm L},{\rm R})
 -X({\rm L},{\slashed\Delta}{\rm L})\\
 &\qquad+\tfrac14X(\gamma_\mu{\slashed\Delta}{\rm L},\gamma^\mu{\rm L})
 +\tfrac14X(\gamma_\mu{\rm L},\gamma^\mu{\slashed\Delta} {\rm R}).
\end{aligned}
\label{app:new-fierz5}
\end{equation}
The terms $(1-\epsilon/4)\langle P_{4,3}\rangle^{(0)}$ in eq.~\eqref{app:new-fierz14} make
$\Pi_H\langle\mathcal F[P_i]\rangle^{(0)}=\langle P_i\rangle^{(0)}+O(\epsilon^2)$ for
$i=3,4$, as for $i=1,2$.

Other orders of the gamma matrices in the second factor are first brought to the order of the table with
$\{\gamma^\mu,\gamma^\nu\}=2g^{\mu\nu}$ in $d$ dimensions. The chiralities $\chi$, $\chi'$ in the table are those next to the core. For example, $\gamma_\mu\gamma_\nu\slashed p\,{\rm L}
=\gamma_\mu\gamma_\nu\,{\rm R}\,\slashed p$ enters the table with ${\rm R}$. We write the average in
eq.~\eqref{eq:scheme-residue} as
\begin{equation}
	\Pi_H(O)
	=\frac12\left[{\Pi}(O)+
	\bigl({\Pi}(O^{\dagger})\bigr)^{\dagger}\right].
	\label{app:new-Greek}
\end{equation}
The average in eq.~\eqref{app:new-Greek} is needed because $\Pi$ is not Hermitian. For
	$S=X(\slashed p_f\gamma_\mu\gamma_\nu{\rm L},\gamma^\mu\gamma^\nu{\rm L})$,
	\begin{equation}
		\Pi(S)-\bigl(\Pi(S^\dagger)\bigr)^\dagger=6\epsilon\,X(\slashed p_f{\rm L},{\rm L}),
		\label{eq:pi-not-hermitian}
	\end{equation}
	and such structures occur in the amplitudes of $P_{1,2}$.

The reference tensors $E_0(P_i^{\rm c})$ and $E_0(\mathcal F[P_i^{\rm c}])$,
$i=1,\ldots,5$, ${\rm c}=1,8$, are the evanescent remainders before the transformation of eq.~\eqref{eq:evanescent-freedom}, which gives
$E(Z)=E_0(Z)+\epsilon\,d_ZO_Z$.  The ancillary
Mathematica file gives all twenty $E_0(Z)$. As an example, with the pairings of
eq.~\eqref{app:new-pairings} and the colour tensors $\mathcal C_b$, write
$S^Y_{k,b}=\mathcal C_b\sum_{\chi={\rm L},{\rm R}}Y(\gamma_{\mu_1}\cdots\gamma_{\mu_k}\chi,\gamma^{\mu_1}\cdots\gamma^{\mu_k}\chi)$
for $Y=D,X$.
The singlet $P_3$ input has the complete reference remainder
\begin{equation}
 \langle E_0(P_3^1)\rangle^{(0)}
 =\frac{m_q}{M}\left[\frac83(d-4)S^X_{0,1}
       +8S^D_{0,8}+8S^X_{0,8}-\frac4dS^X_{2,8}\right].
 \label{eq:reference-evanescent-example}
\end{equation}
At $d=4$, the spinor identity
$S^X_{2,b}=8(S^D_{0,b}+S^X_{0,b})$ makes the complete remainder vanish.
Expanding
$-4/d=-1-\epsilon/2+O(\epsilon^2)$ and applying $\Pi_H$ of eq.~\eqref{app:new-Greek} gives
\begin{equation}
 \Pi_H\langle E_0(P_3^1)\rangle^{(0)}
 =-\epsilon\left[\frac{16}{3}\langle P_3^1\rangle^{(0)}
        +2\langle P_3^8\rangle^{(0)}
        +4\langle P_4^8\rangle^{(0)}\right]+O(\epsilon^2).
 \label{eq:reference-evanescent-example-tail}
\end{equation}
This is the $O(\epsilon)$ image of the remainder of $P_3^1$ alone. The finite insertions
follow from that of the sum ${\cal E}_i$ of eq.~\eqref{eq:scheme-residue} for the pair
$(P_3^1,\mathcal F[P_3^1])$, appendix~\ref{app:uv-poles}.

\section{Decomposition of a loop integral}
\label{app:integral-decomposition}
The integrals of the matching are separated into the UV, IR and finite terms of
eqs.~\eqref{eq:amp-Laurent} and~\eqref{eq:insert-Laurent} as follows, with
$[dg]=e^{\epsilon\gamma_E}\mu^{2\epsilon}d^dg/(i\pi^{d/2})$. An integrand ${\cal I}(g)$ is
expanded at large $g$ and at small $g$. In dimensional regularization a pole at $d=4$
comes only from the terms of degree $-4$ in $g$, which diverge logarithmically, giving the UV pole at large $g$
and the IR pole at small $g$. Terms of other degree give no pole. We
keep the degree-$(-4)$ terms and average their numerators over the directions of $g$ in $d$
dimensions. This
defines the UV residue $U(\epsilon)$ and the soft residue ${\cal R}(\epsilon)$ through the large-$g$
part $U(\epsilon)/(g^2)^2$ and the small-$g$ part ${\cal R}(\epsilon)/(g^2)^2$. We write
$U_0=U(0)$ and, through $O(\epsilon)$, ${\cal R}={\cal R}_0+\epsilon {\cal R}_1$. With
\begin{equation}
 \int[dg]\,\frac{1}{(g^2)^2}=\frac{1}{\epsilon_{\rm UV}}-\frac{1}{\epsilon_{\rm IR}},
 \label{eq:scaleless-integral}
\end{equation}
the integral has the form
\begin{equation}
 \int[dg]\,{\cal I}(g)=\frac{U_0}{\epsilon_{\rm UV}}
 -\frac{{\cal R}_0+\epsilon {\cal R}_1}{\epsilon_{\rm IR}}+{\cal I}_{\rm fin}+O(\epsilon),
 \label{eq:integral-decomposition}
\end{equation}
where ${\cal I}_{\rm fin}$ is finite  and contains the $\epsilon/\epsilon_{\rm UV}$ term of the
UV part. The labels record the origin of each term. As a
number $\epsilon/\epsilon_{\rm IR}=1$, so the $\epsilon^0$ term
of the integral is ${\cal I}_{\rm fin}-{\cal R}_1$, and ${\cal R}_1$ must be taken
from the soft expansion.

For example, up to $O(\epsilon)$
\begin{equation}
\begin{aligned}
 \int[dg]\,\frac{(g\cdot p)^2}{(g^2)^3}
 &=\frac{M^2}{d}\Bigl(\frac{1}{\epsilon_{\rm UV}}-\frac{1}{\epsilon_{\rm IR}}\Bigr),\\
 \int[dg]\,\frac{(g\cdot p)^2}{(g^2)^2(g^2+2g\cdot p)}
 &=\frac{M^2}{d}\,\frac{1}{\epsilon_{\rm UV}}
 +\frac{M^2}{4}\Bigl(l+\frac12\Bigr),\\
 \int[dg]\,\frac{2(g\cdot p)^3}{(g^2)^3(g^2+2g\cdot p)}
 &=-\frac{M^2}{d}\,\frac{1}{\epsilon_{\rm IR}}
 -\frac{M^2}{4}\Bigl(l+\frac12\Bigr).
\end{aligned}
\label{eq:integral-examples}
\end{equation}
The factor $1/d$ is the tensor reduction $(g\cdot p)^2\to g^2M^2/d$. The first
integral is scaleless, with $U={\cal R}$, and vanishes as a number. The second has no term of
degree $-4$ at small $g$, so it has only a UV pole. The third is the first minus the second,   and has only a soft term, with ${\cal R}_0=M^2/4$,
${\cal R}_1=M^2/8$ and ${\cal I}_{\rm fin}=-(M^2/4)(l+1/2)$.  

The IR divergences can also be regulated with a gluon mass $\uplambda$, as discussed in
refs.~\cite{Ciuchini:2001vx,Beneke:2002rj}. The two regulators are related term by
term. With the gluon propagator $1/(g^2-\uplambda^2)$, eq.~\eqref{eq:scaleless-integral} becomes
$\int[dg]\,[g^2(g^2-\uplambda^2)]^{-1}=1/\epsilon_{\rm UV}+\log(\mu^2/\uplambda^2)+1$, so in the IR
term $-({\cal R}_0+\epsilon{\cal R}_1)/\epsilon_{\rm IR}$ the pole $-{\cal R}_0/\epsilon_{\rm IR}$ becomes
${\cal R}_0[\log(\mu^2/\uplambda^2)+1]$ and the finite term $-{\cal R}_1$ becomes zero, while the UV
term and ${\cal I}_{\rm fin}$ are unchanged.
 The UV terms and ${\bf F}_{76}$ do not
depend on the IR regulator, so with a gluon mass the insertions on the tabulated outputs are
\begin{equation}
 {\bf J}-{\bf J}_{\rm IR}\Bigl[\log\frac{\mu^2}{\uplambda^2}+1\Bigr],
 \label{eq:insert-gluon-mass}
\end{equation}
up to $1/\uplambda$ terms in the $67$ block and terms that vanish as $\uplambda\to0$. The
$1/\uplambda$ terms arise because expanding the dimension-six insertions to first order in the
light-quark momentum makes their integrand one power more singular at small $g$, of degree
$-5$. For example, $\int[dg]\,[(g^2-\uplambda^2)^2(2g\cdot p+i0)]^{-1}=-\pi/(2M\uplambda)$, whereas
at $\uplambda=0$ the same integral is scaleless and vanishes in dimensional regularization.
These terms cancel in the matching.  

\section{Ultraviolet poles and finite insertions}
\label{app:uv-poles}
In the order of eq.~\eqref{eq:uv-basis}, with rows labelling the inserted operator, the one-loop anomalous
dimension of eq.~\eqref{eq:insert-Laurent} has the blocks
\begin{subequations}\label{eq:uv-J76-div}
\begin{equation}
 \begingroup\setlength{\arraycolsep}{4pt}\renewcommand{\arraystretch}{1.25}
 \boldsymbol\gamma^{(0)}=\begin{pmatrix}
 \boldsymbol\gamma^{(0)}_{66} & \boldsymbol\gamma^{(0)}_{67}\\
 \boldsymbol\gamma^{(0)}_{76} & \boldsymbol\gamma^{(0)}_{77}
 \end{pmatrix},\qquad
 \boldsymbol\gamma^{(0)}_{66}=\begin{pmatrix}
 0 & 0 & 12 & 0 \\
 0 & -16 & 0 & 0 \\
 \frac{8}{3} & 0 & -4 & 0 \\
 0 & 0 & 0 & 2 \\
 \end{pmatrix},\qquad
 \boldsymbol\gamma^{(0)}_{67}=0,
 \endgroup
\end{equation}
and the remaining block, transposed, reads
\begin{equation}
 \begingroup\setlength{\arraycolsep}{4pt}\renewcommand{\arraystretch}{1.25}
 \boldsymbol\gamma^{(0)T}_{76}=\begin{pmatrix}
\frac{32}{9} & 0 & 0 & 0 & -\frac{32}{9} & \frac{4}{3} & \frac{4}{27} & 0 & 0 & -\frac{56}{27} \\
\frac{16}{9} & 0 & 0 & 0 & -\frac{64}{9} & 0 & -\frac{40}{27} & 0 & 0 & -\frac{112}{27} \\
6 & \frac{2}{3} & 0 & 0 & \frac{8}{3} & -\frac{22}{9} & -\frac{2}{9} & 0 & 0 & -\frac{22}{9} \\
0 & -\frac{20}{3} & 0 & 0 & \frac{16}{3} & -\frac{2}{9} & \frac{20}{9} & 0 & 0 & -\frac{44}{9} \\
 \end{pmatrix}.
 \endgroup
\end{equation}
\end{subequations}
\begin{equation}
 \begingroup\setlength{\arraycolsep}{3pt}\renewcommand{\arraystretch}{1.25}
 \boldsymbol\gamma^{(0)}_{77}=\begin{pmatrix}
0 & 0 & 0 & 0 & \frac{16}{9} & \frac{44}{3} & 0 & -\frac{4}{3} & -4 & -\frac{10}{3} \\
0 & -16 & 0 & 0 & -\frac{16}{27} & 0 & 8 & 0 & -4 & \frac{4}{9} \\
0 & 0 & -8 & 0 & 0 & 0 & 0 & 8 & 16 & 0 \\
0 & 0 & \frac{32}{3} & \frac{40}{3} & 0 & 0 & 0 & 8 & -8 & 0 \\
0 & 0 & 0 & \frac{16}{3} & \frac{8}{3} & 0 & 0 & 0 & -4 & 0 \\
\frac{88}{27} & 0 & -\frac{8}{27} & -\frac{8}{9} & \frac{4}{27} & \frac{100}{9} & 0 & \frac{40}{9} & \frac{4}{3} & \frac{20}{9} \\
0 & \frac{16}{9} & 0 & -\frac{8}{9} & -\frac{28}{81} & 0 & \frac{46}{3} & -4 & -\frac{8}{3} & -\frac{11}{27} \\
0 & 0 & \frac{16}{9} & \frac{32}{9} & 0 & 0 & 0 & \frac{40}{3} & \frac{20}{3} & 0 \\
0 & 0 & \frac{16}{9} & -\frac{16}{9} & 0 & 0 & 0 & -4 & -8 & 0 \\
0 & 0 & 0 & \frac{28}{9} & 0 & 0 & 0 & 0 & \frac{11}{3} & \frac{8}{3} \\
 \end{pmatrix}.
 \endgroup
 \label{eq:uv-J77-div}
\end{equation}

The finite subtraction ${\bf F}_{76}$ equals the finite part of the $76$ insertions,
so that subtracting it imposes ${\bf J}_{76}=0$:
\begin{equation}
 \begingroup\setlength{\arraycolsep}{3.5pt}\renewcommand{\arraystretch}{1.25}
 {\bf F}_{76}=\begin{pmatrix}
- \frac{192l + 623}{108} & - \frac{48l - 31}{54} & - \frac{3  \left(16l + 29\right)}{16} & 0 \\
- \frac{7}{36} & \frac{143}{18} & - \frac{48l + 29}{144} & \frac{240l + 589}{72} \\
0 & 0 & 0 & 0 \\
0 & 0 & 0 & 0 \\
\frac{8 \left(2 l + 5\right)}{9} & \frac{2 \left(16 l + 11\right)}{9} & - \frac{2 \left(2 l + 5\right)}{3} & - \frac{16 l + 11}{6} \\
- \frac{16l + 29}{24} & 0 & \frac{11  \left(96l + 199\right)}{864} & \frac{48l - 31}{432} \\
- \frac{48l + 29}{648} & \frac{240l + 589}{324} & \frac{96l + 79}{864} & - \frac{480l + 1607}{432} \\
0 & 0 & 0 & 0 \\
0 & 0 & 0 & 0 \\
\frac{112 l + 289}{108} & \frac{112 l + 113}{54} & \frac{176 l + 431}{144} & \frac{176 l + 85}{72} \\
 \end{pmatrix}.
 \endgroup
 \label{eq:insert-F76}
\end{equation}

The dimension-six block of ${\bf J}$, eq.~\eqref{eq:J-definition}, is
\begin{equation}
 \begingroup\setlength{\arraycolsep}{3.5pt}\renewcommand{\arraystretch}{1.35}
 {\bf J}_{66}=\begin{pmatrix}
- \frac{16}{3} & \frac{16}{3} & - \frac{3 \left(2 l + 7\right)}{2} & -2 \\
0 & \frac{8 \left(3 l + 4\right)}{3} & - \frac{3}{2} & \frac{3 \left(2 l + 1\right)}{2} \\
- \frac{2 l + 7}{3} & - \frac{4}{9} & \frac{66 l + 113}{12} & -3 \\
- \frac{1}{3} & \frac{2 l + 1}{3} & - \frac{1}{4} & \frac{30 l + 41}{12} \\
 \end{pmatrix}.
 \endgroup
 \label{eq:insert-J66fin}
\end{equation}

\enlargethispage{2\baselineskip}The block ${\bf J}_{67}$, transposed, is
\begin{equation}
 \begingroup\setlength{\arraycolsep}{3.5pt}\renewcommand{\arraystretch}{1.35}
\left(  {\bf J}_{67}\right)  ^ {T}=\begin{pmatrix}
- \frac{4 \left(6 l + 5\right)}{3} & 0 & \frac{16}{9} & \frac{2 l + 7}{9} \\
- \frac{4 \left(2 l + 7\right)}{3} & - \frac{16}{3} & 0 & \frac{6 l + 5}{9} \\
- \frac{4 \left(2 l + 1\right)}{3} & \frac{8}{3} & \frac{8}{9} & - \frac{2 l + 1}{9} \\
\frac{16}{3} & 0 & \frac{8}{9} & - \frac{2 l + 5}{9} \\
- \frac{2 \left(2 l + 9\right)}{9} & - \frac{2 l + 5}{9} & \frac{2 l - 9}{27} & \frac{2 l + 23}{54} \\
8 & \frac{2 l + 7}{2} & \frac{6 l - 11}{6} & - \frac{2 l + 7}{6} \\
0 & \frac{6 l + 5}{2} & \frac{2 l + 7}{6} & - \frac{6 l + 1}{6} \\
4 & - \frac{2 l + 1}{2} & \frac{2 l - 7}{6} & \frac{2 l - 1}{6} \\
4 & - \frac{2 l + 5}{2} & -2 & \frac{2 l + 5}{6} \\
- \frac{8 l + 27}{3} & \frac{2 l + 5}{12} & - \frac{2 l + 9}{36} & - \frac{5 l + 17}{18} \\
 \end{pmatrix}.
 \endgroup
 \label{eq:insert-J67fin}
\end{equation}

${\bf J}_{67}$ does not depend on the dimension-seven evanescent convention.
With columns in the order of $\boldsymbol P$,
${\bf J}_{77}$ is
\begin{equation}
 \begingroup\setlength{\arraycolsep}{2pt}\renewcommand{\arraystretch}{1.35}
 \resizebox{\linewidth}{!}{$\displaystyle {\bf J}_{77}=\begin{pmatrix}
- \frac{16}{3} & - \frac{8}{3} & 0 & - \frac{1}{2} & - \frac{2 \left(24 l + 95\right)}{81} & - \frac{5 l + 16}{3} & \frac{1}{3} & \frac{32 l + 123}{24} & \frac{16 l + 35}{8} & \frac{13 \left(48 l + 37\right)}{432} \\
0 & \frac{8 \left(3 l + 4\right)}{3} & 0 & \frac{1}{2} & \frac{163}{162} & \frac{5}{3} & \frac{5 \left(6 l + 17\right)}{18} & - \frac{2 \left(3 l + 10\right)}{9} & \frac{96 l + 101}{72} & - \frac{515}{432} \\
0 & 0 & \frac{2 \left(12 l + 17\right)}{3} & \frac{4}{3} & 0 & 0 & 0 & \frac{3}{2} & - 6 l - 7 & 0 \\
0 & 0 & - \frac{16 \left(l + 2\right)}{3} & - \frac{2 \left(4 l + 9\right)}{3} & 0 & 0 & 0 & - 2 \left(2 l + 3\right) & \frac{8 l - 3}{2} & 0 \\
\frac{16}{9} & \frac{16}{9} & - \frac{1}{3} & - \frac{8 l + 17}{3} & \frac{2 \left(4 l + 7\right)}{3} & - \frac{4}{3} & - \frac{4}{3} & \frac{1}{4} & \frac{8 l + 17}{4} & - \frac{23}{16} \\
- \frac{2 \left(5 l + 16\right)}{27} & \frac{2}{27} & \frac{32 l + 123}{108} & \frac{16 l + 35}{36} & \frac{624 l - 131}{1944} & \frac{38 l + 205}{36} & \frac{13}{18} & - \frac{52 l + 21}{36} & - \frac{32 l + 67}{48} & - \frac{1392 l + 3215}{1296} \\
\frac{10}{27} & \frac{5 \left(6 l + 17\right)}{81} & - \frac{4 \left(3 l + 10\right)}{81} & \frac{96 l + 101}{324} & - \frac{216 l - 53}{972} & - \frac{1}{18} & - \frac{114 l - 37}{108} & \frac{66 l + 67}{54} & \frac{240 l - 13}{432} & \frac{432 l + 3017}{2592} \\
0 & 0 & \frac{1}{3} & - \frac{2 \left(6 l + 7\right)}{9} & 0 & 0 & 0 & \frac{6 l + 19}{3} & - \frac{6 l - 13}{6} & 0 \\
0 & 0 & - \frac{4 \left(2 l + 3\right)}{9} & \frac{8 l - 3}{9} & 0 & 0 & 0 & \frac{12 l + 29}{6} & 8 l + 5 & 0 \\
\frac{18 l + 37}{27} & - \frac{18 l - 19}{27} & \frac{12 l + 29}{9} & - \frac{16 l + 31}{18} & - \frac{23}{72} & - \frac{18 l - 35}{36} & \frac{18 l + 53}{36} & - \frac{24 l + 67}{24} & - \frac{28 l + 61}{12} & \frac{10 l + 11}{24}
 \end{pmatrix}.$}
 \endgroup
 \label{eq:insert-J77fin}
\end{equation}

\section{A worked example: emission from the light quark}
\label{app:real-example}
For illustration, 
we set $c_2=0$, so that only the term $c_1^2$ of eq.~\eqref{eq:width} contributes.
Consider the forward graph in which the light spectator emits and
reabsorbs the gluon, figure~\ref{fig:real-example}, with $p$, $r$ the momenta of the incoming $c$, $\bar u$,
$k$, $k'$ those of the cut $s$, $\bar d$ and $g$ that of the gluon, and with all quarks except the
charm massless. It is the graph real~$u$--$u$ of the page \texttt{charm\_nlo\_diagrams.html}, which evaluates it
with the three cuts combined. Its discontinuity has three cuts. The
real-gluon cut $C_g$ crosses $k$, $k'$ and the gluon. The light-quark
cuts $C_L$ and $C_R$ cross $k$, $k'$ and one light-quark propagator,
with the gluon uncut.  We follow the graph to first soft order,
which gives its dimension-seven terms.

\begin{figure}[!b]
\centering
\includegraphics[width=.62\textwidth]{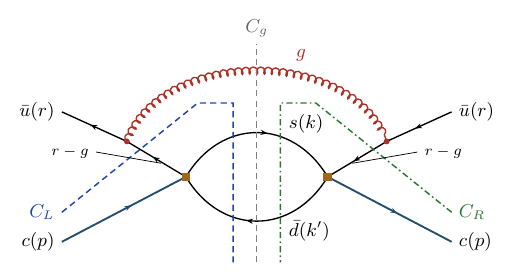}
\caption{Gluon emission and reabsorption by the light spectator, with
the real-gluon cut $C_g$ and the light-quark cuts $C_L$ and $C_R$.}
\label{fig:real-example}
\end{figure}

Up to constant factors, collected in eq.~\eqref{eq:real-example-density} below, the forward amplitude of
the graph is the two-loop integral
\begin{equation}
\begin{aligned}
 {\cal T}_{uu}&=\int d^dk\,d^dg\,
   \frac{\mathcal N}{(k^2+i0)\,(k'^2+i0)\,(t+i0)\,(\Delta_u+i0)^2},\\
 \mathcal N&=-[\bar c\,\gamma^\nu{\rm L}\slashed k\,\gamma^\alpha{\rm L} c]\,
   [\bar q\,\gamma^\rho(\slashed r-\slashed g)\gamma_\alpha{\rm L}\slashed k'
   \gamma_\nu{\rm L}(\slashed r-\slashed g)\gamma_\rho\,q],
\end{aligned}
\label{eq:real-example-two-loop}
\end{equation}
with $k'=K-k-g$, $K=p+r$ and $t=g^2$. Here $\Delta_u=(r-g)^2=t-2r\cdot g$ is the inverse of both light-quark
propagators. Summed over the colours of the gluon and the cut quarks,
the colour factor of the graph is $\tfrac49\,\delta_{ij}\delta_{kl}+\tfrac83\,t^A_{ij}t^A_{kl}$, with $i,j,k,l$ the colours of
$\bar c$, $q$, $\bar q$ and $c$ suppressed in $\mathcal N$. 

To first soft order the propagator factor is
\begin{equation}
 \frac1{(t+i0)(\Delta_u+i0)^2}=\frac1{(t+i0)^3}+\frac{4\,r\cdot g}{(t+i0)^4}+O(r^2).
 \label{eq:real-example-expansion}
\end{equation}
The light-quark cuts, $\Delta_u=0$, then coincide with the gluon cut, $t=0$, so the three cuts of
figure~\ref{fig:real-example} merge into one cut through $k$ and $k'$, where the propagator factor has poles of
third and fourth order, whose discontinuity in $t$ is a distribution given by
\begin{equation}
\begin{aligned}
 \frac{1}{2\pi i}\Bigl[\frac{1}{(t-i0)^{n+1}}-\frac{1}{(t+i0)^{n+1}}\Bigr]
 &=\frac{(-1)^n}{n!}\,\delta^{(n)}(t),\\
 \int dt\,F(t)\,\delta^{(n)}(t)&{}=(-1)^nF^{(n)}(0).
\end{aligned}
\label{eq:real-example-distribution}
\end{equation}
The cut lines $k$ and $k'$ give $\delta^+(k^2)\,\delta^+(k'^2)$. Of the components of $g$ we keep the spatial ones and
replace the energy by $t=g^2$, so that $d^dg=dt\,d^{d-1}g/(2g^0)$ with $g^0=(\vec g^{\,2}+t)^{1/2}$. Then
\begin{equation}
\begin{aligned}
 \int d^dk\,d^dg\,\delta^+(k^2)\,\delta^+(k'^2)\cdots&=(2\pi)^{2d-3}\int dt\int d\Phi_3(t)\cdots,\\
 d\Phi_3(t)&=(2\pi)^d\,\delta^d(K-k-k'-g)\prod_{q=k,k',g}\frac{d^{d-1}q}{(2\pi)^{d-1}\,2q^0},
\end{aligned}
 \label{eq:real-example-phase-space}
\end{equation}
where $d\Phi_3(t)$ is the standard phase space of three particles with momenta $k$, $k'$, $g$ and masses $0$, $0$,
$\sqrt t$. The propagator
factor of eq.~\eqref{eq:real-example-expansion} depends on $g$ only through $t$ and the numerator $r\cdot g$, so by
eq.~\eqref{eq:real-example-distribution} the three cuts together give derivatives in $t$, at $t=0$, of the integral over
$d\Phi_3(t)$. In the rest frame of $K$ we use $a=2k\cdot g$, $b=2k'\cdot g$ and the orientation of $k$, $k'$ and
$g$ as variables, so that
\begin{equation}
\begin{aligned}
 d\Phi_3(t)&=\frac{(4\pi)^{2\epsilon}}{128\,\pi^3\,\Gamma(2-2\epsilon)}\,(K^2)^{\epsilon-1}\,
   \mathcal G(K^2,t;a,b)^{-\epsilon}\,da\,db\,\frac{d\Omega}{\Omega},\\
 \mathcal G(K^2,t;a,b)&=(K^2-t-a-b)\,[ab-t(K^2-t-a-b)],
\end{aligned}
 \label{eq:real-example-gram-polynomial}
\end{equation}
where $d\Omega/\Omega$ is the measure of the orientation, normalized to unit volume, and
$\mathcal G=4\det(p_i\cdot p_j)$ is the Gram determinant of the cut momenta $p_i=k,k',g$.
We write the integrand at fixed $t$ as $F/[(t+i0)(\Delta_u+i0)^2]$, where $F$ contains $\mathcal N$ and
$d\Phi_3(t)$ without its constant prefactor. Then
\begin{equation}
 F=(K^2)^{\epsilon-1}\mathcal G^{-\epsilon}\langle\mathcal N\rangle,
 \label{eq:real-example-F}
\end{equation}
with $\langle\cdots\rangle=\int(d\Omega/\Omega)\cdots$ the average over the
orientation of the three momenta $k$, $k'$ and $g$ in the rest frame of $K=k+k'+g$, at fixed
$a$, $b$ and $t$. The average reduces every tensor in $k$, $k'$ and $g$ to $K^\mu$ and the metric. For example,
$\langle k^\mu\rangle=(k\cdot K)\,K^\mu/K^2$, with $k\cdot K=(K^2-b-t)/2$ fixed by the invariants. We write $F=F_0+F_1$, with $F_1$ linear in $r$.
Setting $r=0$ gives
\begin{equation}
 F_0=(M^2)^{\epsilon-1}\mathcal G^{-\epsilon}\langle\mathcal N_0\rangle,
 \qquad
 \mathcal N_0=-[\bar c\,\gamma^\nu{\rm L}\slashed k\,\gamma^\alpha{\rm L} c]\,
   [\bar q\,\gamma^\rho\slashed g\,\gamma_\alpha{\rm L}\slashed k'\gamma_\nu{\rm L}
   \slashed g\,\gamma_\rho\,q],
 \label{eq:real-example-F0}
\end{equation}
with $\mathcal G$ at $K^2=M^2$, $\mathcal G=(H-t)(ab-tH+t^2)=\mathcal G_0-(ab+H^2)t+2Ht^2-t^3$,
where $H=M^2-a-b$ and $\mathcal G_0=abH$. The terms linear in $r$ come from the numerators
of the light-quark propagators, from the average through $K=p+r$ and from the measure
through $K^2=M^2+2p\cdot r$:
\begin{equation}
\begin{aligned}
 F_1&=(M^2)^{\epsilon-1}\mathcal G^{-\epsilon}\Bigl\{
   \langle\mathcal N_1\rangle+r\cdot\partial_K\langle\mathcal N_0\rangle
   +2p\cdot r\Bigl[\frac{\epsilon-1}{M^2}
   -\epsilon\,\frac{ab-2t(H-t)}{\mathcal G}\Bigr]
   \langle\mathcal N_0\rangle\Bigr\},\\
 \mathcal N_1&=[\bar c\,\gamma^\nu{\rm L}\slashed k\,\gamma^\alpha{\rm L} c]\,
   \bigl[\bar q\,\gamma^\rho\slashed r\,\gamma_\alpha{\rm L}\slashed k'
   \gamma_\nu{\rm L}\slashed g\,\gamma_\rho\,q
   +\bar q\,\gamma^\rho\slashed g\,\gamma_\alpha{\rm L}\slashed k'\gamma_\nu{\rm L}
   \slashed r\,\gamma_\rho\,q\bigr],
\end{aligned}
\label{eq:real-example-F1}
\end{equation}
where $ab-2t(H-t)=\partial\mathcal G/\partial K^2$ at $K^2=M^2$.

At dimension seven the first term of eq.~\eqref{eq:real-example-expansion} multiplies $F_1$ and the second,
linear in $r$, multiplies $F_0$.
With $n=2$ and $n=3$ in eq.~\eqref{eq:real-example-distribution}, the contribution of this graph to ${\cal A}_7^{(1)}$ of
eq.~\eqref{eq:amp-Laurent}, which by eq.~\eqref{eq:matching-full} is the discontinuity,
$2\,\mathrm{Im}$, of its forward amplitude, is
\begin{equation}
 {\cal A}^{(1)}_{7,uu}=-\frac38\Bigl(\frac{\mu^2}{M^2}\Bigr)^{\epsilon}{M^{4\epsilon-2}}\int_{a,b>0,\;a+b<M^2}\!da\,db\,
 \Bigl[\frac12F_1''(0)+\frac23\,\frac{\partial^3}{\partial t^3}\bigl[(r\cdot g)F_0\bigr]_{t=0}\Bigr],
 \label{eq:real-example-density}
\end{equation}
with derivatives in $t$ at fixed Dalitz variables.
The second term comes from the $1/t^4$ term of eq.~\eqref{eq:real-example-expansion} and has one
derivative more. In it, $r\cdot g$ is averaged together with $\mathcal N_0$. The prefactor contains the weak
vertices, the three-body phase space and $g_s^2$ in the $\MSb$ scheme, divided by the
prefactor of eq.~\eqref{eq:width} and by the $d$-dimensional normalization of the Born phase space.
Its sign is the Fermi sign of the antiquark spectator. The pole of ${\cal A}^{(1)}_{7,uu}$ is infrared. Its Dirac structure is projected onto
$\langle P_i\rangle^{(0)}$ with $\Pi_H$ of eq.~\eqref{app:new-Greek}. For the averaged numerators of
eqs.~\eqref{eq:real-example-F0} and~\eqref{eq:real-example-F1}, with $s=2k\cdot k'=M^2-a-b-t$, the projections with $\Pi_H$ through $O(\epsilon)$ are given by
\begin{equation}
\begin{aligned}
 \Pi_H\langle\mathcal N_0\rangle
 &=\frac{8}{3M^2}\Bigl\{(a+t)^2s+2M^2(ab-st)\\
 &\qquad-\frac{\epsilon}{6}\bigl[18(a+b+2t)^2s-10(a+t)^2s+M^2(ab-55st)\bigr]\Bigr\}\langle Q\rangle^{(0)}\\
 &\quad-\frac{16}{3M^2}\Bigl\{2(a+t)^2s+M^2(ab-st)\\
 &\qquad-\frac{\epsilon}{6}\bigl[18(a+b+2t)^2s-8(a+t)^2s-M^2(ab+17st)\bigr]\Bigr\}\langle Q_S\rangle^{(0)},\\
 \Pi_H\langle\mathcal N_1\rangle
 &=\frac{8\epsilon}{3}\,\bigl[ab-(b+2t)s\bigr]\langle P_1\rangle^{(0)}.
\end{aligned}
\label{app:new-A2-projection}
\end{equation}
For $\mu=M$, in units of
$c_1^2\,G_F^2|\lambda|^2M^2/(6\pi)$, the result reads
\begin{equation}
\begin{aligned}
 {\cal A}^{(1)}_{7,uu}
 ={}&\Bigl(-\frac{8}{27\epsilon_{\rm IR}}+\frac{22}{9}\,\frac{\epsilon}{\epsilon_{\rm IR}}-\frac{4}{27}\Bigr)\langle P_1^1\rangle^{(0)}
 +\Bigl(\frac{4}{27}\,\frac{\epsilon}{\epsilon_{\rm IR}}+\frac{4}{27}\Bigr)\langle P_2^1\rangle^{(0)}\\
 &+\Bigl(-\frac{16}{9\epsilon_{\rm IR}}+\frac{44}{3}\,\frac{\epsilon}{\epsilon_{\rm IR}}-\frac{8}{9}\Bigr)\langle P_1^8\rangle^{(0)}
 +\Bigl(\frac{8}{9}\,\frac{\epsilon}{\epsilon_{\rm IR}}+\frac{8}{9}\Bigr)\langle P_2^8\rangle^{(0)},
\end{aligned}
 \label{eq:real-example-result}
\end{equation}
which for $\epsilon/\epsilon_{\rm IR}=1$ are the values of the graph real~$u$--$u$ in \texttt{charm\_nlo\_diagrams.html}.
  Its $\epsilon/\epsilon_{\rm IR}$ terms belong to ${\cal A}^{(1)\prime}_{7,{\rm IR}}$ of eq.~\eqref{eq:amp-Laurent}, which cancels in the
matching.
}
\endgroup
\section{Analytical charm coefficients}
\label{app:charm-analytic-coefficients}
\begingroup
\small
\interdisplaylinepenalty=10000
\let\CharmCoefficientSection\subsection
\let\subsection\subsubsection
\let\section\CharmCoefficientSection
\label{sec:conventions}
We give the nonleptonic coefficients of eq.~\eqref{eq:width} without its prefactor $G_F^2|\lambda|^2M^2/(6\pi)$. The
cut strange mass is in the same scheme as $M$.

In $C_{P_i,w}^{s\mid u\bar d,{\rm c},(n)}$, the flavour before the vertical bar labels
$P_i^{s,{\rm c}}$, and $(u,\bar d)$ is the ordered cut pair.
The labels of the four charm channels are listed in table~\ref{tab:charm-labels}.
The cut masses are fixed by the label and not displayed in the coefficients,
whose $z$ and scale dependence is also left implicit.
The second-current down entry labels the operator flavour even when
the corresponding decay particle is an antiquark.
Only $P_1,P_2,P_5$ contribute for external $u,d$, since
$P_{3,4}^{u,{\rm c}}=P_{3,4}^{d,{\rm c}}=0$. All five families are retained for $s$.

\begin{table}[!htbp]
\centering
\renewcommand{\arraystretch}{1.35}
\begin{tabular}{@{}l c c c@{}}
\toprule
Decay & First-current down & Up & Second-current down\\
\midrule
$c\to su\bar d$ & $s\mid u\bar d:(0,0)$ & $u\mid s\bar d:(z,0)$ & $d\mid su:(z,0)$\\
$c\to du\bar s$ & $d\mid u\bar s:(0,z)$ & $u\mid d\bar s:(0,z)$ & $s\mid du:(0,0)$\\
$c\to su\bar s$ & $s\mid u\bar s:(0,z)$ & $u\mid s\bar s:(z,z)$ & $s\mid su:(z,0)$\\
$c\to du\bar d$ & $d\mid u\bar d:(0,0)$ & $u\mid d\bar d:(0,0)$ & $d\mid du:(0,0)$\\
\bottomrule
\end{tabular}
\caption{Coefficient labels of the four charm channels. The pair after a label gives $(m/M)^2$ of the two cut quarks,
in the order of the label.}
\label{tab:charm-labels}
\end{table}

For external $d$, use
\begin{equation}
\begin{aligned}
 C_{P_i,w}^{d\mid u\bar d,{\rm c},(n)}
 &=C_{P_i,w}^{s\mid u\bar d,{\rm c},(n)},&
 C_{P_i,w}^{d\mid du,{\rm c},(n)}
 &=C_{P_i,w}^{s\mid du,{\rm c},(n)},\\
 C_{P_i,w}^{d\mid u\bar s,{\rm c},(n)}
 &=C_{P_i,w}^{s\mid u\bar s,{\rm c},(n)},&
 C_{P_i,w}^{d\mid su,{\rm c},(n)}
 &=C_{P_i,w}^{s\mid su,{\rm c},(n)},
\end{aligned}
\label{eq:charm-d-map}
\end{equation}
for $n=0,1$. 
The remaining external-$u$ sets follow by weak-row exchange:
\begin{equation}
\begin{aligned}
 C_{P_i,w}^{u\mid d\bar s,{\rm c},(n)}
 &=C_{P_i,\bar w}^{s\mid u\bar s,{\rm c},(n)},\\
 C_{P_i,w}^{u\mid d\bar d,{\rm c},(n)}
 &=C_{P_i,\bar w}^{s\mid u\bar d,{\rm c},(n)},\qquad
 (\overline{22},\overline{12},\overline{11})=(11,12,22),
\end{aligned}
\label{eq:charm-q2-map}
\end{equation}
for $n=0,1$, at fixed cut masses.

The leading-order coefficients use the printed order
\begin{equation}
 (P_1^{s,1},P_2^{s,1},P_3^{s,1},P_4^{s,1},P_5^{s,1};
  P_1^{s,8},P_2^{s,8},P_3^{s,8},P_4^{s,8},P_5^{s,8}).
 \label{eq:lo-order}
\end{equation}
All required LO rows follow from four functions of the squared cut masses, with $h_i(0,z)=h_i(z,0)$ and $\beta=\sqrt{1-4z}$:
\begin{equation}
\begin{aligned}
 h_1(z,0)&=1-z^3, & h_1(z,z)&=\frac{1-2z-2z^2}{\beta},\\
 h_2(z,0)&=z^2(1-z), & h_2(z,z)&=\frac{2z^2}{\beta},\\
 h_3(z,0)&=(1-z)^2(1+2z), & h_3(z,z)&=\beta(1+2z),\\
 \tilde h_1(z,0)&=1-z^2.
\end{aligned}
\label{eq:lo-functions}
\end{equation}
The first set is
\begin{equation}
\begin{aligned}
\mathbf C_{7,22}^{s\mid u\bar d,(0)}&=(6h_1,18h_2,-3h_3,0,0;\ 0,0,0,0,0),\\
\mathbf C_{7,12}^{s\mid u\bar d,(0)}&=(4h_1,12h_2,-2h_3,0,0;\ 0,0,0,0,0),\\
\mathbf C_{7,11}^{s\mid u\bar d,(0)}&=(\tfrac23 h_1,2h_2,-\tfrac13 h_3,0,0;\ 4h_1,12h_2,-2h_3,0,0).
\end{aligned}
\label{eq:lo-s-udbar}
\end{equation}
For $s\mid u\bar d$ the functions are taken at $(0,0)$. The same expressions
with label $s\mid u\bar s$ take them at $(0,z)$.
The second set is
\begin{equation}
\begin{aligned}
\mathbf C_{7,22}^{s\mid du,(0)}&=\mathbf C_{7,11}^{s\mid du,(0)}
 =(2\tilde h_1,0,0,0,0;\ 12\tilde h_1,0,0,0,0),\\
\mathbf C_{7,12}^{s\mid du,(0)}&=(12\tilde h_1,0,0,0,0;\ 0,0,0,0,0).
\end{aligned}
\label{eq:lo-s-du}
\end{equation}
Use $\tilde h_1(0,0)$ for $s\mid du$ and the same expressions with $\tilde h_1(z,0)$
for $s\mid su$. External-$d$ coefficients follow from
eq.~\eqref{eq:charm-d-map}.
For external $u$, use the reduced order
$(P_1^{u,1},P_2^{u,1},P_5^{u,1};P_1^{u,8},P_2^{u,8},P_5^{u,8})$:
\begin{equation}
\begin{aligned}
\mathbf C_{7,22}^{u\mid s\bar d,(0)}&=(\tfrac23 h_1,2h_2,0;\ 4h_1,12h_2,0),\\
\mathbf C_{7,12}^{u\mid s\bar d,(0)}&=(4h_1,12h_2,0;\ 0,0,0),\\
\mathbf C_{7,11}^{u\mid s\bar d,(0)}&=(6h_1,18h_2,0;\ 0,0,0).
\end{aligned}
\label{eq:lo-u-sdbar}
\end{equation}
These rows take the functions at $(z,0)$. The same expressions with label
$u\mid s\bar s$ take them at $(z,z)$. The two remaining external-$u$
sets follow from eq.~\eqref{eq:charm-q2-map}.
Both colours of $P_4$ and $P_5$ have zero Born coefficients but nonzero NLO coefficients.

\section{Next-to-leading-order nonleptonic coefficients}
The displays below give $C^{(1)}$, grouped by the cut masses. One massive cut line has $0<z<1$.
The $s\bar s$ cut has $0<z<1/4$. We use the principal dilogarithm
and its real part on the cut, with
\begin{equation}
 H(z)=2\operatorname{Li}_2(z)+\log z\log(1-z).
 \label{eq:charm-H}
\end{equation}
\footnotesize

\label{sec:charm-zero}
For external $s$, the massless cuts $(u\bar d)$ and $(du)$ give
constant and logarithmic coefficients. In the order of eq.~\eqref{eq:lo-order}, they read
\begin{equation}
\begin{aligned}
\mathbf C_{7,22}^{s\mid u\bar d,(1)}&=\bigl(24,\ 0,\ 12\,l+6,\ 7,\ \tfrac{16}{3}\,l+\tfrac{113}{9};\\
&\qquad 44\,l+\tfrac{331}{6},\ -\tfrac{35}{2},\ -16\,l-\tfrac{601}{12},\ -36\,l-\tfrac{155}{4},\ -10\,l-\tfrac{133}{24}\bigr),\\
\mathbf C_{7,12}^{s\mid u\bar d,(1)}&=\bigl(-32\,l-\tfrac{32}{3},\ \tfrac{32}{3},\ 24\,l+\tfrac{32}{3},\ -30,\ \tfrac{32}{9}\,l+\tfrac{466}{27};\\
&\qquad \tfrac{16}{3}\,l+\tfrac{151}{9},\ -\tfrac{11}{3},\ \tfrac{4}{3}\,l-\tfrac{511}{18},\ -24\,l-\tfrac{311}{6},\ -\tfrac{20}{3}\,l+\tfrac{707}{36}\bigr),\\
\mathbf C_{7,11}^{s\mid u\bar d,(1)}&=\bigl(-\tfrac{112}{27}\,l+\tfrac{158}{81},\ \tfrac{26}{27},\ \tfrac{116}{27}\,l-\tfrac{367}{81},\ -\tfrac{16}{3}\,l-\tfrac{446}{27},\ \tfrac{8}{9}\,l+\tfrac{187}{18};\\
&\qquad \tfrac{316}{9}\,l-8\,\pi^2+\tfrac{2467}{27},\ -\tfrac{146}{9},\ -\tfrac{92}{9}\,l+4\,\pi^2-\tfrac{7535}{108},\ -8\,l-\tfrac{190}{9},\ \tfrac{10}{3}\,l+\tfrac{93}{8}\bigr),
\end{aligned}
\label{eq:charm-s-udbar-nlo}
\end{equation}
and
\begin{equation}
\begin{aligned}
\mathbf C_{7,22}^{s\mid du,(1)}=\mathbf C_{7,11}^{s\mid du,(1)}&=\bigl(-\tfrac{112}{9}\,l-\tfrac{64}{9}\,\pi^2-\tfrac{80}{3},\ \tfrac{4}{3},\ -\tfrac{16}{9}\,l-\tfrac{127}{9},\ -\tfrac{16}{3}\,l-\tfrac{104}{9},\ \tfrac{8}{3}\,l+\tfrac{31}{2};\\
&\qquad \tfrac{316}{3}\,l-\tfrac{56}{3}\,\pi^2+193,\ -4,\ \tfrac{76}{3}\,l-\tfrac{47}{12},\ 4\,l+\tfrac{8}{3},\ 10\,l+\tfrac{263}{8}\bigr),\\
\mathbf C_{7,12}^{s\mid du,(1)}&=\bigl(-\tfrac{64}{3}\,\pi^2+\tfrac{560}{9},\ \tfrac{32}{3},\ -\tfrac{16}{9},\ 6,\ \tfrac{32}{3}\,l+\tfrac{356}{9};\\
&\qquad -56\,l-16\,\pi^2-\tfrac{500}{3},\ -2,\ -8\,l-\tfrac{373}{6},\ -24\,l-\tfrac{113}{2},\ -20\,l+\tfrac{121}{12}\bigr).
\end{aligned}
\label{eq:charm-s-du-nlo}
\end{equation}

\label{sec:charm-xzero}
For external $s$ with the cut $(u\bar s)$, one cut line is massive:
\begin{equation}
\begin{aligned}[b]
C_{P_1,22}^{s\mid u\bar s,1,(1)}={}&32 \left(1-z\right) \left(z^{2} + z + 1\right)\,H(z)\;-16 z \left(2 z + 1\right) \left(3 z - 2\right)\,\log z\\
&{}-48 \left(1-z\right) z \left(2 z + 1\right)\,\log(1-z)\;+8 \left(1-z\right) \left(5 z^{2} + 4 z + 3\right)
\end{aligned}\label{eq:charm-xzero-q1-22-1}
\end{equation}
\begin{equation}
\begin{aligned}[b]
C_{P_2,22}^{s\mid u\bar s,1,(1)}={}&-144 \left(1-z\right) z^{2}\,l\;+96 \left(1-z\right) z^{2}\,H(z)\;+32 z^{2} \cdot \left(3 z^{2} - 13 z + 9\right)\,\log z\\
&{}+16 \left(1-z\right) \left(6 z^{3} - 20 z^{2} + 4 z + 1\right)\,\log(1-z)\;+8 \left(1-z\right) z \left(3 z - 7\right)
\end{aligned}\label{eq:charm-xzero-q1-22-2}
\end{equation}
\begin{equation}
\begin{aligned}[b]
C_{P_3,22}^{s\mid u\bar s,1,(1)}={}&12 \left(1-z\right)^{2} \cdot \left(2 z + 1\right)\,l\;-16 \left(1-z\right)^{2} \cdot \left(2 z + 1\right)\,H(z)\\
&{}+8 z \left(3 z^{3} - 16 z^{2} + 17 z - 2\right)\,\log z\;-8 \left(1-z\right)^{2} \cdot \left(3 z^{2} - 10 z - 2\right)\,\log(1-z)\\
&{}{+2 \left(1-z\right) \left(14 z^{2} - 9 z + 3\right)}
\end{aligned}\label{eq:charm-xzero-q1-22-3}
\end{equation}
\begin{equation}
\begin{aligned}[b]
C_{P_4,22}^{s\mid u\bar s,1,(1)}={}&7 \left(1-z\right)^{2} \cdot \left(2 z + 1\right)
\end{aligned}\label{eq:charm-xzero-q1-22-4}
\end{equation}
\begin{equation}
\begin{aligned}[b]
C_{P_5,22}^{s\mid u\bar s,1,(1)}={}&\frac{16 \left(1-z\right) \left(z + 1\right)}{3}\,l\;+\frac{2 z^{2} \cdot \left(5 z - 13\right)}{3}\,\log z\;-\frac{20 \left(1-z\right)^{2} \cdot \left(2 z + 1\right)}{9}\,\log(1-z)\\
&{}{-\frac{\left(1-z\right) \left(148 z^{2} - 204 z - 113\right)}{9}}
\end{aligned}\label{eq:charm-xzero-q1-22-9}
\end{equation}
\begin{equation}
\begin{aligned}[b]
C_{P_1,22}^{s\mid u\bar s,8,(1)}={}&44 \left(1-z\right) \left(z^{2} + z + 1\right)\,l\;-z^{2} \cdot \left(59 z + 9\right)\,\log z\\
&{}-2 \left(1-z\right) \left(28 z^{2} + 37 z + 37\right)\,\log(1-z)\;+\frac{\left(1-z\right) \left(637 z^{2} + 382 z + 331\right)}{6}
\end{aligned}\label{eq:charm-xzero-q1-22-5}
\end{equation}
\begin{equation}
\begin{aligned}[b]
C_{P_2,22}^{s\mid u\bar s,8,(1)}={}&72 \left(1-z\right) z^{2}\,l\;-3 z^{2} \cdot \left(59 z - 51\right)\,\log z\;-6 \left(1-z\right) \left(28 z^{2} + 3 z + 3\right)\,\log(1-z)\\
&{}{+\frac{\left(1-z\right) \left(459 z^{2} - 200 z - 35\right)}{2}}
\end{aligned}\label{eq:charm-xzero-q1-22-6}
\end{equation}
\begin{equation}
\begin{aligned}[b]
C_{P_3,22}^{s\mid u\bar s,8,(1)}={}&4 \left(1-z\right) \left(5 z^{2} - 4 z - 4\right)\,l\;-z^{2} \cdot \left(47 z - 63\right)\,\log z\\
&{}+22 \left(1-z\right)^{2} \cdot \left(2 z + 1\right)\,\log(1-z)\;+\frac{\left(1-z\right) \left(1169 z^{2} - 1003 z - 601\right)}{12}
\end{aligned}\label{eq:charm-xzero-q1-22-7}
\end{equation}
\begin{equation}
\begin{aligned}[b]
C_{P_4,22}^{s\mid u\bar s,8,(1)}={}&-36 \left(1-z\right) \left(z + 1\right)\,l\;-3 z^{2} \cdot \left(4 z - 9\right)\,\log z\;-6 \left(1-z\right) \left(2 z^{2} - 3 z - 3\right)\,\log(1-z)\\
&{}{-\frac{\left(1-z\right) \left(90 z^{2} + 311 z + 155\right)}{4}}
\end{aligned}\label{eq:charm-xzero-q1-22-8}
\end{equation}
\begin{equation}
\begin{aligned}[b]
C_{P_5,22}^{s\mid u\bar s,8,(1)}={}&-2 \left(1-z\right) \left(3 z^{2} + 5 z + 5\right)\,l\;-\frac{z^{2} \cdot \left(5 z - 13\right)}{2}\,\log z\\
&{}+\frac{5 \left(1-z\right)^{2} \cdot \left(2 z + 1\right)}{3}\,\log(1-z)\;+\frac{\left(1-z\right) \left(578 z^{2} - 315 z - 133\right)}{24}
\end{aligned}\label{eq:charm-xzero-q1-22-10}
\end{equation}
\begin{equation}
\begin{aligned}[b]
C_{P_1,12}^{s\mid u\bar s,1,(1)}={}&-32 \left(1-z\right) \left(z^{2} + z + 1\right)\,l\;-\frac{32 z \left(z + 1\right) \left(2 z - 1\right)}{3}\,H(z)\;-\frac{4 z^{2} \cdot \left(51 z + 44\right)}{3}\,\log z\\
&{}-\frac{4 \left(1-z\right) \left(27 z^{4} + 31 z^{3} - z^{2} + 3 z - 4\right)}{3 z^{2}}\,\log(1-z)\;+\frac{32 z \left(z + 1\right) \left(2 z - 1\right)}{9}\,\pi^2\\
&{}{-\frac{2 \left(1-z\right) \left(91 z^{3} + 57 z^{2} + 18 z - 8\right)}{3 z}}
\end{aligned}\label{eq:charm-xzero-q1-12-1}
\end{equation}
\begin{equation}
\begin{aligned}[b]
C_{P_2,12}^{s\mid u\bar s,1,(1)}={}&-192 \left(1-z\right) z^{2}\,l\;-\frac{32 z \left(6 z^{2} - 3 z - 1\right)}{3}\,H(z)\;+\frac{4 z^{2} \cdot \left(48 z^{2} - 217 z + 56\right)}{3}\,\log z\\
&{}+\frac{4 \left(1-z\right) \left(48 z^{4} - 97 z^{3} - z^{2} - z - 5\right)}{3 z}\,\log(1-z)\;+\frac{32 z \left(6 z^{2} - 3 z - 1\right)}{9}\,\pi^2\\
&{}{-\frac{2 \left(1-z\right) \left(369 z^{2} + 23 z - 6\right)}{3}}
\end{aligned}\label{eq:charm-xzero-q1-12-2}
\end{equation}
\begin{equation}
\begin{aligned}[b]
C_{P_3,12}^{s\mid u\bar s,1,(1)}={}&24 \left(1-z\right)^{2} \cdot \left(2 z + 1\right)\,l\;+\frac{64 \left(1-z\right) z^{2}}{3}\,H(z)\;+\frac{4 z^{2} \cdot \left(12 z^{2} - 67 z + 51\right)}{3}\,\log z\\
&{}-\frac{4 \left(1-z\right)^{2} \cdot \left(12 z^{4} - 19 z^{3} - 5 z^{2} + 5 z - 1\right)}{3 z^{2}}\,\log(1-z)\;-\frac{64 \left(1-z\right) z^{2}}{9}\,\pi^2\\
&{}{-\frac{2 \left(1-z\right) \left(103 z^{3} - 88 z^{2} - 5 z - 2\right)}{3 z}}
\end{aligned}\label{eq:charm-xzero-q1-12-3}
\end{equation}
\begin{equation}
\begin{aligned}[b]
C_{P_4,12}^{s\mid u\bar s,1,(1)}={}&-\frac{32 z}{3}\,H(z)\;+16 z^{2}\,\log z\;+\frac{32 \left(1-z\right) \left(z^{2} + z - 1\right)}{3 z}\,\log(1-z)\;+\frac{32 z}{9}\,\pi^2\\
&{}-\frac{2 \left(1-z\right) \left(14 z^{2} + 5 z + 61\right)}{3}
\end{aligned}\label{eq:charm-xzero-q1-12-4}
\end{equation}
\begin{equation}
\begin{aligned}[b]
C_{P_5,12}^{s\mid u\bar s,1,(1)}={}&\frac{32 \left(1-z\right) \left(z + 1\right)}{9}\,l\;+\frac{32 \left(1-z\right) z}{9}\,H(z)\;+\frac{4 z^{2} \cdot \left(4 z - 17\right)}{9}\,\log z\\
&{}-\frac{4 \left(1-z\right)^{2} \cdot \left(17 z^{3} + 4 z^{2} + 3 z + 12\right)}{27 z^{2}}\,\log(1-z)\;-\frac{32 \left(1-z\right) z}{27}\,\pi^2\\
&{}{-\frac{2 \left(1-z\right) \left(151 z^{3} - 75 z^{2} - 239 z + 24\right)}{27 z}}
\end{aligned}\label{eq:charm-xzero-q1-12-9}
\end{equation}
\begin{equation}
\begin{aligned}[b]
C_{P_1,12}^{s\mid u\bar s,8,(1)}={}&\frac{16 \left(1-z\right) \left(z^{2} + z + 1\right)}{3}\,l\;-8 \left(z^{2} - z + 2\right)\,H(z)\\
&{}-\frac{z \left(127 z^{2} + 174 z + 48\right)}{3}\,\log z\;-\frac{\left(1-z\right) \left(49 z^{4} + 169 z^{3} + 145 z^{2} + 9 z - 12\right)}{3 z^{2}}\,\log(1-z)\\
&{}+\frac{8 z \left(z + 1\right) \left(2 z - 1\right)}{3}\,\pi^2\;+\frac{\left(1-z\right) \left(95 z^{3} - 37 z^{2} + 284 z + 72\right)}{18 z}
\end{aligned}\label{eq:charm-xzero-q1-12-5}
\end{equation}
\begin{equation}
\begin{aligned}[b]
C_{P_2,12}^{s\mid u\bar s,8,(1)}={}&-24 \left(1-z\right) z^{2}\,l\;-8 z \left(3 z - 1\right)\,H(z)\;-z^{2} \cdot \left(127 z - 14\right)\,\log z\\
&{}-\frac{\left(1-z\right) \left(49 z^{3} + 45 z^{2} + 21 z + 5\right)}{z}\,\log(1-z)\;+\frac{8 z \left(6 z^{2} - 3 z - 1\right)}{3}\,\pi^2\\
&{}{-\frac{\left(1-z\right) \left(261 z^{2} + 301 z + 52\right)}{6}}
\end{aligned}\label{eq:charm-xzero-q1-12-6}
\end{equation}
\begin{equation}
\begin{aligned}[b]
C_{P_3,12}^{s\mid u\bar s,8,(1)}={}&-\frac{4 \left(1-z\right) \left(8 z^{2} - z - 1\right)}{3}\,l\;+8 \left(1-z\right) \left(z + 1\right)\,H(z)\\
&{}-\frac{z \left(103 z^{2} - 75 z - 24\right)}{3}\,\log z\;+\frac{\left(1-z\right)^{2} \cdot \left(25 z^{3} + 35 z^{2} - 15 z + 3\right)}{3 z^{2}}\,\log(1-z)\\
&{}-\frac{16 \left(1-z\right) z^{2}}{3}\,\pi^2\;-\frac{\left(1-z\right) \left(10 z^{3} + 49 z^{2} + 610 z - 18\right)}{18 z}
\end{aligned}\label{eq:charm-xzero-q1-12-7}
\end{equation}
\begin{equation}
\begin{aligned}[b]
C_{P_4,12}^{s\mid u\bar s,8,(1)}={}&-24 \left(1-z\right) \left(z + 1\right)\,l\;-8 z\,H(z)\;-2 z^{2} \cdot \left(4 z - 15\right)\,\log z\\
&{}-\frac{4 \left(1-z\right) \left(z + 1\right) \left(2 z^{2} - 7 z + 2\right)}{z}\,\log(1-z)\;+\frac{8 z}{3}\,\pi^2\;-\frac{\left(1-z\right) \left(90 z^{2} + 347 z + 359\right)}{6}
\end{aligned}\label{eq:charm-xzero-q1-12-8}
\end{equation}
\begin{equation}
\begin{aligned}[b]
C_{P_5,12}^{s\mid u\bar s,8,(1)}={}&-\frac{4 \left(1-z\right) \left(3 z^{2} + 5 z + 5\right)}{3}\,l\;+\frac{28 \left(1-z\right) z}{3}\,H(z)\;-\frac{z^{2} \cdot \left(17 z + 2\right)}{6}\,\log z\\
&{}+\frac{\left(1-z\right)^{2} \cdot \left(61 z^{3} + 62 z^{2} - 21 z - 84\right)}{18 z^{2}}\,\log(1-z)\;-\frac{28 \left(1-z\right) z}{9}\,\pi^2\\
&{}{+\frac{\left(1-z\right) \left(557 z^{3} - 1218 z^{2} + 749 z - 168\right)}{36 z}}
\end{aligned}\label{eq:charm-xzero-q1-12-10}
\end{equation}
\begin{equation}
\begin{aligned}[b]
C_{P_1,11}^{s\mid u\bar s,1,(1)}={}&-\frac{112 \left(1-z\right) \left(z^{2} + z + 1\right)}{27}\,l\;-\frac{32 \left(z^{3} + z^{2} - z + 1\right)}{9}\,H(z)\\
&{}-\frac{4 z \left(140 z^{2} + 153 z + 24\right)}{27}\,\log z\;-\frac{4 \left(1-z\right) \left(65 z^{4} + 131 z^{3} + 71 z^{2} + 9 z - 12\right)}{27 z^{2}}\,\log(1-z)\\
&{}+\frac{32 z \left(z + 1\right) \left(2 z - 1\right)}{27}\,\pi^2\;-\frac{2 \left(1-z\right) \left(362 z^{3} + 275 z^{2} - 61 z - 72\right)}{81 z}
\end{aligned}\label{eq:charm-xzero-q1-11-1}
\end{equation}
\begin{equation}
\begin{aligned}[b]
C_{P_2,11}^{s\mid u\bar s,1,(1)}={}&-\frac{112 \left(1-z\right) z^{2}}{3}\,l\;-\frac{32 z \left(3 z^{2} - 1\right)}{9}\,H(z)\;+\frac{4 z^{2} \cdot \left(24 z^{2} - 172 z + 35\right)}{9}\,\log z\\
&{}+\frac{4 \left(1-z\right) \left(24 z^{4} - 73 z^{3} - 23 z^{2} - 11 z - 5\right)}{9 z}\,\log(1-z)\;+\frac{32 z \left(6 z^{2} - 3 z - 1\right)}{27}\,\pi^2\\
&{}{-\frac{2 \left(1-z\right) \left(684 z^{2} + 185 z + 17\right)}{27}}
\end{aligned}\label{eq:charm-xzero-q1-11-2}
\end{equation}
\begin{equation}
\begin{aligned}[b]
C_{P_3,11}^{s\mid u\bar s,1,(1)}={}&-\frac{4 \left(1-z\right) \left(70 z^{2} - 29 z - 29\right)}{27}\,l\;+\frac{16 \left(1-z\right) \left(2 z^{2} + z + 1\right)}{9}\,H(z)\\
&{}+\frac{8 z \left(9 z^{3} - 76 z^{2} + 57 z + 6\right)}{27}\,\log z\;-\frac{4 \left(1-z\right)^{2} \left(18 z^{4} - 41 z^{3} - 25 z^{2} + 15 z - 3\right)}{27 z^{2}}\,\log(1-z)\\
&{}-\frac{64 \left(1-z\right) z^{2}}{27}\,\pi^2\;-\frac{\left(1-z\right) \left(937 z^{3} - 743 z^{2} + 565 z - 36\right)}{81 z}
\end{aligned}\label{eq:charm-xzero-q1-11-3}
\end{equation}
\begin{equation}
\begin{aligned}[b]
C_{P_4,11}^{s\mid u\bar s,1,(1)}={}&-\frac{16 \left(1-z\right) \left(z + 1\right)}{3}\,l\;-\frac{32 z}{9}\,H(z)\;-\frac{4 z^{2} \cdot \left(4 z - 21\right)}{9}\,\log z\\
&{}-\frac{8 \left(1-z\right) \left(2 z^{3} - 7 z^{2} - 7 z + 4\right)}{9 z}\,\log(1-z)\;+\frac{32 z}{27}\,\pi^2\;-\frac{2 \left(1-z\right) \left(66 z^{2} + 181 z + 271\right)}{27}
\end{aligned}\label{eq:charm-xzero-q1-11-4}
\end{equation}
\begin{equation}
\begin{aligned}[b]
C_{P_5,11}^{s\mid u\bar s,1,(1)}={}&\frac{8 \left(1-z\right)^{2} \cdot \left(2 z + 1\right)}{9}\,l\;-\frac{2 z^{2} \cdot \left(8 z - 9\right)}{9}\,\log z\\
&{}-\frac{2 \left(1-z\right)^{2} \cdot \left(17 z^{3} - 2 z^{2} + 3 z + 12\right)}{9 z^{2}}\,\log(1-z)\;-\frac{\left(1-z\right) \left(39 z^{3} + 57 z^{2} - 199 z + 48\right)}{18 z}
\end{aligned}\label{eq:charm-xzero-q1-11-9}
\end{equation}
\begin{equation}
\begin{aligned}[b]
C_{P_1,11}^{s\mid u\bar s,8,(1)}={}&\frac{316 \left(1-z\right) \left(z^{2} + z + 1\right)}{9}\,l\;+\frac{8 \cdot \left(10 z^{3} + z^{2} - z - 8\right)}{3}\,H(z)\\
&{}-\frac{4 z \left(46 z^{2} - 45 z + 48\right)}{9}\,\log z\;-\frac{2 \left(1-z\right) \left(314 z^{4} + 497 z^{3} + 473 z^{2} + 36 z - 48\right)}{9 z^{2}}\,\log(1-z)\\
&{}+\frac{8 \cdot \left(7 z^{3} - z^{2} + z - 9\right)}{9}\,\pi^2\;+\frac{\left(1-z\right) \left(3295 z^{3} + 565 z^{2} + 2395 z + 288\right)}{27 z}
\end{aligned}\label{eq:charm-xzero-q1-11-5}
\end{equation}
\begin{equation}
\begin{aligned}[b]
C_{P_2,11}^{s\mid u\bar s,8,(1)}={}&124 \left(1-z\right) z^{2}\,l\;+\frac{8 z \left(30 z^{2} - 27 z - 1\right)}{3}\,H(z)\;-\frac{4 z^{2} \cdot \left(6 z^{2} + 38 z - 61\right)}{3}\,\log z\\
&{}-\frac{2 \left(1-z\right) \left(12 z^{4} + 310 z^{3} + 65 z^{2} + 17 z + 20\right)}{3 z}\,\log(1-z)\;+\frac{8 z \left(21 z^{2} - 24 z + 1\right)}{9}\,\pi^2\\
&{}{+\frac{2 \left(1-z\right) \left(1764 z^{2} - 496 z - 133\right)}{9}}
\end{aligned}\label{eq:charm-xzero-q1-11-6}
\end{equation}
\begin{equation}
\begin{aligned}[b]
C_{P_3,11}^{s\mid u\bar s,8,(1)}={}&\frac{4 \left(1-z\right) \left(49 z^{2} - 23 z - 23\right)}{9}\,l\;-\frac{16 \left(1-z\right) \left(5 z^{2} - 2 z - 2\right)}{3}\,H(z)\\
&{}-\frac{2 z \left(9 z^{3} + 86 z^{2} - 159 z - 48\right)}{9}\,\log z\\
&{}+\frac{2 \left(1-z\right)^{2} \cdot \left(9 z^{4} + 326 z^{3} + 235 z^{2} - 60 z + 12\right)}{9 z^{2}}\,\log(1-z)\;-\frac{4 \left(1-z\right) \left(14 z^{2} - 9 z - 9\right)}{9}\,\pi^2\\
&{}{+\frac{\left(1-z\right) \left(9991 z^{3} + 49 z^{2} - 9119 z + 288\right)}{108 z}}
\end{aligned}\label{eq:charm-xzero-q1-11-7}
\end{equation}
\begin{equation}
\begin{aligned}[b]
C_{P_4,11}^{s\mid u\bar s,8,(1)}={}&-8 \left(1-z\right) \left(z + 1\right)\,l\;+\frac{8 z}{3}\,H(z)\;+\frac{4 z^{2} \left(z - 3\right)}{3}\,\log z\\
&{}+\frac{2 \left(1-z\right) \left(2 z^{3} + 29 z^{2} + 29 z - 32\right)}{3 z}\,\log(1-z)\;-\frac{8 z}{9}\,\pi^2\;-\frac{2 \left(1-z\right) \left(15 z^{2} - 97 z + 191\right)}{9}
\end{aligned}\label{eq:charm-xzero-q1-11-8}
\end{equation}
\begin{equation}
\begin{aligned}[b]
C_{P_5,11}^{s\mid u\bar s,8,(1)}={}&\frac{2 \left(1-z\right) \left(2 z^{2} + 5 z + 5\right)}{3}\,l\;+4 \left(1-z\right) z\,H(z)\;+\frac{10 z^{2} \left(z - 3\right)}{3}\,\log z\\
&{}-\left(1-z\right)^{2}\,\log(1-z)\;-\frac{4 \left(1-z\right) z}{3}\,\pi^2\;-\frac{\left(1-z\right) \left(263 z^{2} - 207 z - 279\right)}{24}
\end{aligned}\label{eq:charm-xzero-q1-11-10}
\end{equation}

\label{sec:charm-yzero}
With one strange cut line, the coefficients for external $u$ and the cut
$(s\bar d)$ are
\begin{equation}
\begin{aligned}[b]
C_{P_1,22}^{u\mid s\bar d,1,(1)}={}&-\frac{112 \left(1-z\right) \left(z^{2} + z + 1\right)}{27}\,l\;-\frac{32 \left(z^{3} - z^{2} - z - 3\right)}{9}\,H(z)\\
&{}-\frac{4 z \left(149 z^{2} - 69 z - 72\right)}{27}\,\log z\;-\frac{8 \left(1-z\right) \left(37 z^{4} + 31 z^{3} + 19 z^{2} + 9\right)}{27 z^{2}}\,\log(1-z)\\
&{}-\frac{32 z \left(2 z^{2} + z + 1\right)}{27}\,\pi^2\;-\frac{2 \left(1-z\right) \left(551 z^{3} + 500 z^{2} - 25 z + 108\right)}{81 z}
\end{aligned}\label{eq:charm-yzero-q2-22-1}
\end{equation}
\begin{equation}
\begin{aligned}[b]
C_{P_2,22}^{u\mid s\bar d,1,(1)}={}&-\frac{112 \left(1-z\right) z^{2}}{3}\,l\;-\frac{32 z \left(3 z^{2} - 6 z + 1\right)}{9}\,H(z)\;+\frac{4 z^{2} \cdot \left(24 z^{2} - 181 z + 205\right)}{9}\,\log z\\
&{}+\frac{8 \left(1-z\right) \left(12 z^{5} - 41 z^{4} + 5 z^{3} + z^{2} - 1\right)}{9 z^{2}}\,\log(1-z)\;-\frac{32 z \left(6 z^{2} - 3 z - 1\right)}{27}\,\pi^2\\
&{}{-\frac{2 \left(1-z\right) \left(873 z^{3} + 134 z^{2} - 7 z + 12\right)}{27 z}}
\end{aligned}\label{eq:charm-yzero-q2-22-2}
\end{equation}
\begin{equation}
\begin{aligned}[b]
C_{P_5,22}^{u\mid s\bar d,1,(1)}={}&\frac{8 \left(1-z\right)^{2} \cdot \left(2 z + 1\right)}{9}\,l\;+\frac{16 z \left(z + 1\right)}{3}\,H(z)\;-\frac{2 z^{2} \cdot \left(5 z + 39\right)}{9}\,\log z\\
&{}+\frac{8 \left(1-z\right) \left(5 z^{2} - 10 z - 7\right)}{9}\,\log(1-z)\;-\frac{16 z \left(z + 1\right)}{9}\,\pi^2\;-\frac{\left(1-z\right) \left(117 z^{2} - 405 z - 187\right)}{18}
\end{aligned}\label{eq:charm-yzero-q2-22-9}
\end{equation}
\begin{equation}
\begin{aligned}[b]
C_{P_1,22}^{u\mid s\bar d,8,(1)}={}&\frac{316 \left(1-z\right) \left(z^{2} + z + 1\right)}{9}\,l\;+\frac{16 \cdot \left(5 z^{3} + 4 z^{2} + 4 z + 3\right)}{3}\,H(z)\\
&{}-\frac{2 z \left(47 z^{2} - 195 z - 72\right)}{9}\,\log z\;-\frac{2 \left(1-z\right) \left(269 z^{4} + 221 z^{3} + 287 z^{2} - 9\right)}{9 z^{2}}\,\log(1-z)\\
&{}-\frac{8 \cdot \left(7 z^{3} + 8 z^{2} + 8 z + 9\right)}{9}\,\pi^2\;+\frac{\left(1-z\right) \left(3754 z^{3} + 3229 z^{2} + 2494 z + 54\right)}{27 z}
\end{aligned}\label{eq:charm-yzero-q2-22-5}
\end{equation}
\begin{equation}
\begin{aligned}[b]
C_{P_2,22}^{u\mid s\bar d,8,(1)}={}&124 \left(1-z\right) z^{2}\,l\;+\frac{16 z \left(15 z^{2} - 3 z - 4\right)}{3}\,H(z)\;-\frac{2 z^{2} \cdot \left(12 z^{2} + 31 z - 307\right)}{3}\,\log z\\
&{}-\frac{2 \left(1-z\right) \left(12 z^{5} + 265 z^{4} - 67 z^{3} - 17 z^{2} - 1\right)}{3 z^{2}}\,\log(1-z)\;-\frac{8 z \left(21 z^{2} + 3 z - 8\right)}{9}\,\pi^2\\
&{}{+\frac{\left(1-z\right) \left(3987 z^{3} - 896 z^{2} - 143 z + 6\right)}{9 z}}
\end{aligned}\label{eq:charm-yzero-q2-22-6}
\end{equation}
\begin{equation}
\begin{aligned}[b]
C_{P_5,22}^{u\mid s\bar d,8,(1)}={}&\frac{2 \left(1-z\right) \left(2 z^{2} + 5 z + 5\right)}{3}\,l\;+\frac{z^{2} \cdot \left(7 z - 9\right)}{3}\,\log z\\
&{}+\frac{\left(1-z\right)^{2} \left(z^{3} + 2 z^{2} + z - 1\right)}{z^{2}}\,\log(1-z)\;-\frac{\left(1-z\right) \left(323 z^{3} - 399 z^{2} - 315 z + 24\right)}{24 z}
\end{aligned}\label{eq:charm-yzero-q2-22-10}
\end{equation}
\begin{equation}
\begin{aligned}[b]
C_{P_1,12}^{u\mid s\bar d,1,(1)}={}&-32 \left(1-z\right) \left(z^{2} + z + 1\right)\,l\;-\frac{32 \cdot \left(2 z^{3} - z^{2} - z - 4\right)}{3}\,H(z)\\
&{}-\frac{8 z \left(27 z^{2} - 15 z - 16\right)}{3}\,\log z\;-\frac{8 \left(1-z\right) \left(15 z^{4} + 4 z^{3} - 6 z^{2} + 3\right)}{3 z^{2}}\,\log(1-z)\\
&{}-\frac{32 z \left(2 z^{2} + z + 1\right)}{9}\,\pi^2\;-\frac{4 \left(1-z\right) \left(56 z^{3} + 41 z^{2} + 11 z + 6\right)}{3 z}
\end{aligned}\label{eq:charm-yzero-q2-12-1}
\end{equation}
\begin{equation}
\begin{aligned}[b]
C_{P_2,12}^{u\mid s\bar d,1,(1)}={}&-192 \left(1-z\right) z^{2}\,l\;-\frac{32 z \left(6 z^{2} - 9 z + 1\right)}{3}\,H(z)\;+\frac{8 z^{2} \cdot \left(24 z^{2} - 113 z + 113\right)}{3}\,\log z\\
&{}+\frac{8 \left(1-z\right) \left(24 z^{5} - 53 z^{4} + 16 z^{3} + 6 z^{2} - 1\right)}{3 z^{2}}\,\log(1-z)\;-\frac{32 z \left(6 z^{2} - 3 z - 1\right)}{9}\,\pi^2\\
&{}{-\frac{4 \left(1-z\right) \left(216 z^{3} + 3 z^{2} - 7 z + 2\right)}{3 z}}
\end{aligned}\label{eq:charm-yzero-q2-12-2}
\end{equation}
\begin{equation}
\begin{aligned}[b]
C_{P_5,12}^{u\mid s\bar d,1,(1)}={}&\frac{32 \left(1-z\right) \left(z + 1\right)}{9}\,l\;+\frac{32 z \left(z + 1\right)}{9}\,H(z)\;+\frac{4 z^{2} \cdot \left(3 z - 19\right)}{9}\,\log z\\
&{}+\frac{8 \left(1-z\right) \left(7 z^{4} - 23 z^{3} - 11 z^{2} + 6 z - 3\right)}{27 z^{2}}\,\log(1-z)\;-\frac{32 z \left(z + 1\right)}{27}\,\pi^2\\
&{}{-\frac{2 \left(1-z\right) \left(220 z^{3} - 402 z^{2} - 251 z + 12\right)}{27 z}}
\end{aligned}\label{eq:charm-yzero-q2-12-9}
\end{equation}
\begin{equation}
\begin{aligned}[b]
C_{P_1,12}^{u\mid s\bar d,8,(1)}={}&\frac{16 \left(1-z\right) \left(z^{2} + z + 1\right)}{3}\,l\;+8 \left(z^{2} + z + 2\right)\,H(z)\;-\frac{8 z \left(17 z^{2} - 6 z - 6\right)}{3}\,\log z\\
&{}-\frac{2 \left(1-z\right) \left(29 z^{4} + 50 z^{3} + 56 z^{2} + 9\right)}{3 z^{2}}\,\log(1-z)\;-\frac{8 z \left(2 z^{2} + z + 1\right)}{3}\,\pi^2\\
&{}{-\frac{\left(1-z\right) \left(47 z^{3} + 131 z^{2} - 124 z + 54\right)}{9 z}}
\end{aligned}\label{eq:charm-yzero-q2-12-5}
\end{equation}
\begin{equation}
\begin{aligned}[b]
C_{P_2,12}^{u\mid s\bar d,8,(1)}={}&-24 \left(1-z\right) z^{2}\,l\;+8 z \left(3 z - 1\right)\,H(z)\;-8 z^{2} \cdot \left(17 z - 23\right)\,\log z\\
&{}-\frac{2 \left(1-z\right) \left(29 z^{4} + 6 z^{3} + 4 z^{2} + 1\right)}{z^{2}}\,\log(1-z)\;-\frac{8 z \left(6 z^{2} - 3 z - 1\right)}{3}\,\pi^2\\
&{}{-\frac{\left(1-z\right) \left(225 z^{3} + 125 z^{2} + 14 z + 6\right)}{3 z}}
\end{aligned}\label{eq:charm-yzero-q2-12-6}
\end{equation}
\begin{equation}
\begin{aligned}[b]
C_{P_5,12}^{u\mid s\bar d,8,(1)}={}&-\frac{4 \left(1-z\right) \left(3 z^{2} + 5 z + 5\right)}{3}\,l\;+\frac{28 z \left(z + 1\right)}{3}\,H(z)\;-\frac{4 z^{2} \cdot \left(3 z + 2\right)}{3}\,\log z\\
&{}-\frac{\left(1-z\right) \left(41 z^{4} + 116 z^{3} + 32 z^{2} - 42 z + 21\right)}{9 z^{2}}\,\log(1-z)\;-\frac{28 z \left(z + 1\right)}{9}\,\pi^2\\
&{}{+\frac{\left(1-z\right) \left(74 z^{3} + 1071 z^{2} + 833 z - 84\right)}{36 z}}
\end{aligned}\label{eq:charm-yzero-q2-12-10}
\end{equation}
\begin{equation}
\begin{aligned}[b]
C_{P_1,11}^{u\mid s\bar d,1,(1)}={}&32 \left(1-z\right) \left(z^{2} + z + 1\right)\,H(z)\;-16 z \left(2 z + 1\right) \left(3 z - 2\right)\,\log z\\
&{}-48 \left(1-z\right) z \left(2 z + 1\right)\,\log(1-z)\;+8 \left(1-z\right) \left(5 z^{2} + 4 z + 3\right)
\end{aligned}\label{eq:charm-yzero-q2-11-1}
\end{equation}
\begin{equation}
\begin{aligned}[b]
C_{P_2,11}^{u\mid s\bar d,1,(1)}={}&-144 \left(1-z\right) z^{2}\,l\;+96 \left(1-z\right) z^{2}\,H(z)\;+32 z^{2} \cdot \left(3 z^{2} - 13 z + 9\right)\,\log z\\
&{}+16 \left(1-z\right) \left(6 z^{3} - 20 z^{2} + 4 z + 1\right)\,\log(1-z)\;+8 \left(1-z\right) z \left(3 z - 7\right)
\end{aligned}\label{eq:charm-yzero-q2-11-2}
\end{equation}
\begin{equation}
\begin{aligned}[b]
C_{P_5,11}^{u\mid s\bar d,1,(1)}={}&\frac{16 \left(1-z\right) \left(z + 1\right)}{3}\,l\;+\frac{2 z^{2} \cdot \left(5 z - 13\right)}{3}\,\log z\;-\frac{20 \left(1-z\right)^{2} \cdot \left(2 z + 1\right)}{9}\,\log(1-z)\\
&{}{-\frac{\left(1-z\right) \left(148 z^{2} - 204 z - 113\right)}{9}}
\end{aligned}\label{eq:charm-yzero-q2-11-9}
\end{equation}
\begin{equation}
\begin{aligned}[b]
C_{P_1,11}^{u\mid s\bar d,8,(1)}={}&44 \left(1-z\right) \left(z^{2} + z + 1\right)\,l\;-z^{2} \cdot \left(59 z + 9\right)\,\log z\\
&{}-2 \left(1-z\right) \left(28 z^{2} + 37 z + 37\right)\,\log(1-z)\;+\frac{\left(1-z\right) \left(637 z^{2} + 382 z + 331\right)}{6}
\end{aligned}\label{eq:charm-yzero-q2-11-5}
\end{equation}
\begin{equation}
\begin{aligned}[b]
C_{P_2,11}^{u\mid s\bar d,8,(1)}={}&72 \left(1-z\right) z^{2}\,l\;-3 z^{2} \cdot \left(59 z - 51\right)\,\log z\;-6 \left(1-z\right) \left(28 z^{2} + 3 z + 3\right)\,\log(1-z)\\
&{}{+\frac{\left(1-z\right) \left(459 z^{2} - 200 z - 35\right)}{2}}
\end{aligned}\label{eq:charm-yzero-q2-11-6}
\end{equation}
\begin{equation}
\begin{aligned}[b]
C_{P_5,11}^{u\mid s\bar d,8,(1)}={}&-2 \left(1-z\right) \left(3 z^{2} + 5 z + 5\right)\,l\;-\frac{z^{2} \cdot \left(5 z - 13\right)}{2}\,\log z\\
&{}+\frac{5 \left(1-z\right)^{2} \cdot \left(2 z + 1\right)}{3}\,\log(1-z)\;+\frac{\left(1-z\right) \left(578 z^{2} - 315 z - 133\right)}{24}
\end{aligned}\label{eq:charm-yzero-q2-11-10}
\end{equation}
For external $s$ with the cut $(su)$ they are
\begin{equation}
\begin{aligned}[b]
C_{P_1,22}^{s\mid su,1,(1)}={}&-\frac{112 \left(1-z\right) \left(z + 1\right)}{9}\,l\;+\frac{32 \cdot \left(3 z^{2} - 1\right)}{3}\,H(z)\;-\frac{8 z \left(8 z^{2} - 17 z + 4\right)}{3}\,\log z\\
&{}-\frac{16 \left(1-z\right) \left(12 z^{3} + 7 z^{2} + z + 8\right)}{9 z}\,\log(1-z)\;-\frac{64}{9}\,\pi^2\;+\frac{8 \left(1-z\right) \left(5 z^{2} - 45 z - 138\right)}{27}
\end{aligned}\label{eq:charm-yzero-q3-22-1}
\end{equation}
\begin{equation}
\begin{aligned}[b]
C_{P_2,22}^{s\mid su,1,(1)}={}&\frac{8 z^{2}}{9}\,\log z\;+\frac{16 \left(1-z\right)^{2}}{3}\,\log(1-z)\;+\frac{4 \left(1-z\right) \left(z + 1\right) \left(10 z + 3\right)}{9}
\end{aligned}\label{eq:charm-yzero-q3-22-2}
\end{equation}
\begin{equation}
\begin{aligned}[b]
C_{P_3,22}^{s\mid su,1,(1)}={}&-\frac{16 \left(1-z\right) \left(z + 1\right)}{9}\,l\;-\frac{40 z^{2}}{9}\,\log z\;-\frac{16 \left(1-z\right) \left(3 z^{2} + 1\right)}{9 z}\,\log(1-z)\\
&{}{+\frac{\left(1-z\right) \left(40 z^{2} - 357 z - 429\right)}{27}}
\end{aligned}\label{eq:charm-yzero-q3-22-3}
\end{equation}
\begin{equation}
\begin{aligned}[b]
C_{P_4,22}^{s\mid su,1,(1)}={}&-\frac{16 \left(1-z\right) \left(z + 1\right)}{3}\,l\;-\frac{16 z^{2}}{3}\,\log z\;-\frac{16 \left(1-z\right)^{2} \cdot \left(2 z + 1\right)}{9 z}\,\log(1-z)\\
&{}-\frac{40 \left(1-z\right) \left(z + 1\right)}{3}
\end{aligned}\label{eq:charm-yzero-q3-22-4}
\end{equation}
\begin{equation}
\begin{aligned}[b]
C_{P_5,22}^{s\mid su,1,(1)}={}&\frac{8 \left(1-z\right) \left(z + 1\right)}{3}\,l\;-4 z^{2}\,\log z\;-\frac{32 \left(1-z\right) \left(z + 1\right)}{3}\,\log(1-z)\\
&{}-\frac{\left(1-z\right) \left(88 z^{2} - 255 z - 279\right)}{18}
\end{aligned}\label{eq:charm-yzero-q3-22-9}
\end{equation}
\begin{equation}
\begin{aligned}[b]
C_{P_1,22}^{s\mid su,8,(1)}={}&\frac{316 \left(1-z\right) \left(z + 1\right)}{3}\,l\;+16 \cdot \left(3 z^{2} + 5\right)\,H(z)\;+8 z \left(2 z^{2} - 5 z + 10\right)\,\log z\\
&{}+\frac{4 \left(1-z\right) \left(12 z^{3} - 119 z^{2} - 125 z + 8\right)}{3 z}\,\log(1-z)\;-\frac{8 \cdot \left(9 z^{2} + 7\right)}{3}\,\pi^2\\
&{}{+\frac{\left(1-z\right) \left(44 z^{2} + 2349 z + 1833\right)}{9}}
\end{aligned}\label{eq:charm-yzero-q3-22-5}
\end{equation}
\begin{equation}
\begin{aligned}[b]
C_{P_2,22}^{s\mid su,8,(1)}={}&\frac{16 z^{2}}{3}\,\log z\;-4 \left(1-z\right)^{2}\,\log(1-z)\;+\frac{4 \left(1-z\right) \left(11 z^{2} - 10 z - 3\right)}{3}
\end{aligned}\label{eq:charm-yzero-q3-22-6}
\end{equation}
\begin{equation}
\begin{aligned}[b]
C_{P_3,22}^{s\mid su,8,(1)}={}&\frac{76 \left(1-z\right) \left(z + 1\right)}{3}\,l\;-\frac{80 z^{2}}{3}\,\log z\;-\frac{4 \left(1-z\right) \left(15 z^{2} + 18 z - 1\right)}{3 z}\,\log(1-z)\\
&{}{+\frac{\left(1-z\right) \left(176 z^{2} - 381 z - 93\right)}{36}}
\end{aligned}\label{eq:charm-yzero-q3-22-7}
\end{equation}
\begin{equation}
\begin{aligned}[b]
C_{P_4,22}^{s\mid su,8,(1)}={}&4 \left(1-z\right) \left(z + 1\right)\,l\;-20 z^{2}\,\log z\;+\frac{4 \left(1-z\right)^{2} \cdot \left(2 z + 1\right)}{3 z}\,\log(1-z)\\
&{}-4 \left(1-z\right) \left(2 z - 1\right)
\end{aligned}\label{eq:charm-yzero-q3-22-8}
\end{equation}
\begin{equation}
\begin{aligned}[b]
C_{P_5,22}^{s\mid su,8,(1)}={}&10 \left(1-z\right) \left(z + 1\right)\,l\;+4 z^{2}\,\log z\;+6 \left(1-z\right)^{2}\,\log(1-z)\\
&{}{-\frac{\left(1-z\right) \left(112 z^{2} - 741 z - 789\right)}{24}}
\end{aligned}\label{eq:charm-yzero-q3-22-10}
\end{equation}
\begin{equation}
\begin{aligned}[b]
C_{P_1,12}^{s\mid su,1,(1)}={}&64 z^{2}\,H(z)\;+64 z^{2}\,\log z\;-\frac{64 \left(1-z\right) \left(5 z^{2} - 4 z + 2\right)}{3 z}\,\log(1-z)\;-\frac{64}{3}\,\pi^2\\
&{}{+\frac{16 \left(1-z\right) \left(2 z^{2} + 131 z + 11\right)}{9}}
\end{aligned}\label{eq:charm-yzero-q3-12-1}
\end{equation}
\begin{equation}
\begin{aligned}[b]
C_{P_2,12}^{s\mid su,1,(1)}={}&32 \left(1-z\right)^{2}\,\log(1-z)\;+\frac{32 \left(1-z\right) \left(z^{2} + 4 z + 1\right)}{3}
\end{aligned}\label{eq:charm-yzero-q3-12-2}
\end{equation}
\begin{equation}
\begin{aligned}[b]
C_{P_3,12}^{s\mid su,1,(1)}={}&-\frac{32 z^{2}}{3}\,\log z\;+\frac{16 \left(1-z\right)^{2} \cdot \left(4 z - 1\right)}{3 z}\,\log(1-z)\;+\frac{32 \left(1-z\right) \left(z^{2} + 7 z - 2\right)}{9}
\end{aligned}\label{eq:charm-yzero-q3-12-3}
\end{equation}
\begin{equation}
\begin{aligned}[b]
C_{P_4,12}^{s\mid su,1,(1)}={}&-\frac{64 z^{2}}{3}\,\log z\;-\frac{16 \left(1-z\right)^{2} \cdot \left(2 z + 1\right)}{3 z}\,\log(1-z)\;+\frac{2 \left(1-z\right) \left(33 z + 1\right)}{3}
\end{aligned}\label{eq:charm-yzero-q3-12-4}
\end{equation}
\begin{equation}
\begin{aligned}[b]
C_{P_5,12}^{s\mid su,1,(1)}={}&\frac{32 \left(1-z\right) \left(z + 1\right)}{3}\,l\;+\frac{8 z^{2}}{9}\,\log z\;-\frac{16 \left(1-z\right) \left(7 z + 1\right)}{9}\,\log(1-z)\\
&{}{-\frac{4 \left(1-z\right) \left(50 z^{2} - 249 z - 267\right)}{27}}
\end{aligned}\label{eq:charm-yzero-q3-12-9}
\end{equation}
\begin{equation}
\begin{aligned}[b]
C_{P_1,12}^{s\mid su,8,(1)}={}&-56 \left(1-z\right) \left(z + 1\right)\,l\;+48 \cdot \left(2 z^{2} - 1\right)\,H(z)\;-12 z \left(8 z^{2} - 13 z + 4\right)\,\log z\\
&{}-\frac{8 \left(1-z\right) \left(12 z^{3} - 3 z^{2} + 9 z + 4\right)}{z}\,\log(1-z)\;-16\,\pi^2\;+\frac{4 \left(1-z\right) \left(3 z^{2} - 176 z - 149\right)}{3}
\end{aligned}\label{eq:charm-yzero-q3-12-5}
\end{equation}
\begin{equation}
\begin{aligned}[b]
C_{P_2,12}^{s\mid su,8,(1)}={}&4 z^{2}\,\log z\;+2 \left(1-z\right) \left(6 z^{2} - 3 z - 1\right)
\end{aligned}\label{eq:charm-yzero-q3-12-6}
\end{equation}
\begin{equation}
\begin{aligned}[b]
C_{P_3,12}^{s\mid su,8,(1)}={}&-8 \left(1-z\right) \left(z + 1\right)\,l\;-12 z^{2}\,\log z\;-\frac{4 \left(1-z\right) \left(2 z^{2} + 5 z + 1\right)}{z}\,\log(1-z)\\
&{}{+\frac{\left(1-z\right) \left(24 z^{2} - 469 z - 397\right)}{6}}
\end{aligned}\label{eq:charm-yzero-q3-12-7}
\end{equation}
\begin{equation}
\begin{aligned}[b]
C_{P_4,12}^{s\mid su,8,(1)}={}&-24 \left(1-z\right) \left(z + 1\right)\,l\;-8 z^{2}\,\log z\;-\frac{4 \left(1-z\right)^{2} \cdot \left(2 z + 1\right)}{z}\,\log(1-z)\\
&{}-\frac{\left(1-z\right) \left(153 z + 121\right)}{2}
\end{aligned}\label{eq:charm-yzero-q3-12-8}
\end{equation}
\begin{equation}
\begin{aligned}[b]
C_{P_5,12}^{s\mid su,8,(1)}={}&-20 \left(1-z\right) \left(z + 1\right)\,l\;+\frac{70 z^{2}}{3}\,\log z\;-\frac{16 \left(1-z\right) \left(5 z - 7\right)}{3}\,\log(1-z)\\
&{}{-\frac{\left(1-z\right) \left(808 z^{2} - 3 z - 363\right)}{36}}
\end{aligned}\label{eq:charm-yzero-q3-12-10}
\end{equation}
For weak row $11$, seven rows equal those of weak row $22$:
\begin{equation}
 C_{P_i,11}^{s\mid su,{\rm c},(1)}=C_{P_i,22}^{s\mid su,{\rm c},(1)},\qquad
 \begin{cases}i=1,\ldots,5,&{\rm c}=1,\\ i=2,5,&{\rm c}=8.\end{cases}
 \label{eq:s-su-weak11}
\end{equation}
The remaining three octet rows are
\begin{equation}
\begin{aligned}[b]
C_{P_1,11}^{s\mid su,8,(1)}={}&\frac{316 \left(1-z\right) \left(z + 1\right)}{3}\,l\;+16 \cdot \left(3 z^{2} - 4\right)\,H(z)\;+8 z \left(2 z^{2} - 17 z - 8\right)\,\log z\\
&{}+\frac{4 \left(1-z\right) \left(12 z^{3} - 191 z^{2} - 197 z - 64\right)}{3 z}\,\log(1-z)\;+\frac{8 \cdot \left(9 z^{2} - 7\right)}{3}\,\pi^2\\
&{}{+\frac{\left(1-z\right) \left(44 z^{2} + 2349 z + 969\right)}{9}}
\end{aligned}\label{eq:charm-yzero-q3-11-5}
\end{equation}
\begin{equation}
\begin{aligned}[b]
C_{P_3,11}^{s\mid su,8,(1)}={}&\frac{76 \left(1-z\right) \left(z + 1\right)}{3}\,l\;-\frac{8 z^{2}}{3}\,\log z\;+\frac{4 \left(1-z\right) \left(3 z^{2} - 27 z - 8\right)}{3 z}\,\log(1-z)\\
&{}{+\frac{\left(1-z\right) \left(176 z^{2} - 1677 z - 525\right)}{36}}
\end{aligned}\label{eq:charm-yzero-q3-11-7}
\end{equation}
\begin{equation}
\begin{aligned}[b]
C_{P_4,11}^{s\mid su,8,(1)}={}&4 \left(1-z\right) \left(z + 1\right)\,l\;+4 z^{2}\,\log z\;-\frac{32 \left(1-z\right)^{2} \cdot \left(2 z + 1\right)}{3 z}\,\log(1-z)\\
&{}-4 \left(1-z\right) \left(11 z + 2\right)
\end{aligned}\label{eq:charm-yzero-q3-11-8}
\end{equation}
\begingroup\setlength{\abovedisplayskip}{5pt}\setlength{\belowdisplayskip}{5pt}\setlength{\abovedisplayshortskip}{3pt}\setlength{\belowdisplayshortskip}{4pt}\setlength{\parskip}{2pt}\setlength{\jot}{2pt}

\label{sec:charm-equal}
For external $u$ with the equal-mass cut $(s\bar s)$, two common functions   collect the repeated logarithm--dilogarithm terms:
\begin{equation}
\begin{aligned}
\Phi(z)={}&3\log\!\left(\frac{z}{\beta^2}\right)\log\!\left(\frac{1+\beta}{1-\beta}\right)
 -6\left[\operatorname{Li}_2\!\left(-\frac{2\beta}{1-\beta}\right)
             -\operatorname{Li}_2\!\left(\frac{2\beta}{1+\beta}\right)\right]\\
&+4\bigl[\operatorname{Li}_2\bigl(\tfrac{1-\beta}{2}\bigr)-\operatorname{Li}_2\bigl(\tfrac{1+\beta}{2}\bigr)\bigr]
 -2\left[\operatorname{Re}\operatorname{Li}_2\!\left(\frac{1+\beta}{2\beta}\right)
             -\operatorname{Li}_2\!\left(-\frac{1-\beta}{2\beta}\right)\right]-\pi^2,
\end{aligned}
\label{eq:equal-Phi}
\end{equation}
\begin{equation}
\begin{aligned}
\Psi(z)={}&4\log\beta\log\!\left(\frac{1+\beta}{1-\beta}\right)
 -3\log^2\!\left(\frac{1+\beta}{1-\beta}\right)
 +6\left[\operatorname{Li}_2\!\left(-\frac{2\beta}{1-\beta}\right)
             -\operatorname{Li}_2\!\left(\frac{2\beta}{1+\beta}\right)\right]\\
&-4\bigl[\operatorname{Li}_2\bigl(\tfrac{1-\beta}{2}\bigr)-\operatorname{Li}_2\bigl(\tfrac{1+\beta}{2}\bigr)\bigr]
 -4\left[\operatorname{Re}\operatorname{Li}_2\!\left(\frac{(1+\beta)^2}{(1-\beta)^2}\right)
             -\operatorname{Re}\operatorname{Li}_2\!\left(\frac{1+\beta}{1-\beta}\right)\right].
\end{aligned}
\label{eq:equal-Psi}
\end{equation}
The real parts specify the boundary values on the dilogarithm cut. The
coefficients are
\begin{equation}
\begin{aligned}[b]
C_{P_1,22}^{u\mid s\bar s,1,(1)}={}&\frac{112 \cdot \left(2 z^{2} + 2 z - 1\right)}{27 \beta}\,l\;+\frac{16 \cdot \left(2 z^{2} - 1\right)}{9}\,\Phi(z)\;+\frac{16 z}{9}\,\Psi(z)\\
&{}-\frac{4 \cdot \left(616 z^{4} + 256 z^{3} - 104 z^{2} - 18 z + 3\right)}{27 \beta z^{2}}\,\log z\;+\frac{8 \cdot \left(376 z^{3} + 220 z^{2} - 92 z - 3\right)}{27 \beta z}\,\log\beta\\
&{}+\frac{4 \cdot \left(480 z^{6} + 192 z^{5} - 174 z^{4} - 536 z^{3} + 74 z^{2} + 24 z - 3\right)}{27 \beta^{2} z^{2}}\,\log\!\left(\frac{1+\beta}{1-\beta}\right)\\
&{}{+\frac{2 \left(1440 z^{4} - 1412 z^{3} - 1298 z^{2} + 169 z - 36\right)}{81 \beta z}}
\end{aligned}\label{eq:charm-equal-q2-22-1}
\end{equation}
\begin{equation}
\begin{aligned}[b]
C_{P_2,22}^{u\mid s\bar s,1,(1)}={}&-\frac{224 z^{2}}{3 \beta}\,l\;-\frac{64 z^{2}}{9}\,\Phi(z)\;+\frac{4 \cdot \left(424 z^{4} - 104 z^{3} + 6 z - 1\right)}{9 \beta z^{2}}\,\log z\\
&{}-\frac{8 \cdot \left(216 z^{3} - 36 z^{2} - 8 z + 3\right)}{9 \beta z}\,\log\beta\\
&{}+\frac{4 \cdot \left(480 z^{6} - 608 z^{5} + 378 z^{4} - 112 z^{3} - 10 z^{2} + 8 z - 1\right)}{9 \beta^{2} z^{2}}\,\log\!\left(\frac{1+\beta}{1-\beta}\right)\\
&{}{+\frac{2 \left(1440 z^{4} - 2080 z^{3} - 182 z^{2} - 5 z - 12\right)}{27 \beta z}}
\end{aligned}\label{eq:charm-equal-q2-22-2}
\end{equation}
\begin{equation}
\begin{aligned}[b]
C_{P_5,22}^{u\mid s\bar s,1,(1)}={}&\frac{8 \beta \left(2 z + 1\right)}{9}\,l\;+\frac{4 z}{3}\,\Psi(z)\;+\frac{4 \beta \left(16 z^{3} + 5 z^{2} + 9 z - 3\right)}{9 z^{2}}\,\log z\\
&{}-\frac{4 \beta \left(32 z^{2} + 22 z - 3\right)}{9 z}\,\log\beta\;-\frac{4 \cdot \left(8 z^{5} + 21 z^{4} + 7 z^{2} - 15 z + 3\right)}{9 z^{2}}\,\log\!\left(\frac{1+\beta}{1-\beta}\right)\\
&{}{+\frac{128 z^{4} - 1202 z^{3} - 910 z^{2} + 499 z - 48}{18 \beta z}}
\end{aligned}\label{eq:charm-equal-q2-22-9}
\end{equation}
\begin{equation}
\begin{aligned}[b]
C_{P_1,22}^{u\mid s\bar s,8,(1)}={}&-\frac{316 \cdot \left(2 z^{2} + 2 z - 1\right)}{9 \beta}\,l\;-\frac{4 \cdot \left(2 z^{2} - 1\right)}{3}\,\Phi(z)\;+\frac{2 \cdot \left(7 z + 9\right)}{3}\,\Psi(z)\\
&{}-\frac{2120 z^{4} + 3020 z^{3} - 1642 z^{2} + 396 z - 57}{9 \beta z^{2}}\,\log z\;+\frac{4 \cdot \left(856 z^{3} + 1150 z^{2} - 395 z - 12\right)}{9 \beta z}\,\log\beta\\
&{}+\frac{4416 z^{6} + 1392 z^{5} + 768 z^{4} - 5692 z^{3} + 2320 z^{2} - 510 z + 57}{9 \beta^{2} z^{2}}\,\log\!\left(\frac{1+\beta}{1-\beta}\right)\\
&{}{+\frac{2 \left(3312 z^{4} - 5137 z^{3} - 2074 z^{2} + 158 z + 171\right)}{27 \beta z}}
\end{aligned}\label{eq:charm-equal-q2-22-5}
\end{equation}
\begin{equation}
\begin{aligned}[b]
C_{P_2,22}^{u\mid s\bar s,8,(1)}={}&\frac{248 z^{2}}{\beta}\,l\;+\frac{16 z^{2}}{3}\,\Phi(z)\;-6 z\,\Psi(z)\;+\frac{2312 z^{4} - 508 z^{3} - 54 z^{2} + 12 z + 1}{3 \beta z^{2}}\,\log z\\
&{}-\frac{4 \cdot \left(936 z^{3} - 126 z^{2} - 41 z + 12\right)}{3 \beta z}\,\log\beta\\
&{}+\frac{4416 z^{6} - 5584 z^{5} + 2448 z^{4} - 428 z^{3} - 80 z^{2} + 10 z + 1}{3 \beta^{2} z^{2}}\,\log\!\left(\frac{1+\beta}{1-\beta}\right)\\
&{}{+\frac{6624 z^{4} + 7700 z^{3} - 1034 z^{2} - 353 z + 6}{9 \beta z}}
\end{aligned}\label{eq:charm-equal-q2-22-6}
\end{equation}
\begin{equation}
\begin{aligned}[b]
C_{P_5,22}^{u\mid s\bar s,8,(1)}={}&-\frac{2 \cdot \left(16 z^{2} + 10 z - 5\right)}{3 \beta}\,l\;+z\,\Psi(z)\;-\frac{\beta \left(4 z^{3} + 2 z^{2} - 6 z + 1\right)}{2 z^{2}}\,\log z\\
&{}+2 \beta \left(2 z - 1\right)\,\log\beta\;+\frac{32 z^{5} + 96 z^{4} - 24 z^{3} - 36 z^{2} + 24 z - 3}{6 z^{2}}\,\log\!\left(\frac{1+\beta}{1-\beta}\right)\\
&{}{-\frac{256 z^{4} + 1498 z^{3} + 966 z^{2} - 483 z + 24}{24 \beta z}}
\end{aligned}\label{eq:charm-equal-q2-22-10}
\end{equation}
\begin{equation}
\begin{aligned}[b]
C_{P_1,12}^{u\mid s\bar s,1,(1)}={}&\frac{32 \cdot \left(2 z^{2} + 2 z - 1\right)}{\beta}\,l\;+\frac{32 \cdot \left(2 z^{2} - 1\right)}{3}\,\Phi(z)\;+\frac{16 z}{3}\,\Psi(z)\\
&{}+\frac{4 \beta \left(60 z^{3} + 18 z^{2} + 2 z - 1\right)}{3 z^{2}}\,\log z\;-\frac{8 \beta \left(32 z^{2} + 10 z + 1\right)}{3 z}\,\log\beta\\
&{}+\frac{4 \cdot \left(336 z^{6} + 28 z^{5} - 188 z^{4} - 72 z^{3} + 8 z - 1\right)}{3 \beta^{2} z^{2}}\,\log\!\left(\frac{1+\beta}{1-\beta}\right)\\
&{}{+\frac{4 \left(168 z^{4} - 54 z^{3} - 49 z^{2} - 3 z - 2\right)}{3 \beta z}}
\end{aligned}\label{eq:charm-equal-q2-12-1}
\end{equation}
\begin{equation}
\begin{aligned}[b]
C_{P_2,12}^{u\mid s\bar s,1,(1)}={}&-\frac{384 z^{2}}{\beta}\,l\;-\frac{128 z^{2}}{3}\,\Phi(z)\;-\frac{4 \beta \left(132 z^{3} + 10 z^{2} - 2 z + 1\right)}{3 z^{2}}\,\log z\\
&{}+\frac{8 \beta \left(64 z^{2} + 10 z - 3\right)}{3 z}\,\log\beta\\
&{}+\frac{4 \cdot \left(1008 z^{6} - 1036 z^{5} + 452 z^{4} - 64 z^{3} - 28 z^{2} + 8 z - 1\right)}{3 \beta^{2} z^{2}}\,\log\!\left(\frac{1+\beta}{1-\beta}\right)\\
&{}{+\frac{4 \left(504 z^{4} - 594 z^{3} - 15 z^{2} + 5 z - 2\right)}{3 \beta z}}
\end{aligned}\label{eq:charm-equal-q2-12-2}
\end{equation}
\begin{equation}
\begin{aligned}[b]
C_{P_5,12}^{u\mid s\bar s,1,(1)}={}&-\frac{32 \cdot \left(4 z^{2} + 2 z - 1\right)}{9 \beta}\,l\;+\frac{16 z}{9}\,\Psi(z)\;+\frac{4 \beta \left(20 z^{3} + 4 z^{2} + 36 z - 9\right)}{27 z^{2}}\,\log z\\
&{}-\frac{8 \beta \left(2 z + 3\right) \left(10 z - 1\right)}{27 z}\,\log\beta\;+\frac{4 \cdot \left(16 z^{5} + 54 z^{4} - 24 z^{3} - 50 z^{2} + 54 z - 9\right)}{27 z^{2}}\,\log\!\left(\frac{1+\beta}{1-\beta}\right)\\
&{}{-\frac{2 \left(64 z^{4} + 1350 z^{3} + 938 z^{2} - 491 z + 36\right)}{27 \beta z}}
\end{aligned}\label{eq:charm-equal-q2-12-9}
\end{equation}
\begin{equation}
\begin{aligned}[b]
C_{P_1,12}^{u\mid s\bar s,8,(1)}={}&-\frac{16 \cdot \left(2 z^{2} + 2 z - 1\right)}{3 \beta}\,l\;+4 z\,\Psi(z)\;-\frac{512 z^{4} + 476 z^{3} - 178 z^{2} - 18 z + 3}{3 \beta z^{2}}\,\log z\\
&{}+\frac{2 \cdot \left(368 z^{3} + 416 z^{2} - 166 z - 3\right)}{3 \beta z}\,\log\beta\;+\frac{12 z^{5} - 72 z^{4} - 72 z^{3} + 196 z^{2} + 12 z - 3}{3 z^{2}}\,\log\!\left(\frac{1+\beta}{1-\beta}\right)\\
&{}{-\frac{72 z^{4} + 926 z^{3} + 857 z^{2} - 196 z + 18}{9 \beta z}}
\end{aligned}\label{eq:charm-equal-q2-12-5}
\end{equation}
\begin{equation}
\begin{aligned}[b]
C_{P_2,12}^{u\mid s\bar s,8,(1)}={}&-\frac{48 z^{2}}{\beta}\,l\;+\frac{320 z^{4} - 116 z^{3} + 18 z^{2} + 6 z - 1}{\beta z^{2}}\,\log z\\
&{}-\frac{2 \cdot \left(176 z^{3} - 48 z^{2} + 6 z + 3\right)}{\beta z}\,\log\beta\;+\frac{12 z^{5} + 48 z^{4} - 64 z^{3} + 24 z^{2} + 4 z - 1}{z^{2}}\,\log\!\left(\frac{1+\beta}{1-\beta}\right)\\
&{}{-\frac{72 z^{4} + 298 z^{3} + 137 z^{2} + 20 z + 6}{3 \beta z}}
\end{aligned}\label{eq:charm-equal-q2-12-6}
\end{equation}
\begin{equation}
\begin{aligned}[b]
C_{P_5,12}^{u\mid s\bar s,8,(1)}={}&\frac{4 \cdot \left(14 z^{2} + 10 z - 5\right)}{3 \beta}\,l\;+\frac{14 z}{3}\,\Psi(z)\;-\frac{\beta \left(40 z^{3} + 62 z^{2} - 252 z + 63\right)}{18 z^{2}}\,\log z\\
&{}+\frac{\beta \left(40 z^{2} - 106 z + 21\right)}{9 z}\,\log\beta\;+\frac{292 z^{5} + 216 z^{4} - 168 z^{3} - 440 z^{2} + 378 z - 63}{18 z^{2}}\,\log\!\left(\frac{1+\beta}{1-\beta}\right)\\
&{}{-\frac{1168 z^{4} - 1704 z^{3} + 6356 z^{2} - 2513 z + 252}{36 \beta z}}
\end{aligned}\label{eq:charm-equal-q2-12-10}
\end{equation}
\begin{equation}
\begin{aligned}[b]
C_{P_1,11}^{u\mid s\bar s,1,(1)}={}&16 \cdot \left(2 z^{2} - 1\right)\,\Phi(z)\;+96 \beta z\,\log z\;-128 \beta z\,\log\beta\;+\frac{32 z \left(24 z^{3} - 2 z^{2} - 17 z + 1\right)}{\beta^{2}}\,\log\!\left(\frac{1+\beta}{1-\beta}\right)\\
&{}{-8 \beta \left(12 z^{2} - 8 z - 3\right)}
\end{aligned}\label{eq:charm-equal-q2-11-1}
\end{equation}
\begin{equation}
\begin{aligned}[b]
C_{P_2,11}^{u\mid s\bar s,1,(1)}={}&-\frac{288 z^{2}}{\beta}\,l\;-64 z^{2}\,\Phi(z)\;-48 \beta \left(4 z + 1\right)\,\log z\;+64 \beta \left(4 z + 1\right)\,\log\beta\\
&{}+\frac{48 \cdot \left(2 z - 1\right) \left(24 z^{3} - 6 z^{2} + 1\right)}{\beta^{2}}\,\log\!\left(\frac{1+\beta}{1-\beta}\right)\;+\frac{16 z \left(72 z^{2} - 8 z - 7\right)}{\beta}
\end{aligned}\label{eq:charm-equal-q2-11-2}
\end{equation}
\begin{equation}
\begin{aligned}[b]
C_{P_5,11}^{u\mid s\bar s,1,(1)}={}&-\frac{16 \cdot \left(4 z^{2} + 2 z - 1\right)}{3 \beta}\,l\;+\frac{20 \beta \left(2 z + 1\right)}{9}\,\log z\;-\frac{40 \beta \left(2 z + 1\right)}{9}\,\log\beta\\
&{}-\frac{4 \cdot \left(10 z^{3} - 9 z^{2} - 5\right)}{9}\,\log\!\left(\frac{1+\beta}{1-\beta}\right)\;+\frac{80 z^{3} - 1218 z^{2} - 44 z + 113}{9 \beta}
\end{aligned}\label{eq:charm-equal-q2-11-9}
\end{equation}
\begin{equation}
\begin{aligned}[b]
C_{P_1,11}^{u\mid s\bar s,8,(1)}={}&-\frac{44 \left(2 z^{2} + 2 z - 1\right)}{\beta}\,l\;-\frac{2 \left(92 z^{2} + 74 z - 37\right)}{\beta}\,\log z\;+\frac{4 \left(92 z^{2} + 74 z - 37\right)}{\beta}\,\log\beta\\
&{}-2 \left(6 z^{3} + 9 z^{2} - 37\right)\,\log\!\left(\frac{1+\beta}{1-\beta}\right)\;+\frac{144 z^{3} - 2024 z^{2} - 560 z + 331}{6 \beta}
\end{aligned}\label{eq:charm-equal-q2-11-5}
\end{equation}
\begin{equation}
\begin{aligned}[b]
C_{P_2,11}^{u\mid s\bar s,8,(1)}={}&\frac{144 z^{2}}{\beta}\,l\;+\frac{6 \cdot \left(44 z^{2} - 6 z + 3\right)}{\beta}\,\log z\;-\frac{12 \cdot \left(44 z^{2} - 6 z + 3\right)}{\beta}\,\log\beta\\
&{}-6 \cdot \left(6 z^{3} + z^{2} - 3\right)\,\log\!\left(\frac{1+\beta}{1-\beta}\right)\;+\frac{144 z^{3} + 1292 z^{2} - 260 z - 35}{2 \beta}
\end{aligned}\label{eq:charm-equal-q2-11-6}
\end{equation}
\begin{equation}
\begin{aligned}[b]
C_{P_5,11}^{u\mid s\bar s,8,(1)}={}&\frac{2 \cdot \left(14 z^{2} + 10 z - 5\right)}{\beta}\,l\;-\frac{5 \beta \left(2 z + 1\right)}{3}\,\log z\;+\frac{10 \beta \left(2 z + 1\right)}{3}\,\log\beta\\
&{}+\frac{10 z^{3} - 9 z^{2} - 5}{3}\,\log\!\left(\frac{1+\beta}{1-\beta}\right)\;-\frac{160 z^{3} - 2628 z^{2} + 98 z + 133}{24 \beta}
\end{aligned}\label{eq:charm-equal-q2-11-10}
\end{equation}
\endgroup
\endgroup
\subsection{Semileptonic coefficients with a massive charged lepton}
\label{app:sl-massive-coefficients}
For $c\to q\ell^+\nu_\ell$, with $q=d,s$ and $m_{\nu_\ell}=0$, define
$\rho=m_\ell^2/M^2$.  The result below
is for one charged-lepton species and uses the same operator basis and
fixed prescription as the nonleptonic coefficients, with ${\bf J}_{76}=0$.
Our normalization is
\begin{equation}
\begin{aligned}
 \mathcal T_{7,\ell}^{\mathrm{SL},q}
 &=\frac{G_F^2|V_{cq}|^2M^2}{6\pi}
  \sum_{{\rm c}=1,8}\sum_{i=1}^{5}
  C_{i,\ell}^{{\rm c}}(\rho,\mu)\,P_i^{q,{\rm c}}. 
\end{aligned}
\label{eq:sl-massive-normalization}
\end{equation}
The coefficients are independent of $q$ at this order in the external
soft momentum. The dependence on the mass of $q$ resides in the operators.
With $\boldsymbol C_\ell^{{\rm c},(n)}=(C_{1,\ell}^{{\rm c},(n)},\ldots,C_{5,\ell}^{{\rm c},(n)})$,
the LO coefficients are
\begin{equation}
\begin{aligned}
 \boldsymbol C_\ell^{1,(0)}
  &=\bigl(2(1-\rho^3),\;6\rho^2(1-\rho),\;
       -(1-\rho)^2(1+2\rho),\;0,\;0\bigr),\\
 \boldsymbol C_\ell^{8,(0)}&=(0,0,0,0,0).
\end{aligned}
\label{eq:sl-massive-born}
\end{equation}
The NLO expressions are valid for $0<\rho<1$. Their massless limits are
continuous.
\begingroup
\small
\setlength{\abovedisplayskip}{6pt}
\setlength{\belowdisplayskip}{6pt}
\setlength{\jot}{2pt}
\begin{equation}
\begin{aligned}[b]
C_{1,\ell}^{1,(1)}(\rho)={}&0
\end{aligned}\label{eq:sl-massive-P1-1}
\end{equation}
\begin{equation}
\begin{aligned}[b]
C_{2,\ell}^{1,(1)}(\rho)={}&-48 \left(1-\rho\right) \rho^{2}\,l\;-64 \left(1-\rho\right) \rho^{2}
\end{aligned}\label{eq:sl-massive-P2-1}
\end{equation}
\begin{equation}
\begin{aligned}[b]
C_{3,\ell}^{1,(1)}(\rho)={}&4 \left(1-\rho\right)^{2} \cdot \left(2 \rho + 1\right)\,l\;+6 \left(1-\rho\right)^{2} \left(2 \rho + 1\right)
\end{aligned}\label{eq:sl-massive-P3-1}
\end{equation}
\begin{equation}
\begin{aligned}[b]
C_{4,\ell}^{1,(1)}(\rho)={}&\frac{7 \left(1-\rho\right)^{2} \left(2 \rho + 1\right)}{3}
\end{aligned}\label{eq:sl-massive-P4-1}
\end{equation}
\begin{equation}
\begin{aligned}[b]
C_{5,\ell}^{1,(1)}(\rho)={}&\frac{16 \left(1-\rho\right) \left(\rho + 1\right)}{9}\,l\;+\frac{2 \rho^{2} \cdot \left(5 \rho - 13\right)}{9}\,\log\rho\\
&{}-\frac{20 \left(1-\rho\right)^{2} \cdot \left(2 \rho + 1\right)}{27}\,\log(1-\rho)\;-\frac{\left(1-\rho\right) \left(148 \rho^{2} - 204 \rho - 113\right)}{27}
\end{aligned}\label{eq:sl-massive-P5-1}
\end{equation}
\begin{equation}
\begin{aligned}[b]
C_{1,\ell}^{8,(1)}(\rho)={}&\frac{44 \left(1-\rho\right) \left(\rho^{2} + \rho + 1\right)}{3}\,l\;-\frac{\rho^{2} \cdot \left(59 \rho + 9\right)}{3}\,\log\rho\\
&{}-\frac{2 \left(1-\rho\right) \left(28 \rho^{2} + 37 \rho + 37\right)}{3}\,\log(1-\rho)\;+\frac{\left(1-\rho\right) \left(637 \rho^{2} + 382 \rho + 331\right)}{18}
\end{aligned}\label{eq:sl-massive-P1-8}
\end{equation}
\begin{equation}
\begin{aligned}[b]
C_{2,\ell}^{8,(1)}(\rho)={}&24 \left(1-\rho\right) \rho^{2}\,l\;-\rho^{2} \cdot \left(59 \rho - 51\right)\,\log\rho\\
&{}-2 \left(1-\rho\right) \left(28 \rho^{2} + 3 \rho + 3\right)\,\log(1-\rho)\;+\frac{\left(1-\rho\right) \left(459 \rho^{2} - 200 \rho - 35\right)}{6}
\end{aligned}\label{eq:sl-massive-P2-8}
\end{equation}
\begin{equation}
\begin{aligned}[b]
C_{3,\ell}^{8,(1)}(\rho)={}&\frac{4 \left(1-\rho\right) \left(5 \rho^{2} - 4 \rho - 4\right)}{3}\,l\;-\frac{\rho^{2} \cdot \left(47 \rho - 63\right)}{3}\,\log\rho\\
&{}+\frac{22 \left(1-\rho\right)^{2} \cdot \left(2 \rho + 1\right)}{3}\,\log(1-\rho)\;+\frac{\left(1-\rho\right) \left(1169 \rho^{2} - 1003 \rho - 601\right)}{36}
\end{aligned}\label{eq:sl-massive-P3-8}
\end{equation}
\begin{equation}
\begin{aligned}[b]
C_{4,\ell}^{8,(1)}(\rho)={}&-12 \left(1-\rho\right) \left(\rho + 1\right)\,l\;-\rho^{2} \cdot \left(4 \rho - 9\right)\,\log\rho\\
&{}-2 \left(1-\rho\right) \left(2 \rho^{2} - 3 \rho - 3\right)\,\log(1-\rho)\;-\frac{\left(1-\rho\right) \left(90 \rho^{2} + 311 \rho + 155\right)}{12}
\end{aligned}\label{eq:sl-massive-P4-8}
\end{equation}
\begin{equation}
\begin{aligned}[b]
C_{5,\ell}^{8,(1)}(\rho)={}&-\frac{2 \left(1-\rho\right) \left(3 \rho^{2} + 5 \rho + 5\right)}{3}\,l\;-\frac{\rho^{2} \cdot \left(5 \rho - 13\right)}{6}\,\log\rho\\
&{}+\frac{5 \left(1-\rho\right)^{2} \cdot \left(2 \rho + 1\right)}{9}\,\log(1-\rho)\;+\frac{\left(1-\rho\right) \left(578 \rho^{2} - 315 \rho - 133\right)}{72}
\end{aligned}\label{eq:sl-massive-P5-8}
\end{equation}
\endgroup

For a massless charged lepton, $\rho=0$, the NLO coefficients reduce to
\begin{equation}
\begin{aligned}
 \boldsymbol C_\ell^{1,(1)}(0)&=\bigl(0,\ 0,\ 4\,l+6,\ \tfrac{7}{3},\ \tfrac{16}{9}\,l+\tfrac{113}{27}\bigr),\\
 \boldsymbol C_\ell^{8,(1)}(0)&=\bigl(\tfrac{44}{3}\,l+\tfrac{331}{18},\ -\tfrac{35}{6},\ -\tfrac{16}{3}\,l-\tfrac{601}{36},\
   -12\,l-\tfrac{155}{12},\ -\tfrac{10}{3}\,l-\tfrac{133}{72}\bigr).
\end{aligned}
\label{eq:sl-massless}
\end{equation}

\section{Hadronic inputs}
\label{app:hadronic}
\begingroup
\let\section\subsection
\let\subsection\subsubsection
\section{Meson matrix elements and coefficient contraction}
\label{app:meson-literature-bridge}
In both heavy--light colour channels the operators $P_{i,\rm lit}^{\rm QCD}$ of ref.~\cite{Gratrex:2022xpm}, eq.~(2.33), give
those of eq.~\eqref{eq:operators} as
\begin{equation}
 \begin{aligned}
 P_1&=\frac{1}{2M}\bigl(P_{2,\rm lit}^{\rm QCD}+P_{2,\rm lit}^{{\rm QCD}\dagger}\bigr),\qquad
 P_2=-\frac1M\bigl(P_{3,\rm lit}^{\rm QCD}+P_{3,\rm lit}^{{\rm QCD}\dagger}\bigr),\\
 P_3&=\frac1M\bigl(P_{1,\rm lit}^{\rm QCD}+P_{1,\rm lit}^{{\rm QCD}\dagger}\bigr).
 \end{aligned}
 \label{eq:meson-tree-map}
\end{equation}
Expressed through our operators, its eq.~(B.13) has the prefactor $G_F^2M^2/(12\pi)$, half of ours.

The VIA matrix elements, in the
operator order of eq.~\eqref{eq:uv-basis}, are given by
\begin{equation}
\begin{aligned}
 \langle\boldsymbol P\rangle_H^{(0)}&=\frac{B_H}{M}\,\bigl(-\bar\Lambda_q,2\bar\Lambda_q,-2m_q,m_q,\tfrac23(\bar\Lambda_q-m_q),
                 0,0,0,0,\tfrac89(\bar\Lambda_q-m_q)\bigr)^T,\\
 \langle\boldsymbol Q\rangle_H^{(0)}\big|_{M^{-1}}&=\frac{B_H}{M}\,\bigl(0,2(\bar\Lambda_q-m_q),0,0\bigr)^T.
\end{aligned}
\label{eq:meson-via-columns}
\end{equation}
The factor two in $\langle\boldsymbol Q\rangle_H^{(0)}|_{M^{-1}}$ is the descendant $2C_{Q_S}R_2/M$ of ref.~\cite{Gratrex:2022xpm}, appendix~C.
For $P_5$, the heavy-quark EOM gives
\begin{equation}
 P_5^1=\Bigl[\frac iM\,\bar c\overleftrightarrow D^\nu c-2\,\bar c\gamma^\nu c\Bigr](\bar q\gamma_\nu{\rm L}q),\qquad
 \overleftrightarrow D=D-\overleftarrow D,
 \label{eq:p5-gordon}
\end{equation}
and likewise $P_5^8$ with $t^A$ in both bilinears. In VIA, we have
\begin{equation}
\begin{aligned}
 \bigl\langle-2(\bar c\gamma^\nu c)(\bar q\gamma_\nu{\rm L}q)\bigr\rangle_H&=\frac{2\langle Q_S\rangle_H-\langle Q\rangle_H}{N_c},\\
 \Bigl\langle\frac iM(\bar c\overleftrightarrow D^\nu c)(\bar q\gamma_\nu{\rm L}q)\Bigr\rangle_H&=-\frac{\langle Q_S\rangle_H}{N_c}
 +O\Bigl(\frac{\bar\Lambda_q^2}{M^2}\langle Q\rangle_H\Bigr).
\end{aligned}
\end{equation}
With $\langle Q_S\rangle_H=\langle Q\rangle_HM_H^2/(M+m_q)^2$ and $M_H=M+\bar\Lambda_q$, we obtain
\begin{equation}
 \langle P_5^1\rangle_H
 =\frac{\langle Q_S\rangle_H-\langle Q\rangle_H}{N_c}
 =\frac{2(\bar\Lambda_q-m_q)}{N_cM}\langle Q\rangle_H,\qquad
 \langle P_5^8\rangle_H
 =\frac{2C_F(\bar\Lambda_q-m_q)}{N_cM}\langle Q\rangle_H.
 \label{eq:meson-p5-via}
\end{equation}
We set $m_q=0$, so that $P_3$ and $P_4$ vanish.

The dimension-six rows follow from the appendix of ref.~\cite{Franco:2002fc} and, for the semileptonic channel,
appendix~A of ref.~\cite{Lenz:2013aua}, with the transport of eqs.~(15) and (16) of the former at $\mu=M$, apart from
the penguin-like contributions omitted in section~\ref{sec:method}. For a lepton,
$C_{Q_S}^{(1)}=-\frac83[3l+4](1-\rho)^2(1+2\rho)$.

\begin{table}[!tb]
\centering\small
\begin{tabular}{l r r r r}\toprule
 & $\Lambda_c:u,d$ & $\Xi_c:u,d$ & $\Xi_c:s$ & $\Omega_c:s$\\\midrule
$\langle Q\rangle$ & $-5.62^{+0.34}_{-0.61}$ & $-5.54^{+0.33}_{-0.60}$ & $-6.36^{+0.38}_{-0.69}$ & $-36.72^{+2.20}_{-3.99}$\\
$\langle Q_S\rangle$ & $3.40^{+1.17}_{-0.49}$ & $3.35^{+1.15}_{-0.49}$ & $3.86^{+1.33}_{-0.56}$ & $-4.79^{+0.70}_{-1.65}$\\
$\langle T_S\rangle$ & $-1.52^{+0.08}_{-0.04}$ & $-1.50^{+0.08}_{-0.04}$ & $-1.73^{+0.10}_{-0.05}$ & $2.15^{+0.06}_{-0.12}$\\\midrule
$\langle P_1^1\rangle$ & $-1.20^{+0.09}_{-0.16}$ & $-1.18^{+0.09}_{-0.16}$ & $-2.08^{+0.15}_{-0.25}$ & $-11.59^{+0.67}_{-1.06}$\\
$\langle P_2^1\rangle$ & $-1.47^{+0.24}_{-0.53}$ & $-1.45^{+0.24}_{-0.52}$ & $-2.56^{+0.41}_{-0.87}$ & $3.28^{+1.26}_{-0.56}$\\
$\langle P_3^1\rangle$ & $0$ & $0$ & $0.20^{+0.02}_{-0.07}$ & $-1.45^{+0.38}_{-0.93}$\\
$\langle P_4^1\rangle$ & $0$ & $0$ & $-0.46^{+0.10}_{-0.05}$ & $-0.78^{+0.53}_{-0.22}$\\
$\langle P_5^1\rangle$ & $0.28^{+0.01}_{-0.01}$ & $0.27^{+0.01}_{-0.01}$ & $0.52^{+0.07}_{-0.03}$ & $-2.87^{+0.30}_{-0.17}$\\\midrule
$\langle P_1^8\rangle$ & $0.59^{+0.02}_{-0.03}$ & $0.59^{+0.02}_{-0.03}$ & $1.05^{+0.05}_{-0.06}$ & $5.82^{+0.37}_{-0.53}$\\
$\langle P_2^8\rangle$ & $0.51^{+0.06}_{-0.07}$ & $0.51^{+0.05}_{-0.07}$ & $0.91^{+0.10}_{-0.12}$ & $-1.07^{+0.22}_{-0.15}$\\
$\langle P_3^8\rangle$ & $0$ & $0$ & $-0.17^{+0.01}_{-0.02}$ & $0.40^{+0.04}_{-0.10}$\\
$\langle P_4^8\rangle$ & $0$ & $0$ & $0.48^{+0.11}_{-0.05}$ & $1.17^{+0.44}_{-0.18}$\\
$\langle P_5^8\rangle$ & $-0.18^{+0.01}_{-0.00}$ & $-0.18^{+0.01}_{-0.00}$ & $-0.28^{+0.00}_{-0.01}$ & $2.04^{+0.06}_{-0.15}$\\
\bottomrule\end{tabular}
\caption{Baryon matrix elements $\langle O^q\rangle_B$ at $\mu=M=1.59$~GeV in $10^{-3}$~GeV$^3$, for one valence quark $q$, evolved
from $\mu_H=1.0\pm0.2$~GeV with eqs.~\eqref{eq:baryon-QS-evolution} and~\eqref{eq:baryon-P-evolution}, the errors giving the range in $\mu_H$. We use $\langle T \rangle=-\frac23\langle Q \rangle$, and the
$\Omega_c:s$ column counts both strange quarks.}
\label{tab:baryon-missing}
\end{table}
\section{Baryon inputs}
\label{app:baryon-inputs}
Table~\ref{tab:baryon-missing} gives the baryon matrix elements $\langle O\rangle_B=\langle B|O|B\rangle/(2M_B)$ at $\mu=M$, with
$M_B$ the baryon mass, evolved from $\mu_H=1.0\pm0.2$~GeV with eqs.~\eqref{eq:baryon-QS-evolution} and~\eqref{eq:baryon-P-evolution}. At
$\mu_H$, those of $Q$, $Q_S$, $T$, $T_S$ and $P_{1,2,3}$ follow from the bag model of ref.~\cite{Cheng:2023jpz}, appendix~A, with its
energies $E_{u,d}=0.32$~GeV and $E_s=0.50$~GeV. Those of $P_4$ and $P_5$ are new tree overlaps of the same bag
model, with
$(m_{u,d},m_s,m_c)_{\rm bag}=(0,0.279,1.641)$~GeV and radii $(4.86,4.88,4.93)$~GeV$^{-1}$ for $(\Lambda_c,\Xi_c,\Omega_c)$. In $P_{3,4}$, $m_s=\overline m_s(\mu_H)$, run at LO from
$\overline m_s(2~\mathrm{GeV})=0.09$~GeV.
They use the bag spinors and overlaps of ref.~\cite{Cheng:2023jpz}, appendix~A. The derivative in $P_5$ acts on the
complete compact-support spinors, wall included, and its time components cancel. At $\mu_H$ the antisymmetric colour
wavefunction gives $\langle P_i^{q,8}\rangle=-\frac23\langle P_i^{q,1}\rangle$. The evolution changes this ratio, but keeps
$\langle T\rangle=-\frac23\langle Q\rangle$.

\endgroup

\bibliographystyle{JHEP}
\bibliography{refs}
\end{document}